\documentclass[useAMS,usenatbib]{mn2e}
\usepackage{graphicx}
\usepackage{amssymb}
\usepackage{amsmath}
\usepackage{color}
\usepackage{ulem}
\usepackage{pdflscape}
\newcommand{\hii}{H\,{\sc ii}}
\newcommand{\hi}{H\,{\sc i}}
\newcommand{\halpha}{H${\alpha}$}
\newcommand{\km}{km\,s$^{-1}$}
\newcommand{\msolar}{M$_{\odot}$}

\title[Metal-poor gas accretion in UM 461 and Mrk 600] 
{Detecting metal-poor gas accretion in the star-forming dwarf galaxies UM 461 and Mrk 600} %
\author[P. Lagos et al.]{P. Lagos$^{1,2}$\thanks{E-mail: plagos@astro.up.pt}, T. C. Scott$^{1}$, A. Nigoche-Netro$^{3}$, R. Demarco$^{4}$, A. Humphrey$^{1}$, 
\newauthor and P. Papaderos$^{1}$\\
$^{1}$Instituto de Astrof\'isica e Ci\^encias do Espa\c{c}o, Universidade do Porto, CAUP, Rua das Estrelas, 4150-762 Porto, Portugal\\
$^{2}$Centre for Space Research, North-West University, Potchefstroom 2520, South Africa\\
$^{3}$Instituto de Astronom\'{\i}a y Meteorolog\'{\i}a, Av, Vallarta 2602. Col. Arcos Vallarta. Guadalajara, Jalisco. C.P. 
44130 M\'exico\\
$^{4}$Department of Astronomy, Universidad de Concepci\'on, Casilla 160-C, Concepci\'on, Chile\\
}
\begin{document}

\date{Accepted 1988 December 15. Received 1988 December 14; in original form 1988 October 11}

\pagerange{\pageref{firstpage}--\pageref{lastpage}} \pubyear{2002}

\maketitle

\label{firstpage}

\begin{abstract}

Using VIMOS-IFU observations, we study the interstellar medium (ISM) of two star-forming dwarf galaxies, 
UM 461 and Mrk 600. Our aim was to search for the existence of metallicity inhomogeneities 
that might arise from infall of nearly pristine gas feeding ongoing localized star-formation.
The IFU data allowed us to study the impact of external gas accretion 
on the chemical evolution as well as the ionised gas kinematics and morphologies of these
galaxies. Both systems show signs of morphological distortions, including cometary-like morphologies. 
We analysed the spatial variation of 12 + log(O/H) abundances within both galaxies 
using the direct method (T$_e$), the widely applied HII--CHI--mistry code, as well as by employing different standard calibrations.
For  UM 461 our results show that the ISM  is fairly well mixed, at large scales, 
however we find an off-centre and low-metallicity region with 12 + log(O/H) $<$ 7.6 in the SW part of the brightest H\,{\sc ii} region, 
using the direct method. This result is consistent with the recent infall of a low mass metal-poor dwarf or H\,{\sc i} 
cloud into the region now exhibiting the lowest metallicity, which also displays localized perturbed neutral 
and ionized gas kinematics. Mrk 600 in contrast, appears to be chemically homogeneous on both large and small scales. 
The intrinsic differences in the spatially resolved properties of the ISM in our analysed galaxies are
consistent with these systems being at different evolutionary stages.

\end{abstract}


\section{Introduction}\label{sect_intro}

The structure of the starbursting regions and underlying stellar component of star-forming 
dwarf galaxies can provide significant information on the mechanical energy input from
and photoionization by the newly born stars. The distribution of these regions across a galaxy  
can also provide  information on the effect of external interactions or mergers. 
The morphology currently displayed by a dwarf galaxy could have arisen via several alternative 
evolutionary pathways \cite[e.g.][]{Tolstoy2009}. In particular, star-forming dwarf galaxies 
with cometary morphology are commonly observed in high-redshift surveys, such as  
the Hubble Deep Field \cite[HDF; e.g.][]{vandenbergh1996,Straughn06,Windhorst06}. 
This cometary morphology has been  interpreted for high redshift galaxies in the HDF as: 
1) the result of weak tidal interactions; 2) gravitational 
instabilities in gas-rich and turbulent galactic disks in the process of forming \citep{Bournaud2009} and 
3) stream-like accretion of metal-poor gas from the cosmic web \citep[e.g.][]{DekelBirnboim2006,Dekel2009}.
Interestingly, at low redshift, a significant fraction of low-mass ($\sim$10$^8$--10$^9$M$_{\odot}$), 
low-luminosity (10$^7$ $\lesssim$ L/L$_{\odot} \lesssim$ 10$^9$) and low-metallicity 
(Z$_{\odot}$/40 $\lesssim$ Z $\lesssim$ Z$_{\odot}$/3) H\,{\sc ii} or blue compact dwarf (BCD) galaxies also
have cometary or elongated stellar morphologies \citep{Papaderos2008}. 

The study of star-formation feedback and the role played by galaxy interactions
in low redshift dwarfs may offer important insights into galaxy evolution processes in the young Universe.
Recently, some studies \citep[e.g.][]{Sanchez2014,Sanchez2015} highlighted the existence of spatially resolved
chemical inhomogeneities in the ISM of some local H\,{\sc ii}/BCDs and extremely metal-poor\footnote{Defined 
as systems with an 12 + log(O/H) $\la$ 7.6} (XMP) BCD galaxies, which possibly originated 
from the accretion of nearly pristine cold gas. The same mechanism has been invoked by \cite{Creci2010}
to interpret the radial metallicity gradient of massive z $\sim$ 3 galaxies as evidence of accretion of primordial gas, 
which in turn is sustaining the high star-formation activity predicted by cold flow models.
\cite{Sanchez2015} interpret their result, in the case of nearby XMP BCDs, as arising from gas-cloud infall 
from the cosmic web. However, if we use the oxygen abundance (12 + log(O/H)) as a spatially resolved metallicity tracer, 
most of the H\,{\sc ii}/BCD \citep[e.g.][]{Lagos2009,Lagos2012} and XMP \citep[e.g.][]{Lagos2014,Lagos2016,Kehrig2016} 
galaxies studied so far turn out to be chemically homogeneous at large scales ($\sim$0.5 - 1 kpc).
This suggest the presence of global hydro-dynamical effects being responsible for efficient gas transport 
and mixing across the galaxies \citep[e.g.][ and references therein]{LagosPapaderos2013}.
Also, the N/O ratio in most of these galaxies has been found to be homogeneous within the uncertainties. 
Even so, the slight anti-correlation between star-formation and metallicity in the cometary galaxy Tol 65 \citep{Lagos2016} 
indicates that the infall/accretion of metal-poor gas or minor merger/interactions, in the recent past, may have produced 
its moderate abundance gradient and cometary stellar morphology. 
In this sense, \cite{OlmoGarcia2017} argue that if the accretion of metal-poor
gas is fueling the star-formation the metallicity (O/H) of the pre-enriched gas is reduced but it cannot modify
the pre-existing ratio between the metals, then keeping the N/O ratio constant across the ISM.

\begin{figure*}
\includegraphics[width=180mm]{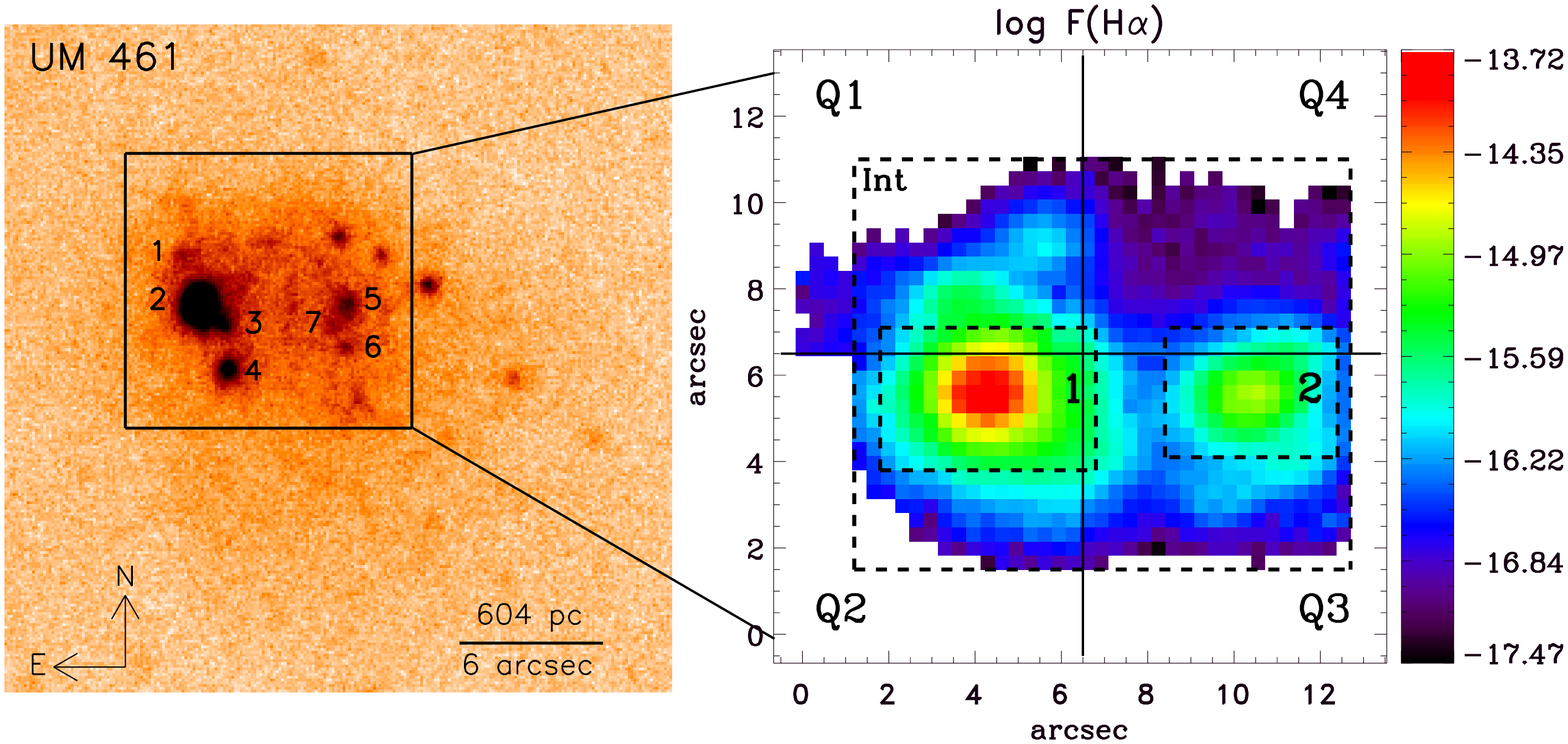}
\includegraphics[width=180mm]{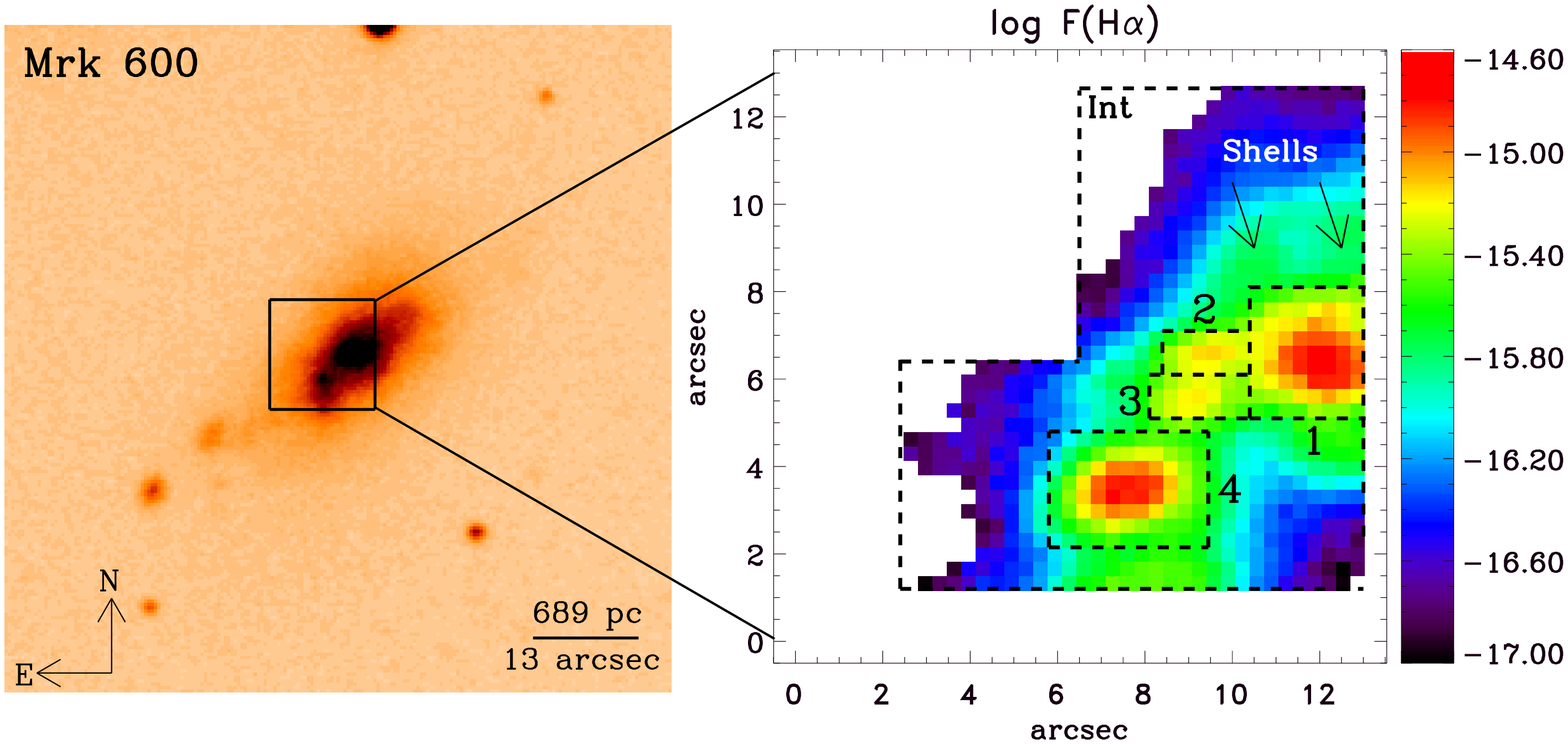}
 \caption{UM\,461 and Mrk 600 optical/NIR images (left) and VIMOS \halpha\ maps (right). 
 \textit{Top left:} UM 461, K$_{p}$  band image \citep{Lagos2011}. The numbers on  image  show the positions 
 of star clusters identified in \citet{Lagos2011} using the notation from that paper. 
 \textit{Bottom left:} Mrk 600, R band image  \citep{GildePaz2003}. \textit{Right :} H$\alpha$ emission line   
 flux (logarithmic scale) maps for  the VIMOS-IFU 13\arcsec $\times$13 \arcsec FoV for each galaxy. 
 The dotted black lines on the images indicate the apertures used in our analysis; regions 1--2 for UM 461, 1--4 in Mrk 600 
 and the integrated (Int) ones. For the UM 461 map the VIMOS  CCD quadrants (Q1$\ldots$Q4) are overlaid on image.  
 Further details are given in Section \ref{sect_obser_reduc}. The H$\alpha$ fluxes are in units of ergs cm$^{-2}$ s$^{-1}$.}
\label{figure_image_campo_Halpha}
\end{figure*}

Here we present new Very Large Telescope (VLT) VIsible MultiObject Spectrograph \citep[VIMOS;][]{LeFevre2003}
observations of two star-forming dwarf galaxies using the integral field unit (IFU)
spectroscopy mode (hereafter VIMOS-IFU).
UM\,461 (upper panel in Figure \ref{figure_image_campo_Halpha}) is a well studied H\,{\sc ii}/BCD galaxy
\citep[e.g.][]{Taylor1995,vanZee1998,Lagos2011}. 
This galaxy has been described as formed by two compact and off-centre giant \hii\  regions (GH\,{\sc ii}R), 
some smaller star-forming regions spread across the galaxy disk and an external stellar envelope that is strongly 
skewed  towards the south-west \citep{Lagos2011}. 
It has been classified as having a cometary-like morphology with
an integrated subsolar metallicity of 12 + log(O/H) = 7.73 - 7.78 \citep{Masegosa1994,IzotovThuan1998,Perez-MonteroDiaz2003}. 
As in most H\,{\sc ii}/BCD galaxies, UM 461 has an underlying component of old stars 
\citep{TellesTerlevich1997,Lagos2011} that exhibits an elliptical outer morphology.
Deep Near-Infrared observations with the Gemini/NIRI camera \citep{Lagos2011} revealed that the
star-formation activity in this galaxy is taking place in several star clusters with  masses typically 
between $\sim$10$^4$ M$_{\odot}$ and $\sim$10$^6$ M$_{\odot}$. 
Figure \ref{figure_image_campo_Halpha} shows the K$_{p}$ band image of UM 461 obtained by \cite{Lagos2011}.
Using the same notation as \cite{Lagos2011}, the main GH\,{\sc ii}R (the brightest one in Figure \ref{figure_image_campo_Halpha}) 
in our study is composed of the star clusters no. 2 and  no. 3, while the faintest one is formed by star clusters 
no. 5, no. 6 and no. 7. \cite{Taylor1995} proposed that the SE tail  in their \hi\  image of \ UM\,461 was formed 
as a result of a tidal interaction with UM 462. However, higher resolution \hi\ maps of UM\,461 by \cite{vanZee1998} 
did not show the extended SE \hi\ tail seen in the Taylor \hi\ map. 
This discrepancy is attributed to solar interference in the Taylor map \citep{vanZee1998}.  
Moreover, the age distribution of the star cluster population in UM 461 indicates 
that the current star burst has began within the last few million years \citep{Lagos2011}. 
This current star burst time scale is too short to realistically be attributed to  a UM\,461/UM\,462 interaction.

Mrk 600 (lower panel in Figure \ref{figure_image_campo_Halpha}) was classified as an iE BCD according 
to the \cite{LooseThuan1986} classification scheme. 
However, the elongated shape and the presence of several fainter regions beyond the main body 
of the galaxy indicate a tadpole or cometary-like stellar morphology. 
The ongoing star-forming activity in this object \citep[see][]{Cairos2001a} is mainly concentrated in the two principal
GH\,{\sc ii}Rs. Spatially resolved colours of those regions \citep{Cairos2001b} are consistent with 
a young starburst. The distribution of these H\,{\sc ii} regions may be the result of a recent interaction 
given the presence of a nearby H\,{\sc i} companion \citep{Taylor1993} as suggested by \cite{Noeske2005}. 
\cite{IzotovThuan1998} and \cite{Guseva2011} derived an integrated oxygen abundance of 12 + log(O/H) = 7.83 - 7.88 for Mrk\,600. 
Basic properties for both galaxies are compiled in Table \ref{table_parameters}.

\begin{table*}
 \centering
 \begin{minipage}{150mm}
  \caption{Basic properties of UM 461 and Mrk 600.}
  \label{table_parameters}
  \begin{tabular}{@{}lccccccrlr@{}}
  \hline

Parameter                            &     Value                  & Reference      \\       
\hline 
UM 461 \\
\hline
RA   (J2000)                         & 11$^{h}$51$^{m}$33.3$^{s}$ &Obtained from NED\\                      
DEC  (J2000)                         & -02$^{o}$22$^{m}$22$^{s}$  &Obtained from NED\\
Distance (3K CMB) Mpc                & 19.2                       &Obtained from NED\\
Pixel scale (pc/arcsec)              & 93                         &Obtained from NED\\
z                                    & 0.003465                   &Obtained from NED\\
E(B-V)$_{Gal}$ mag          & 0.014                      &\cite{SchlaflyFinkbeiner2011}\\
c(H$\beta$)                          & 0.12                       &\cite{IzotovThuan1998}\\
12 + log(O/H)                        & 7.73$\pm$0.03, 7.78$\pm$0.03  &\cite{Masegosa1994,IzotovThuan1998}\\
M$_{*}$ ($\times$10$^8$M$_{\odot}$)  & $\sim$0.76                 &\cite{Lagos2011}\\ 
M$_{HI}$ ($\times$10$^8$M$_{\odot}$) & 0.98, 1.71                 &\cite{Smoker2000},\cite{vanZee1998}\\
\hline 
Mrk 600 \\
\hline
RA   (J2000)                         & 02$^{h}$51$^{m}$04.06$^{s}$ &Obtained from NED\\                      
DEC  (J2000)                         & +04$^{o}$27$^{m}$14$^{s}$   &Obtained from NED\\
Distance (3K CMB) Mpc                & 10.9                        &Obtained from NED\\
Pixel scale (pc/arcsec)              & 53                          &Obtained from NED\\
z                                    & 0.003362                    &Obtained from NED\\
E(B-V)$_{Gal}$ mag                   & 0.058                       &\cite{SchlaflyFinkbeiner2011}\\
c(H$\beta$)                          & 0.24, 0.225                 &\cite{IzotovThuan1998,Guseva2011}\\
12 + log(O/H)                        & 7.83$\pm$0.01, 7.88$\pm$0.01&\cite{IzotovThuan1998,Guseva2011}\\
M$_{*}$ ($\times$10$^8$M$_{\odot}$)  & $\sim$0.64                  &\cite{ZhaoGaoGu2013}\\ 
M$_{HI}$ ($\times$10$^8$M$_{\odot}$) & 2.68                        &\cite{Smoker2000}\\
\hline
\end{tabular}
\end{minipage}
\end{table*}

In this paper, we investigate the relation between the properties and structure of the ISM and the 
star-formation activity in the H\,{\sc ii}/BCD galaxies UM\,461 and Mrk\,600 using VIMOS-IFU spectroscopy. 
Single aperture spectroscopic observations often suffer from limited spatial sampling 
and incomplete coverage. In contrast, IFU observations cover a large fraction of the ISM, allowing us to spatially 
resolve the presence of metallicity inhomogeneities \citep[e.g.][]{Creci2010,Monreal2012,Kumari2017}. 
The paper is organized as follows: Section 2 contains the technical details regarding the data reduction
and measurement of line fluxes; Section 3 describes the structure as well as the physical and kinematic properties 
of the ionized gas; Section 4 discusses the results. Finally, in Section 5 we summarise our conclusions.

\section[]{Observations, data reduction and emission line measurement}\label{sect_obser_reduc}

\subsection{Observations and data reduction}
The observations were obtained using VIMOS-IFU on the 8.2 m VLT UT3/Melipal telescope in Chile, 
using the new high resolution blue (HRB; 0.71 $\rm \AA$ pixel$^{-1}$) and high resolution 
orange (HRO; 0.62 $\rm \AA$ pixel$^{-1}$) gratings. 
The VIMOS-IFU consists of four CCD quadrants (Q1$\ldots$Q4) covered by a pattern of 1600 elements.
In Figure \ref{figure_image_campo_Halpha} (upper right panel) we show the numbering scheme of those quadrants.
We used a projected size per element of 0.33$\arcsec$, covering a total field of view (FoV) 
of 13$\arcsec\times$13$\arcsec$. The data were obtained at low airmass ($<$1.5) during the nights listed 
in Table~\ref{table_observing_log}. The observations were obtained under clear atmospheric conditions. 
Two science exposures were taken per Observing Blok (OB).
A third dithered exposure of 120 s within each set of OBs was taken, after the observation of 
each target, in order to obtain a night sky background exposure. 
One arc-line and three flat-field calibration frames were taken for every OB.
All observations were obtained with a rotator angle of PA = 0. 

\begin{table}
 \centering
 \begin{minipage}{140mm}
  \caption{Observing log.}
  \label{table_observing_log}
  \begin{tabular}{@{}lcccccccccc@{}}
\hline
 Grating &OB & Date      &Exp. time    & Airmass\footnote{Mean of the starting and ending value of the  exposures} & Seeing\footnote{Mean value during the observation} \\
         &   &           & (s)         &                                                     &             ($\arcsec$)                                                   \\
\hline
 UM 461 \\
\hline
HRO      & 1 & 2013-01-24&2$\times$932  & 1.271-1.233  & 0.83\\ 
         & 2 & 2013-02-21&              & 1.335-1.289  & 0.94\\ 
         & 3 & 2013-01-24&              & 1.091-1.084  & 0.52\\ 
HRB      & 1 & 2013-01-24&              & 1.085-1.090  & 0.56\\
         & 2 & 2013-02-10&              & 1.163-1.185  & 0.70\\
         & 3 & 2013-03-16&              & 1.087-1.092  & 0.68\\     
\hline
Mrk 600 \\
\hline
HRO      & 1 & 2012-10-08&2$\times$932  & 1.157-1.145  & 1.00\\ 
         & 2 & 2012-10-08&              & 1.151-1.156  & 1.05\\ 
         & 3 & 2012-10-08&              & 1.207-1.227  & 0.86\\ 
HRB      & 1 & 2012-10-09&              & 1.265-1.295  & 0.74\\
         & 2 & 2012-10-18&              & 1.471-1.530  & 0.90\\
         & 3 & 2012-11-07&              & 1.303-1.338  & 0.65\\     
\hline
\end{tabular}
\end{minipage}
\end{table}

\begin{figure*}
\includegraphics[width=150mm]{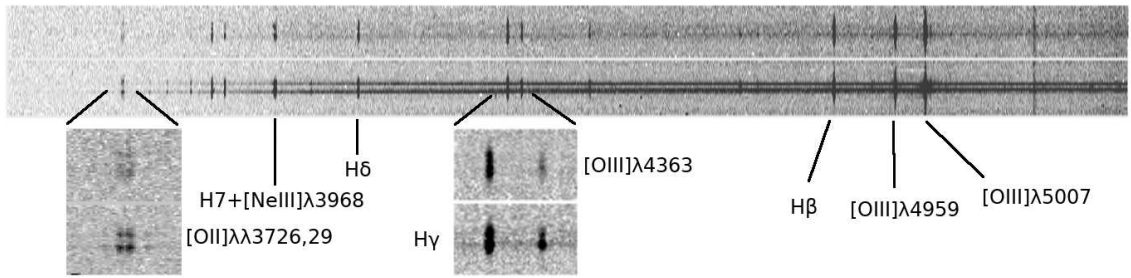}
 \caption{Reduced central block of fibres of quadrant Q2 for the observation of UM 461, OB HRB 3 
 (see Table \ref{table_observing_log}). We labelled the most important emission lines.}     
  \label{figure_2D_fibers_blocks}
\end{figure*}

The data reduction was carried out using the ESOREX software, version 3.10.2. 
This included bias subtraction, flat-field correction, spectra extraction, wavelength and flux calibration.
The master bias was created using the recipe \textit{vmbias}. The spectral extraction mask, wavelength calibration
and the relative fibre transmission correction were obtained, for each quadrant, using the recipe \textit{vmifucalib}.
The instrumental FWHM resolution was obtained by fitting a single Gaussian to
isolated arc lines in the HRB and HRO wavelength calibrated arc exposures.
We found the resolution to be FWHM = $\sim$2.19$\pm$0.05 $\rm \AA$ ($\sim$133.39 km s$^{-1}$) 
and FWHM = $\sim$1.92$\pm$0.04 $\rm \AA$ ($\sim$88.47 km s$^{-1}$) for HRB and HRO, respectively.
From the HRB observations, we found a mean wavelength variation of Q2, Q3, and Q4 relative 
to quadrant Q1 to be 0.01 $\rm \AA$, 0.02 $\rm \AA$ and 0.02 $\rm \AA$. 
While for the HRO we found mean wavelength variations for Q2, Q3, and Q4 of 0.02 $\rm \AA$, 0.01 $\rm \AA$ 
and 0.01 $\rm \AA$, respectively. In addition, we found, for every quadrant and grating, a standard deviation 
of the centroid of the lines of $\lesssim$0.03 $\rm \AA$, 
which implies velocity uncertainties of $\sim$2 km s$^{-1}$ and $\sim$1 km s$^{-1}$ for the HRB and HRO 
gratings, respectively.
Sky subtraction was performed by averaging the spectra from night sky observations
of the same quadrant and subtracting that scaled spectrum from each spaxel.
Those spectra were properly scaled in order to minimize the residuals. 
Since the sky vary with time this method could not be optimal. However 
we are not interested in the continuum level and the residuals do not affect the measurement
of the spatially resolved emission lines. In Figure \ref{figure_2D_fibers_blocks} we show the 2D
reduced central block of fibres (spectra) in quadrant Q2 for the observation of UM 461 HRB 3.


\begin{figure*}
\includegraphics[width=150mm]{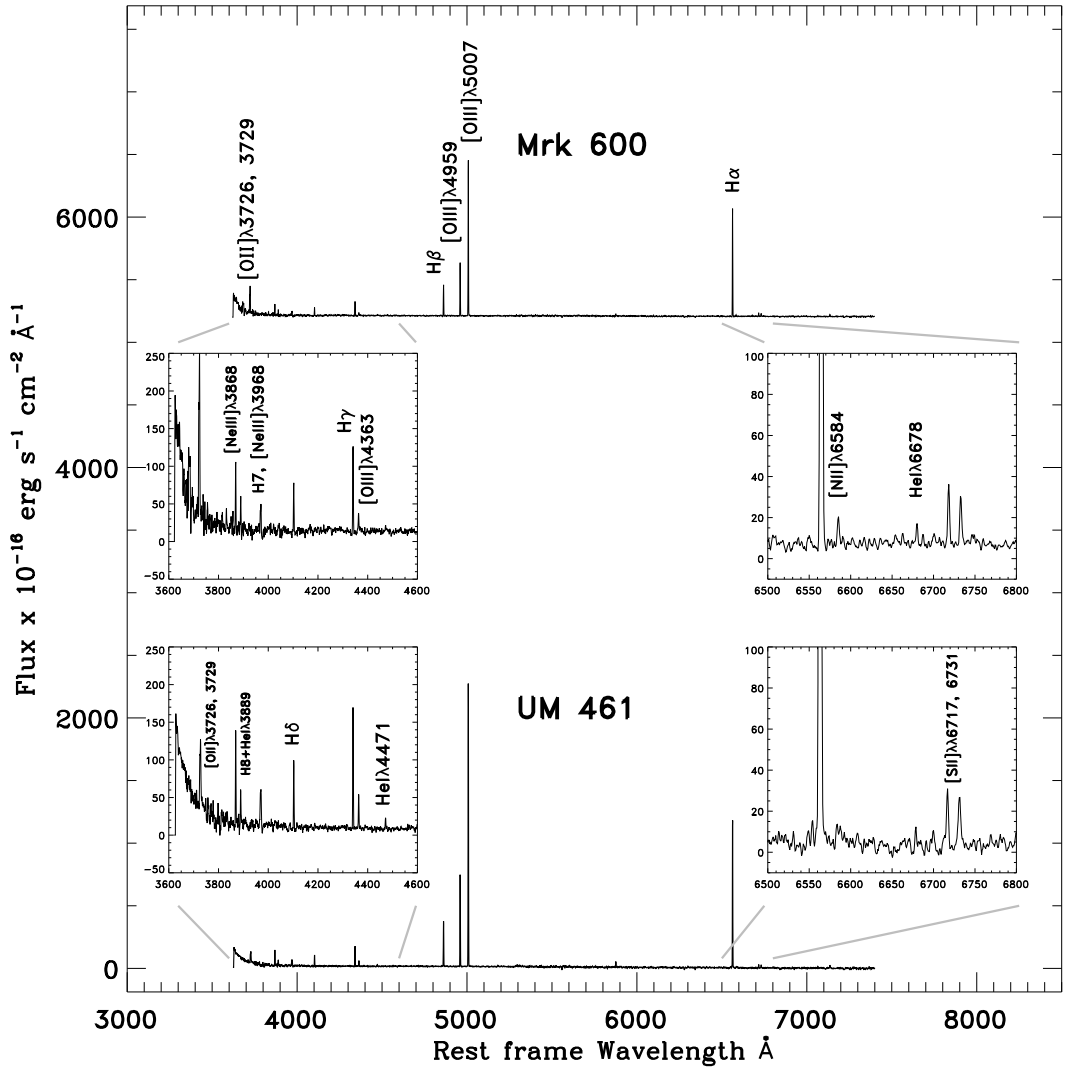}
 \caption{Integrated spectra for UM 461 and Mrk 600, with the inset panels showing wavelength ranges containing  
          important named emission lines.}
\label{figure_integrated_spectra_galaxies}
\end{figure*}

The flux calibration was performed using the sensitivity function derived from
observations of spectrophotometric standard stars included in the VIMOS-IFU calibration plan.
The 2D data images were transformed into 3D data cubes, re-sampled to a 
0.33$\arcsec$ spatial resolution.
We correct for the quadrant-to-quadrant intensity differences following the procedure
applied by \cite{Lagerholm2012} assuming that the intensity
correction is uniform within each quadrant. Therefore, we renormalized the quadrants by comparing 
the intensity levels of the neighbouring pixels at the quadrant borders. 
When comparing the mean intensity value in quadrants Q1, Q3, and Q4 with respect to quadrant Q2 we found
values of $\sim$0.2, $\sim$0.8 and $\sim$0.1, respectively.
We checked the effects of the differential atmospheric refraction (DAR) in each  data cube by calculating
the centroid of the main emission regions in several monochromatic maps. We found in our worst case with an airmass of $\sim$1.5 
(see Table \ref{table_observing_log}) which had an offset of $\sim$2.2 spaxels near [O\,{\sc ii}]$\lambda\lambda$3726,3729. 
While near [O\,{\sc iii}]$\lambda$4363 (the critical emission line to determine O/H abundances)
we found an offset of $\sim$1.41 spaxels. 
Therefore, we applied an IRAF\footnote{Image Reduction and Analysis Facility.}-based script \citep{WalshRoy1990} to correct for DAR.
Finally, we note an offset in the pointing for some of those observations. 
The data cubes obtained using the  HRB and HRO gratings were shifted and combined (using a sigma
clipping algorithm to remove the cosmic rays) into a final data cube covering a useful spectral 
range from $\sim$3700 $\rm\AA$ to $\sim$7400 $\rm\AA$. 
This procedure also remove the dead fibres when averaging exposures.
We scaled the HRB and HRO data cubes by comparing the integrated 
spectra of the galaxies with those obtained by \cite{MoustakasKennicutt2006}.
This is a reasonable method given that there are no telluric lines in common in our HRB and HRO data cubes.
Below, in this Section and Section \ref{sect_results} we find  that the selected emission line ratios and other properties 
derived from the data cubes are in agreement, within the uncertainties, with those found in the literature. 
Finally, in Figure \ref{figure_integrated_spectra_galaxies} we show the integrated spectra for each galaxy, 
obtained summing the spectra from all spaxels within the ``Int"  areas  in Figure \ref{figure_image_campo_Halpha}.

\subsection{Emission line measurement}
Line fluxes were measured, using IRAF tasks \textit{fitprofs} and \textit{splot} 
([O\,{\sc ii}]$\lambda\lambda$3726,3729), from a single Gaussian profile fit to each line. 
The spectral resolution of the VIMOS-IFU observations allowed us to resolve the [O\,{\sc ii}]$\lambda\lambda$3726,3729
(see Figure \ref{figure_2D_fibers_blocks})
with  a $\Delta\lambda$=2.95 $\rm \AA$ separation between the peaks of the lines in the integrated spectrum.
The logarithmic reddening parameter c(H$\beta$) was calculated from the de-reddened raw flux data 
assuming case B \citep{OsterbrockFerland2006} for the Balmer decrement ratio, H$\alpha$/H$\beta$=2.86 at 10,000 K. 
Then, the de-reddened emission line fluxes were calculated as

\begin{equation}\label{c(hb)}
\frac{I(\lambda)}{I(H\beta)}=\frac{F(\lambda)}{F(H\beta)} \times 10^{c(H\beta)f(\lambda)}
\end{equation}\noindent

where I($\lambda$) and F($\lambda$) are the de-reddened flux and observed flux
at a given wavelength, respectively, and f($\lambda$) is the reddening
function given by \cite{Cardelli1989}. In Table \ref{table_integrated_fluxes_UM461_MRK600}
we present for each galaxy the integrated observed F($\lambda$) and corrected emission-line fluxes I($\lambda$) 
relative to H$\beta$ (including uncertainties) multiplied by a factor of 100, 
the observed flux of the H$\beta$ emission line and the extinction coefficient c(H$\beta$). 
In Tables \ref{table_regions_fluxes_UM461} and \ref{table_regions_fluxes_MRK600}
we present the values for the resolved H\,{\sc ii} regions (see Figure \ref{figure_image_campo_Halpha}) 
within UM 461 and Mrk 600, respectively. 

  \begin{table*}
  \centering
 \begin{minipage}{120mm}
  \caption{UM 461 and Mrk 600: observed and de-reddened integrated emission line fluxes. 
  The fluxes are relative to F(H$\beta$)=100.}
  \label{table_integrated_fluxes_UM461_MRK600}
  \begin{tabular}{@{}lccccccccr@{}}
  \hline

 & \multicolumn{2}{c}{UM 461} & \multicolumn{2}{c}{Mrk 600} \\
 &  F($\lambda$)/F(H$\beta$)  & I($\lambda$)/I(H$\beta$)&  F($\lambda$)/F(H$\beta$)  & I($\lambda$)/I(H$\beta$) \\
 \hline
$\left[OII\right] \lambda$3726    & 13.13$\pm$4.37 & 13.53$\pm$6.41  & 42.69$\pm$4.53 & 46.32$\pm$8.64\\
$\left[OII\right] \lambda$3729    & 26.53$\pm$4.49 & 27.33$\pm$6.47  & 86.15$\pm$9.15 & 93.46$\pm$17.44\\
$\left[Ne III\right] \lambda$3868 & 33.48$\pm$1.97 & 34.39$\pm$2.86  & 31.61$\pm$3.04 & 34.03$\pm$4.6\\
H8+He I $\lambda$3889             & 13.79$\pm$0.81 & 14.16$\pm$1.18  & 14.72$\pm$1.42 & 15.83$\pm$2.15\\
$\left[Ne III\right] \lambda$3968 & 11.71$\pm$0.70 & 11.00$\pm$1.74  & 14.08$\pm$0.23 & 15.06$\pm$0.35\\
H7  $\lambda$3970                 & 10.93$\pm$0.64 & 11.20$\pm$0.93  & 14.85$\pm$1.43 & 15.88$\pm$2.16\\
H$\delta  \lambda$4101            & 23.24$\pm$1.37 & 23.74$\pm$1.98  & 24.57$\pm$2.36 & 26.04$\pm$3.54\\
H$\gamma  \lambda$4340            & 42.15$\pm$2.48 & 42.76$\pm$3.56  & 45.94$\pm$4.88 & 47.80$\pm$7.18\\
$\left[OIII\right] \lambda$4363   & 12.17$\pm$0.68 & 12.34$\pm$0.97  &  8.79$\pm$0.62 &  9.13$\pm$0.91\\
HeI $\lambda$4471                 &  3.49$\pm$0.20 &  3.53$\pm$0.28  & $\cdots$    & $\cdots$\\
H$\beta \lambda$4861              &100.00$\pm$1.78 &100.00$\pm$2.52  &100.00$\pm$1.24 &100.00$\pm$1.75 \\
$\left[OIII\right] \lambda$4959   &204.16$\pm$4.05 &203.68$\pm$5.71  &168.80$\pm$6.41 &167.71$\pm$9.00\\
$\left[OIII\right] \lambda$5007   &643.80$\pm$7.76 &641.56$\pm$10.94 &496.16$\pm$6.68 &491.44$\pm$9.36\\
H$\alpha \lambda$6563             &294.97$\pm$3.62 &286.99$\pm$4.98  &311.56$\pm$7.12 &288.94$\pm$9.34\\
$\left[NII\right] \lambda$6584    &  1.58$\pm$0.76 &  1.54$\pm$1.04  &  2.98$\pm$0.34 &  2.76$\pm$0.44\\
HeI $\lambda$6678                 &  2.50$\pm$0.15 &  2.43$\pm$0.15  &  3.82$\pm$0.37 &  3.53$\pm$0.48\\
$\left[SII\right] \lambda$6717    &  8.18$\pm$0.28 &  7.94$\pm$0.38  & 11.24$\pm$0.49 & 10.37$\pm$0.64\\
$\left[SII\right] \lambda$6731    &  7.16$\pm$0.29 &  7.00$\pm$0.40  & 10.46$\pm$0.45 &  9.65$\pm$0.59\\
                                  &       \\
F(H$\beta$)\footnote{In units of $\times$10$^{-14}$ erg cm$^{-2}$ s$^{-1}$} & 11.53$\pm$0.10   & & 7.69$\pm$0.05   \\
c(H$\beta$)                                                                 & 0.04$\pm$0.04  &  &  0.11 $\pm$0.07 \\
\hline
log($\left[O III\right]\lambda$5007/H$\beta$)            & 0.81$\pm$0.02 & & 0.69$\pm$0.02 \\
log($\left[N II\right]\lambda$6584/H$\alpha$)            &-2.27$\pm$0.69 & &-2.02$\pm$0.19 \\
log($\left[S II\right]\lambda\lambda$6717,6731/H$\alpha$)&-1.28$\pm$0.08 & &-1.16$\pm$0.09 \\
\hline

\end{tabular}
\end{minipage}
\end{table*}

  \begin{table*}
\centering
  \begin{minipage}{115mm}
  \caption{UM 461: observed and de-reddened emission line fluxes for regions no. 1 and no. 2. 
The fluxes are relative to F(H$\beta$)=100.}
  \label{table_regions_fluxes_UM461}
  \begin{tabular}{@{}lccccccccr@{}}
  \hline
 & \multicolumn{2}{c}{Region no. 1} & \multicolumn{2}{c}{Region no. 2} \\
 &  F($\lambda$)/F(H$\beta$)  & I($\lambda$)/I(H$\beta$)&  F($\lambda$)/F(H$\beta$)  & I($\lambda$)/I(H$\beta$) \\
 \hline
$\left[OII\right] \lambda$3726    &  8.42$\pm$0.93  &$\ldots$ & 20.74$\pm$0.55 &$\ldots$\\
$\left[OII\right] \lambda$3729    & 21.37$\pm$1.09  &$\ldots$ & 38.31$\pm$5.10 &$\ldots$\\
$\left[Ne III\right] \lambda$3868 & 32.72$\pm$1.75  &$\ldots$ & 30.48$\pm$1.75 &$\ldots$\\
H8+He I $\lambda$3889             & 12.34$\pm$0.66  &$\ldots$ & 13.27$\pm$0.76 &$\ldots$\\
$\left[Ne III\right] \lambda$3968 & 12.11$\pm$0.65  &$\ldots$ &  9.71$\pm$0.56 &$\ldots$\\
H7  $\lambda$3970                 & 12.34$\pm$0.66  &$\ldots$ & 13.77$\pm$0.79 &$\ldots$\\
H$\delta  \lambda$4101            & 23.87$\pm$1.28  &$\ldots$ & 21.53$\pm$1.24 &$\ldots$\\
H$\gamma  \lambda$4340            & 43.66$\pm$2.34  &$\ldots$ & 41.10$\pm$2.36 &$\ldots$\\
$\left[OIII\right] \lambda$4363   & 13.94$\pm$0.39  &$\ldots$ &  8.87$\pm$0.62 &$\ldots$\\
HeI $\lambda$4471                 &  3.35$\pm$0.18  &$\ldots$ &  2.63$\pm$0.15 &$\ldots$\\
H$\beta \lambda$4861              &100.00$\pm$0.73  &$\ldots$ &100.00$\pm$1.48 &$\ldots$\\
$\left[OIII\right] \lambda$4959   &219.03$\pm$3.04  &$\ldots$ &186.82$\pm$4.37 &$\ldots$\\
$\left[OIII\right] \lambda$5007   &696.32$\pm$4.24  &$\ldots$ &564.41$\pm$6.40 &$\ldots$\\
H$\alpha \lambda$6563             &280.86$\pm$1.52  &$\ldots$ &253.88$\pm$2.91 &$\ldots$\\
$\left[NII\right] \lambda$6584    &  1.16$\pm$0.60  &$\ldots$ & $\ldots$       &$\ldots$\\
HeI $\lambda$6678                 &  2.77$\pm$0.15  &$\ldots$ & $\ldots$       & $\ldots$\\
$\left[SII\right] \lambda$6717    &  4.48$\pm$0.10  &$\ldots$ &  6.24$\pm$0.24 &$\ldots$\\
$\left[SII\right] \lambda$6731    &  3.63$\pm$0.18  &$\ldots$ &  6.20$\pm$0.26 &$\ldots$\\
                                  &       \\
F(H$\beta$)\footnote{In units of $\times$10$^{-14}$ erg cm$^{-2}$ s$^{-1}$} &$ 8.59\pm$0.03   & & 1.02$\pm$0.01 \\
c(H$\beta$)                                                                 & 0.00$\pm$0.02  &  & 0.00$\pm$0.03 \\
\hline
log($\left[O III\right]\lambda$5007/H$\beta$)            & 0.84$\pm$0.01 & & 0.75$\pm$0.01\\
log($\left[N II\right]\lambda$6584/H$\alpha$)            &-2.38$\pm$0.38 & &$\ldots$ \\
log($\left[S II\right]\lambda\lambda$6717,6731/H$\alpha$)&-1.54$\pm$0.04 & &-1.31$\pm$0.06\\
\hline

\end{tabular}
\end{minipage}
\end{table*}

  \begin{table*}
  \centering
 \begin{minipage}{180mm}
  \caption{Mrk 600: observed and de-reddened emission line fluxes for regions no. 1, no. 2, no. 3 and no. 4. 
The fluxes are relative to F(H$\beta$)=100.}
  \label{table_regions_fluxes_MRK600}
  \begin{tabular}{@{}lccccccccr@{}}
  \hline
 & \multicolumn{2}{c}{\scriptsize{Region no. 1}} & \multicolumn{2}{c}{\scriptsize{Region no. 2}} & \multicolumn{2}{c}{\scriptsize{Region no. 3}} & \multicolumn{2}{c}{\scriptsize{Region no. 4}} \\
 &  \scriptsize{F($\lambda$)/F(H$\beta$)}  & \scriptsize{I($\lambda$)/I(H$\beta$)}&  \scriptsize{F($\lambda$)/F(H$\beta$)}  & \scriptsize{I($\lambda$)/I(H$\beta$)} &  \scriptsize{F($\lambda$)/F(H$\beta$)}  & \scriptsize{I($\lambda$)/I(H$\beta$)}
 &  \scriptsize{F($\lambda$)/F(H$\beta$)}  & \scriptsize{I($\lambda$)/I(H$\beta$)}\\
 \hline
\scriptsize{$\left[OII\right] \lambda$3726}    & 18.05$\pm$1.90 & 20.48$\pm$3.05 & 41.41$\pm$4.35 & 47.33$\pm$7.03  & 51.60$\pm$5.41  &$\ldots$ & 34.98$\pm$3.66  & 37.95$\pm$5.62\\
\scriptsize{$\left[OII\right] \lambda$3729}    & 33.93$\pm$3.57 & 38.48$\pm$5.73 & 70.90$\pm$7.44 & 81.01$\pm$12.02 &104.44$\pm$10.96 &$\ldots$ & 72.46$\pm$7.58  & 78.61$\pm$11.63\\
\scriptsize{$\left[Ne III\right] \lambda$3868} & 40.43$\pm$3.85 & 45.31$\pm$6.10 & 22.43$\pm$2.13 & 25.31$\pm$3.40  & 29.04$\pm$2.76  &$\ldots$ & 26.14$\pm$2.48  & 28.14 $\pm$3.77\\
\scriptsize{H8+He I $\lambda$3889}             & 14.78$\pm$1.41 & 16.53$\pm$2.23 & 12.27$\pm$1.16 & 13.82$\pm$1.85  & 17.42$\pm$1.65  &$\ldots$ & 13.51$\pm$1.28  & 14.53 $\pm$1.95\\
\scriptsize{$\left[Ne III\right] \lambda$3968} & 12.55$\pm$0.19 & 13.93$\pm$0.30 &  8.58$\pm$0.13 &  9.58$\pm$0.20  & 11.63$\pm$0.17  &$\ldots$ & 11.35$\pm$0.17  & 12.14 $\pm$0.26\\
\scriptsize{H7  $\lambda$3970}                 & 11.15$\pm$1.06 & 12.37$\pm$1.66 & 11.67$\pm$1.11 & 13.03$\pm$1.75  & 16.51$\pm$1.57  &$\ldots$ & 12.83$\pm$1.21  & 13.72 $\pm$1.83\\
\scriptsize{H$\delta  \lambda$4101}            & 23.98$\pm$2.28 & 26.24$\pm$3.53 & 19.84$\pm$1.88 & 21.82$\pm$2.92  & 25.83$\pm$2.45  &$\ldots$ & 22.90$\pm$2.17  & 24.27 $\pm$3.25\\
\scriptsize{H$\gamma  \lambda$4340}            & 46.47$\pm$4.89 & 49.41$\pm$7.35 &  4.06$\pm$0.43 &  4.33$\pm$0.65  & 47.68$\pm$5.00  &$\ldots$ & 44.82$\pm$4.69  & 46.63 $\pm$6.90\\
\scriptsize{$\left[OIII\right] \lambda$4363}   & 12.00$\pm$0.53 & 12.72$\pm$0.79 &  6.58$\pm$0.45 &  7.00$\pm$0.68  &  7.21$\pm$0.46  &$\ldots$ &  7.78$\pm$0.40  &  8.08 $\pm$0.59\\
\scriptsize{HeI $\lambda$4471}                 &  3.46$\pm$0.33 &  3.62$\pm$0.49 &  3.28$\pm$0.31 &  3.44$\pm$0.46  &  3.19$\pm$0.30  &$\ldots$ &  2.97$\pm$0.28  &  3.06 $\pm$0.41\\
\scriptsize{H$\beta \lambda$4861}              &100.00$\pm$1.03 &100.00$\pm$1.46 &100.00$\pm$0.98 &100.00$\pm$1.38  &100.00$\pm$0.98  &$\ldots$ &100.00$\pm$0.94  &100.00$\pm$1.33\\
\scriptsize{$\left[OIII\right] \lambda$4959}   &213.15$\pm$6.00 &211.02$\pm$8.40 &144.81$\pm$4.86 &143.28$\pm$6.80  &146.61$\pm$4.78  &$\ldots$ &163.64$\pm$4.01  &162.58$\pm$5.63\\
\scriptsize{$\left[OIII\right] \lambda$5007}   &631.97$\pm$3.89 &622.70$\pm$5.42 &432.27$\pm$3.18 &425.56$\pm$4.43  &427.79$\pm$5.51  &$\ldots$ &489.78$\pm$4.89  &485.12$\pm$6.85\\
\scriptsize{H$\alpha \lambda$6563}             &325.87$\pm$7.88 &290.04$\pm$9.92 &328.54$\pm$7.12 &292.42$\pm$9.34  &241.73$\pm$5.85  &$\ldots$ &311.14$\pm$6.75  &288.55$\pm$8.85\\
\scriptsize{$\left[NII\right] \lambda$6584}    &  1.59$\pm$0.24 &  1.41$\pm$0.30 &  3.17$\pm$0.27 &  2.80$\pm$0.34  &  2.61$\pm$0.26  &$\ldots$ &  2.50$\pm$0.24  &  2.32$\pm$0.31\\
\scriptsize{HeI $\lambda$6678}                 &  3.12$\pm$0.30 &  2.76$\pm$0.37 &  3.02$\pm$0.29 &  2.65$\pm$0.36  &  2.59$\pm$0.25  &$\ldots$ &  3.63$\pm$0.34  &  3.35$\pm$0.44\\
\scriptsize{$\left[SII\right] \lambda$6717}    &  7.07$\pm$0.33 &  6.24$\pm$0.41 & 14.35$\pm$0.43 & 12.58$\pm$0.53  & 12.29$\pm$0.43  &$\ldots$ & 11.80$\pm$0.36  &  10.89$\pm$0.47\\
\scriptsize{$\left[SII\right] \lambda$6731}    &  5.15$\pm$0.29 &  4.54$\pm$0.36 &  9.48$\pm$0.35 &  8.30$\pm$0.43  &  9.75$\pm$0.35  &$\ldots$ &  9.65$\pm$0.32  &   8.90$\pm$0.42\\
                                  &       \\
\scriptsize{F(H$\beta$)}\footnote{In units of $\times$10$^{-14}$ erg cm$^{-2}$ s$^{-1}$} &2.12$\pm$0.01 & & 0.57$\pm$0.01 & & 0.64$\pm$0.01 & & 2.00$\pm$0.01  \\
\scriptsize{c(H$\beta$)}                                                                 &0.17$\pm$0.07 & & 0.18$\pm$0.06 & & 0.00$\pm$0.07 & & 0.11$\pm$0.06\\
\hline
\scriptsize{log($\left[O III\right]$)\footnote{log($\left[O III\right]\lambda$5007/H$\beta$)}}           & 0.79$\pm$0.01 & & 0.63$\pm$0.01 & & 0.63$\pm$0.01 & & 0.68$\pm$0.01\\
\scriptsize{log($\left[N II\right]$)\footnote{log($\left[N II\right]\lambda$6584/H$\alpha$)}}            &-2.31$\pm$0.25 & &-2.02$\pm$0.15 & &-1.97$\pm$0.12 & &-2.09$\pm$0.16\\
\scriptsize{log($\left[S II\right]$)\footnote{log($\left[S II\right]\lambda\lambda$6717,6731/H$\alpha$)}}&-1.43$\pm$0.10 & & -1.15$\pm$0.08 & &-1.04$\pm$0.06 & &-1.16$\pm$0.08\\

\hline

\end{tabular}
\end{minipage}
\end{table*}

\section{Results}\label{sect_results}

\subsection[]{Emission lines, morphology and emission-line ratios}\label{sect_emission_lines}

\subsubsection{Emission lines and morphology}
\begin{figure*}
\includegraphics[width=58mm]{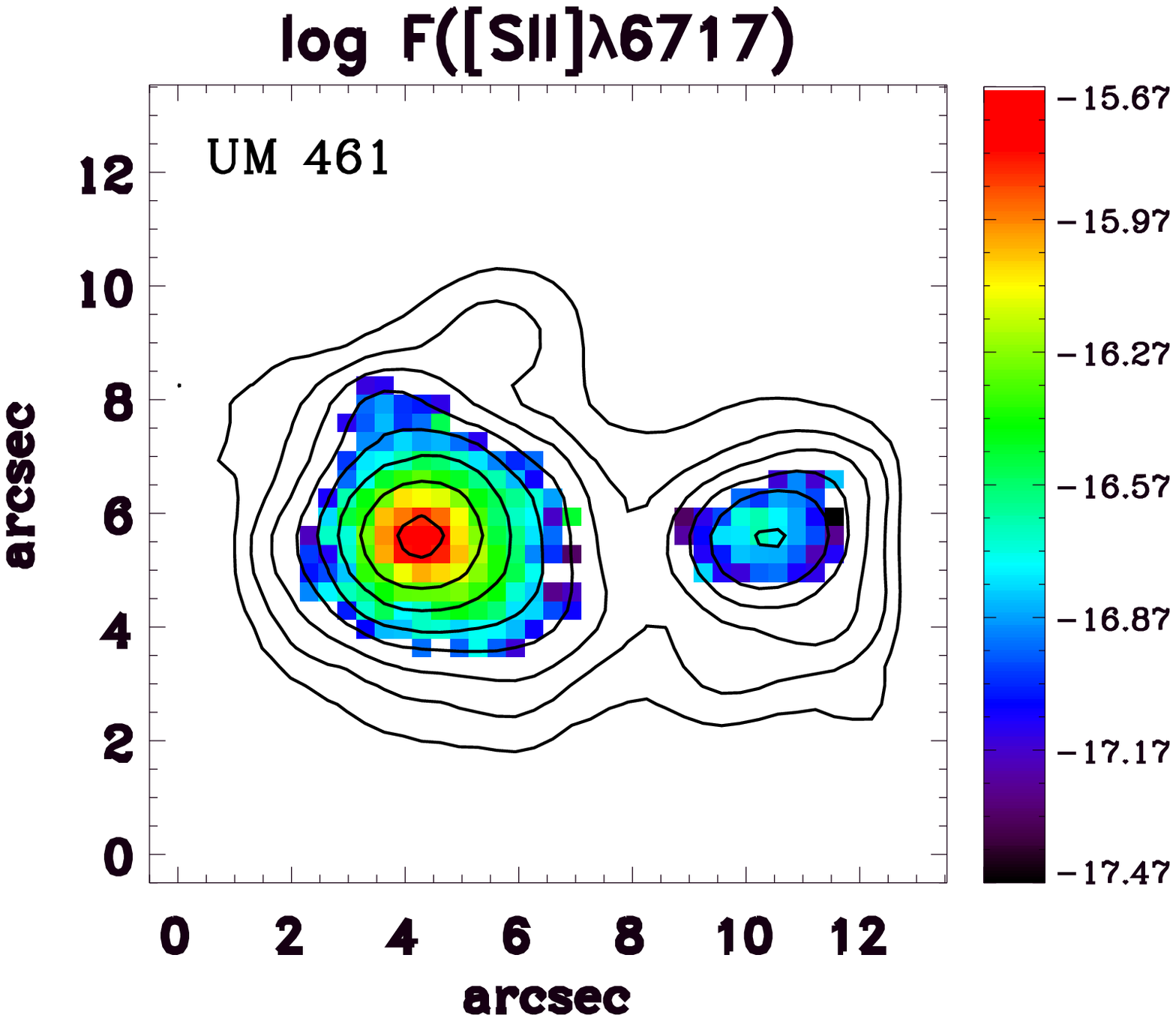}
\includegraphics[width=58mm]{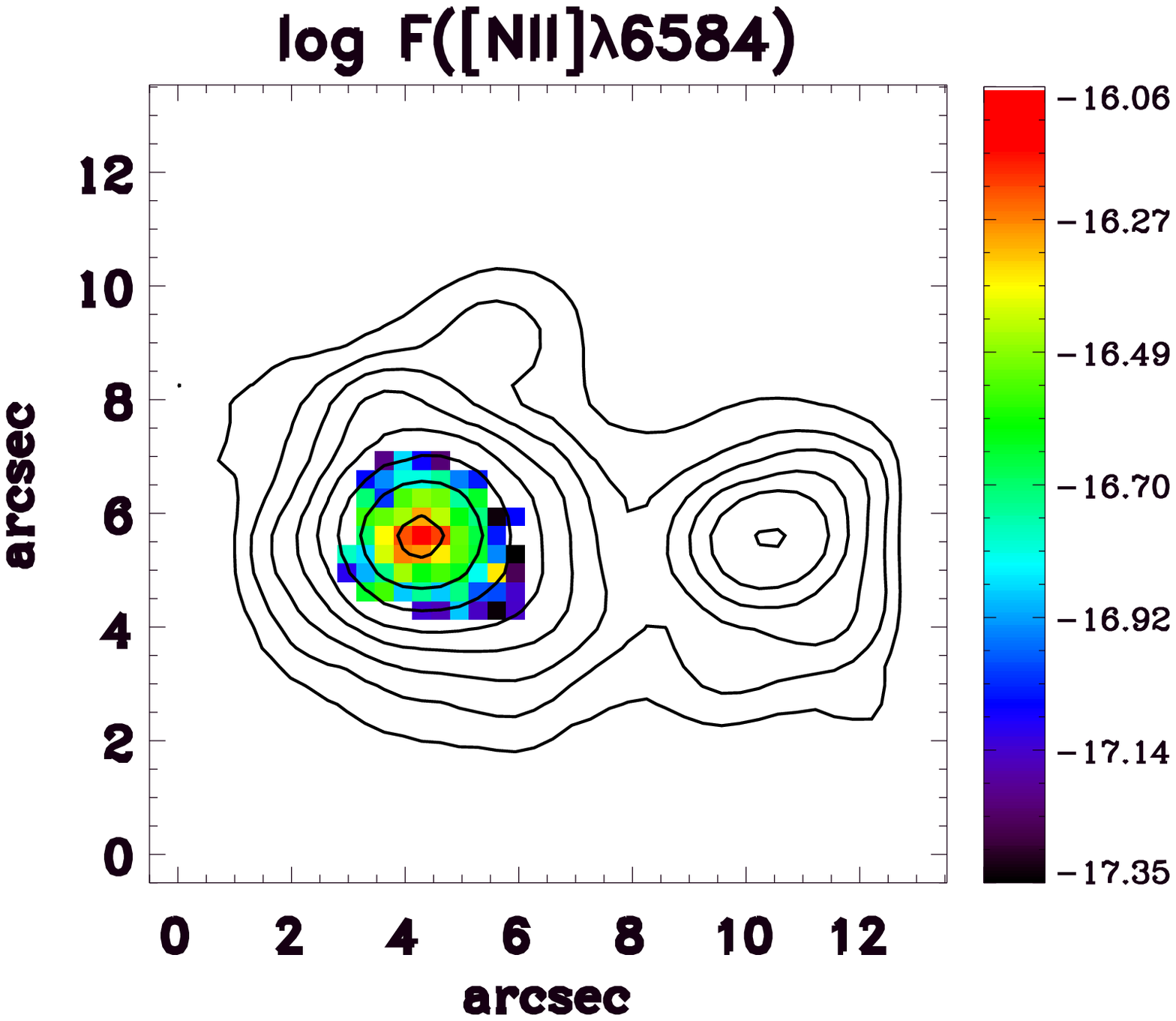}
\includegraphics[width=58mm]{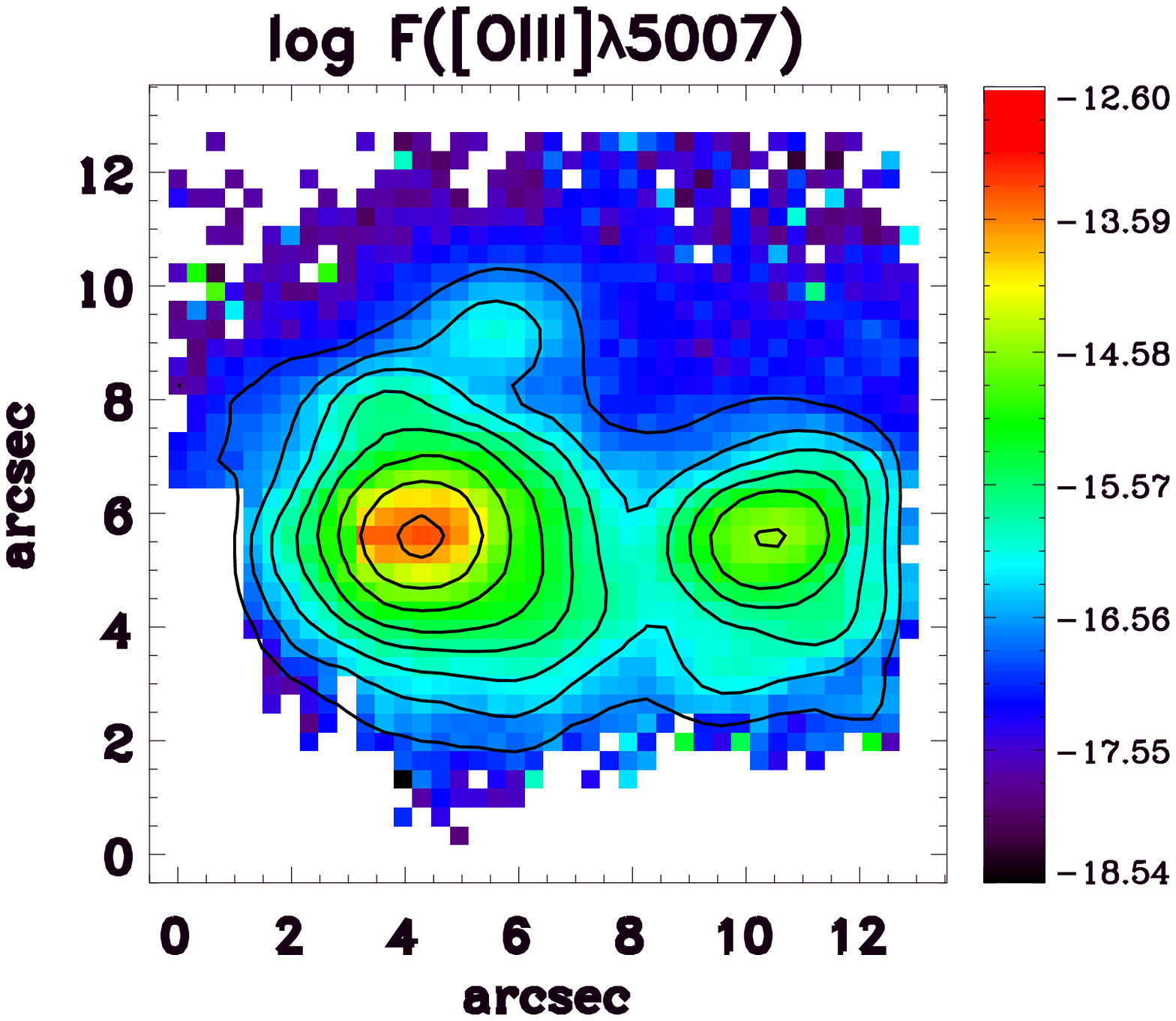}\\
\includegraphics[width=58mm]{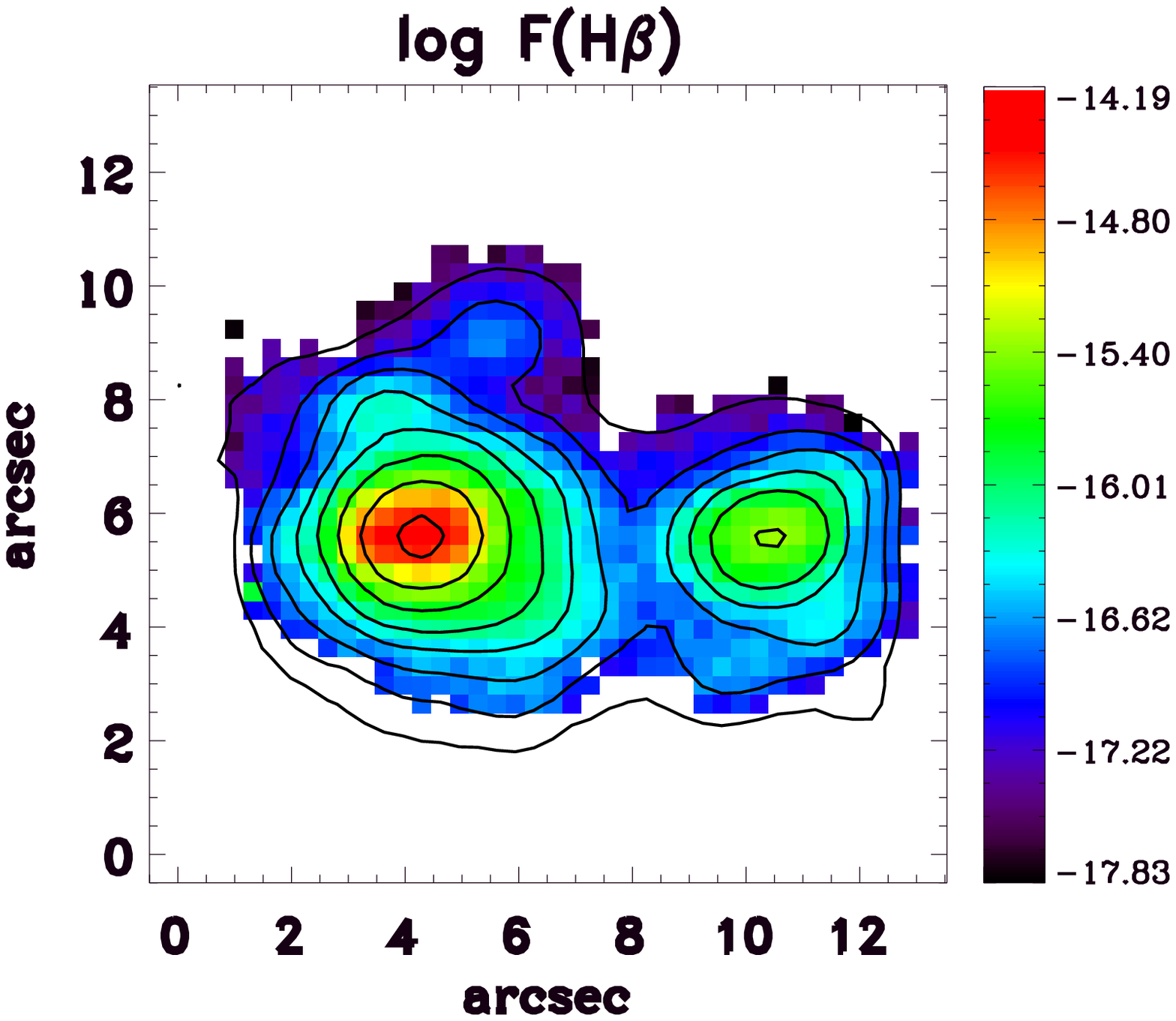}
\includegraphics[width=58mm]{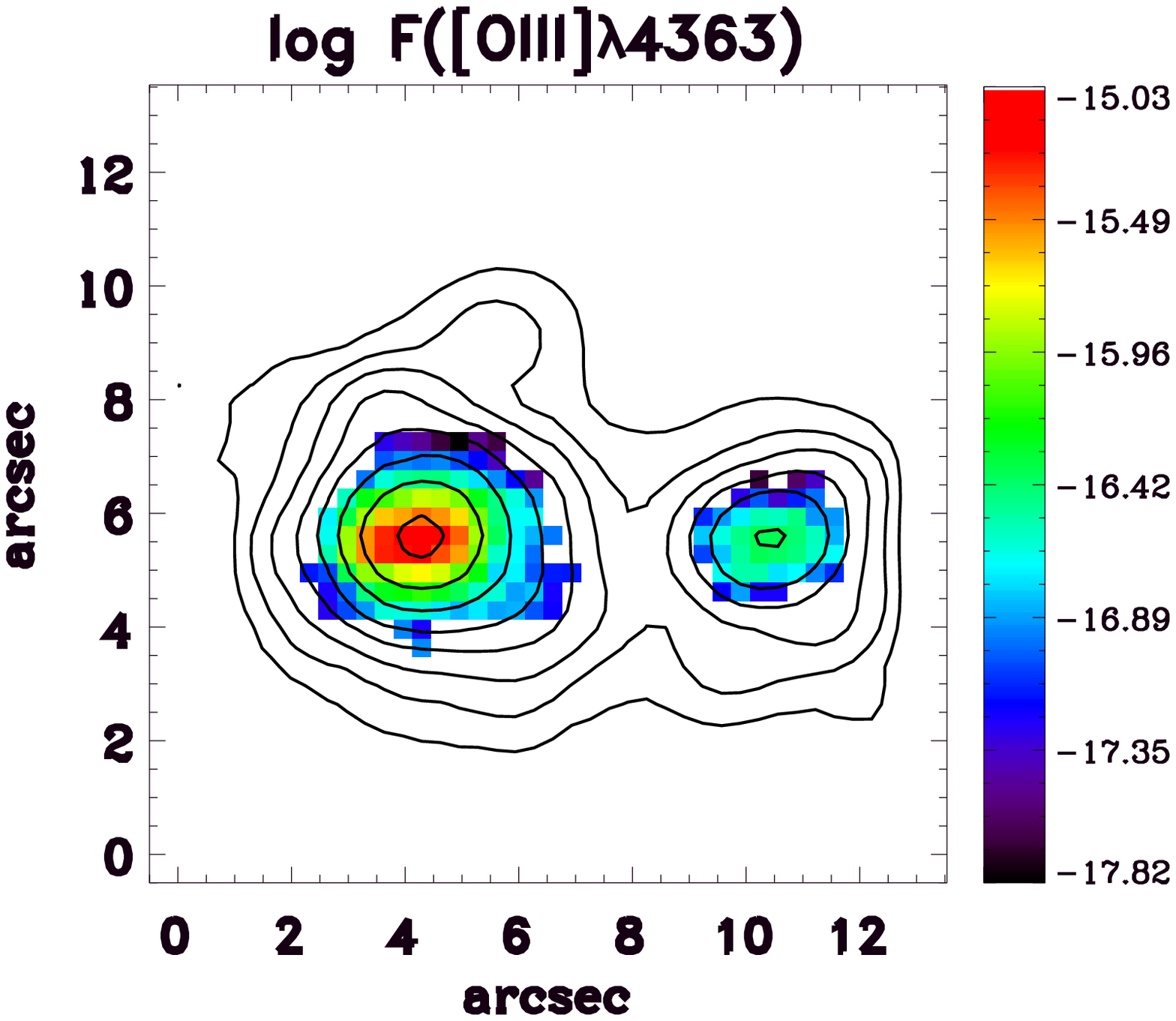}
\includegraphics[width=58mm]{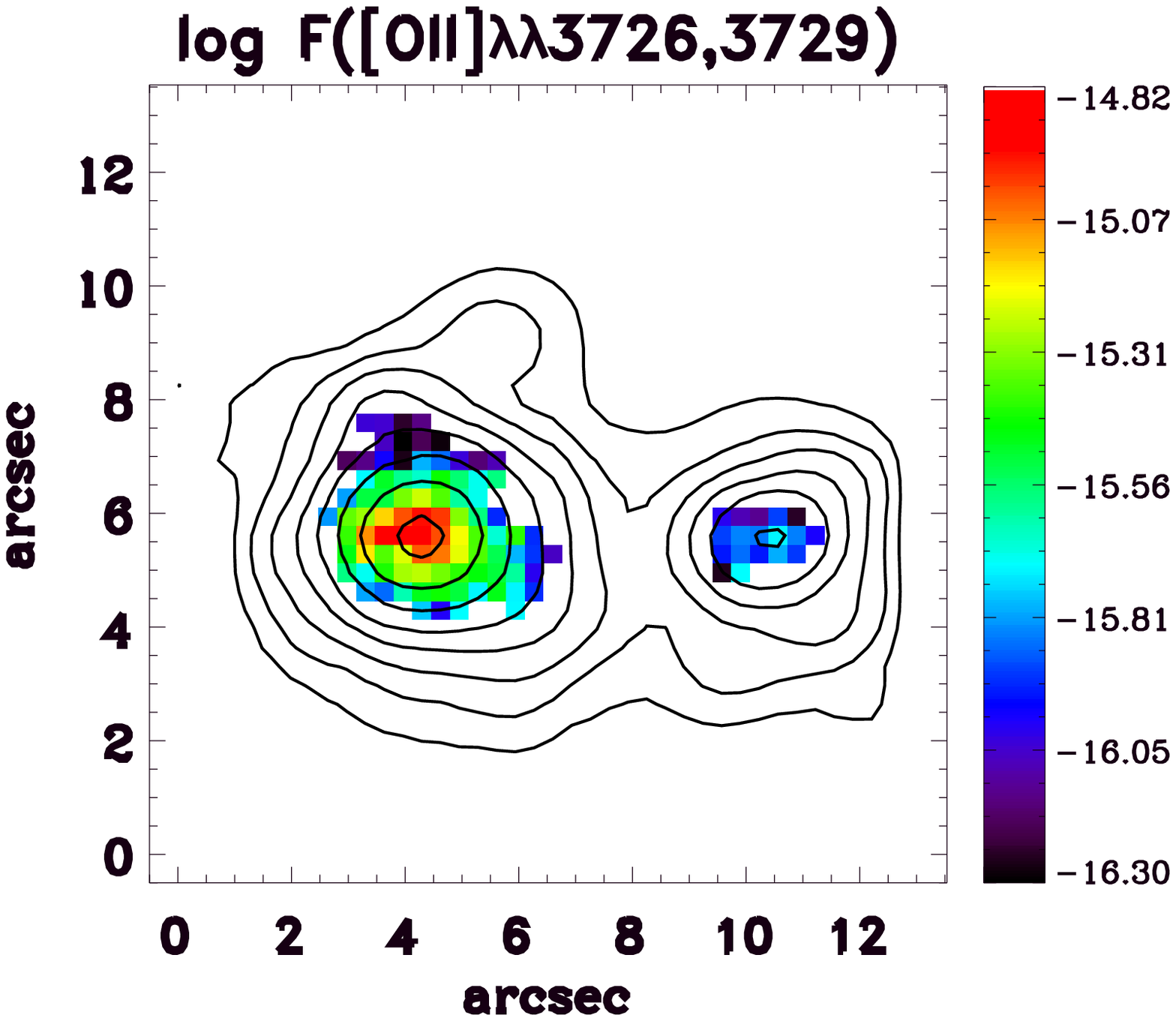}
  \caption{UM 461 - Emission line maps: [S II]$\lambda$6717, [N II]$\lambda$6584, [O III]$\lambda$5007, H$\beta$,
  [O III]$\lambda$4363 and [O II]$\lambda\lambda$3726,3729. 
  H$\alpha$ emission line contours are overlaid on each map. North is up and east is to the left.
}
\label{figure_emission_lines461}
\end{figure*}

\begin{figure*}
\includegraphics[width=58mm]{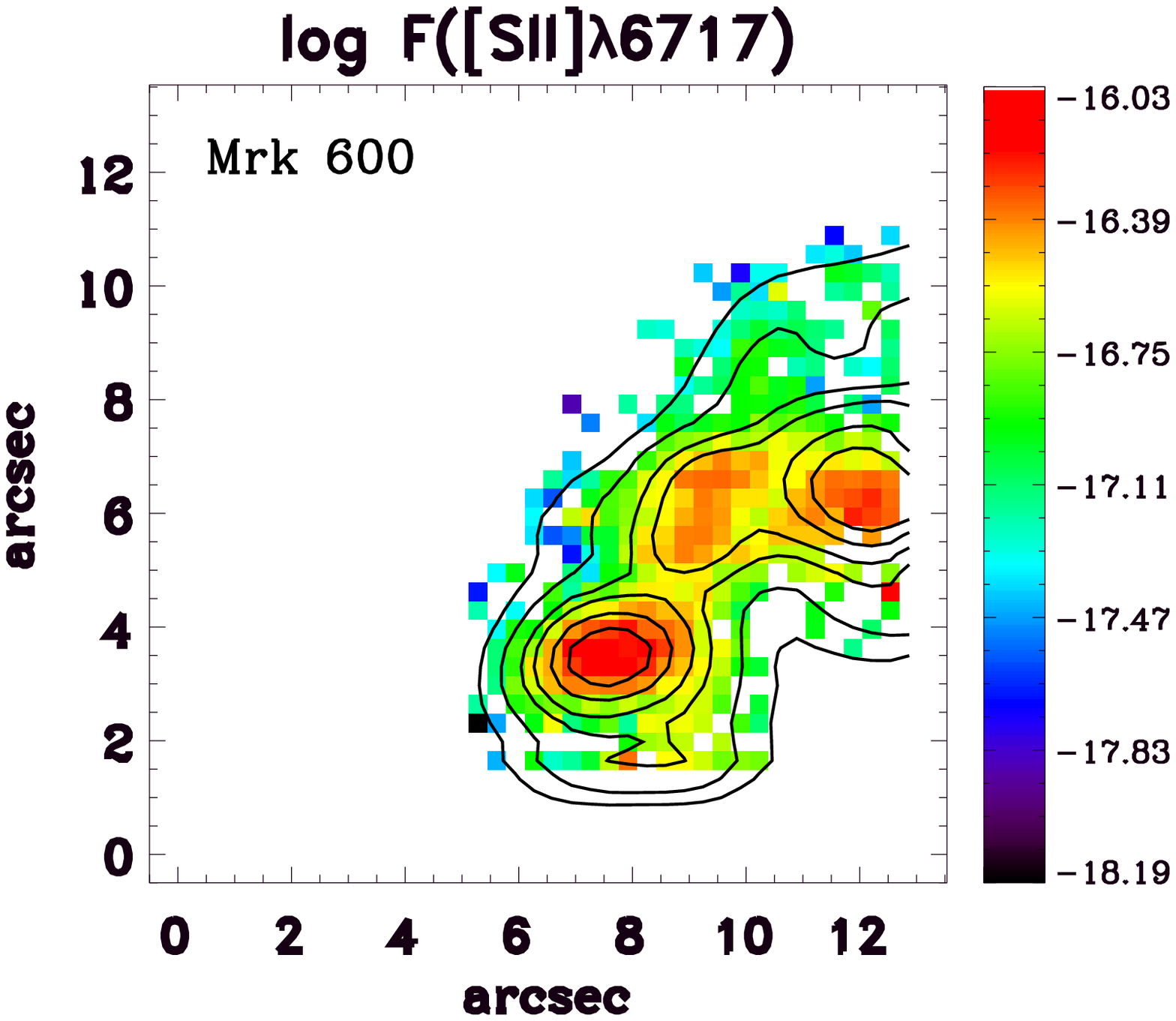}
\includegraphics[width=58mm]{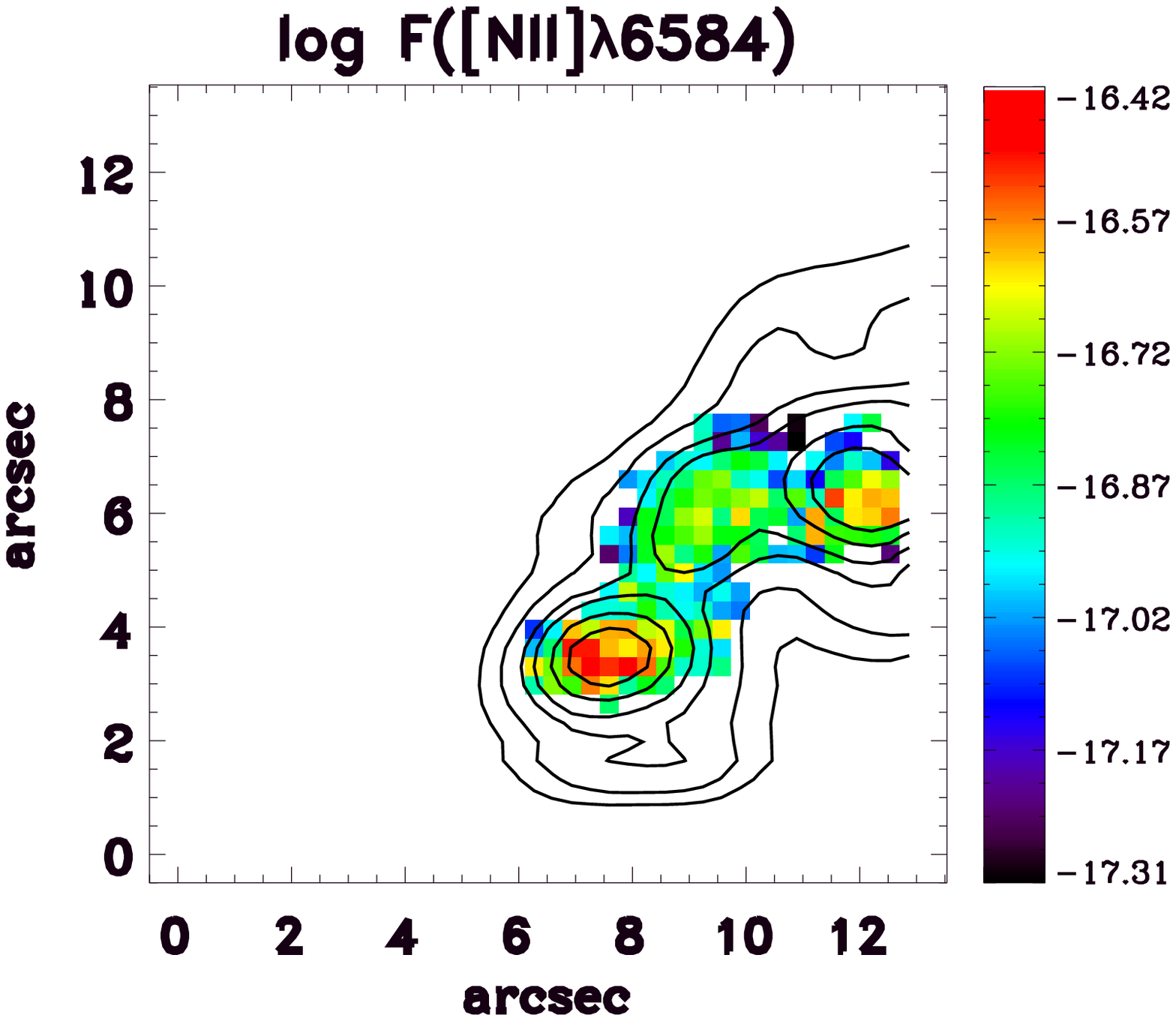}
\includegraphics[width=58mm]{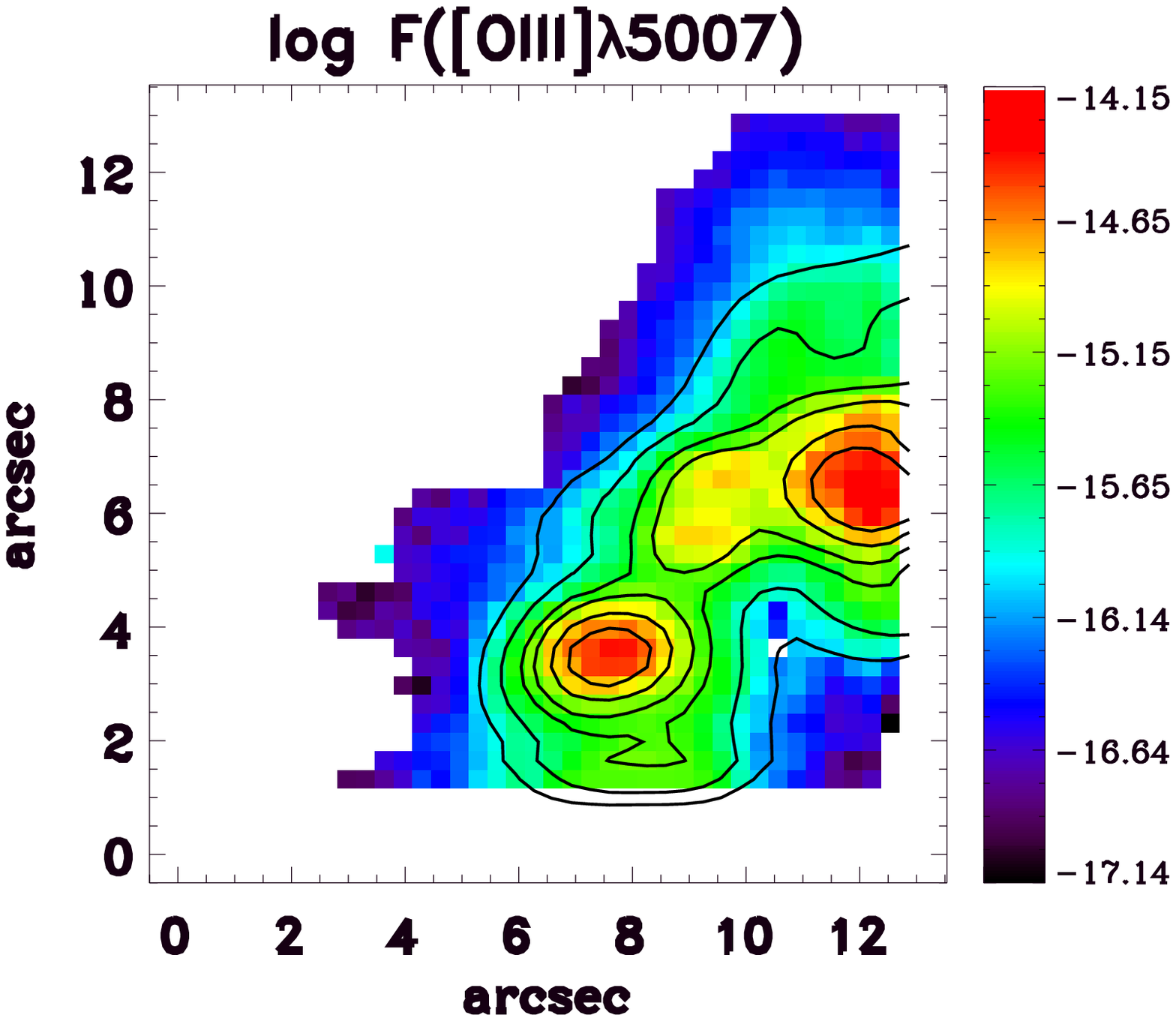}\\
\includegraphics[width=58mm]{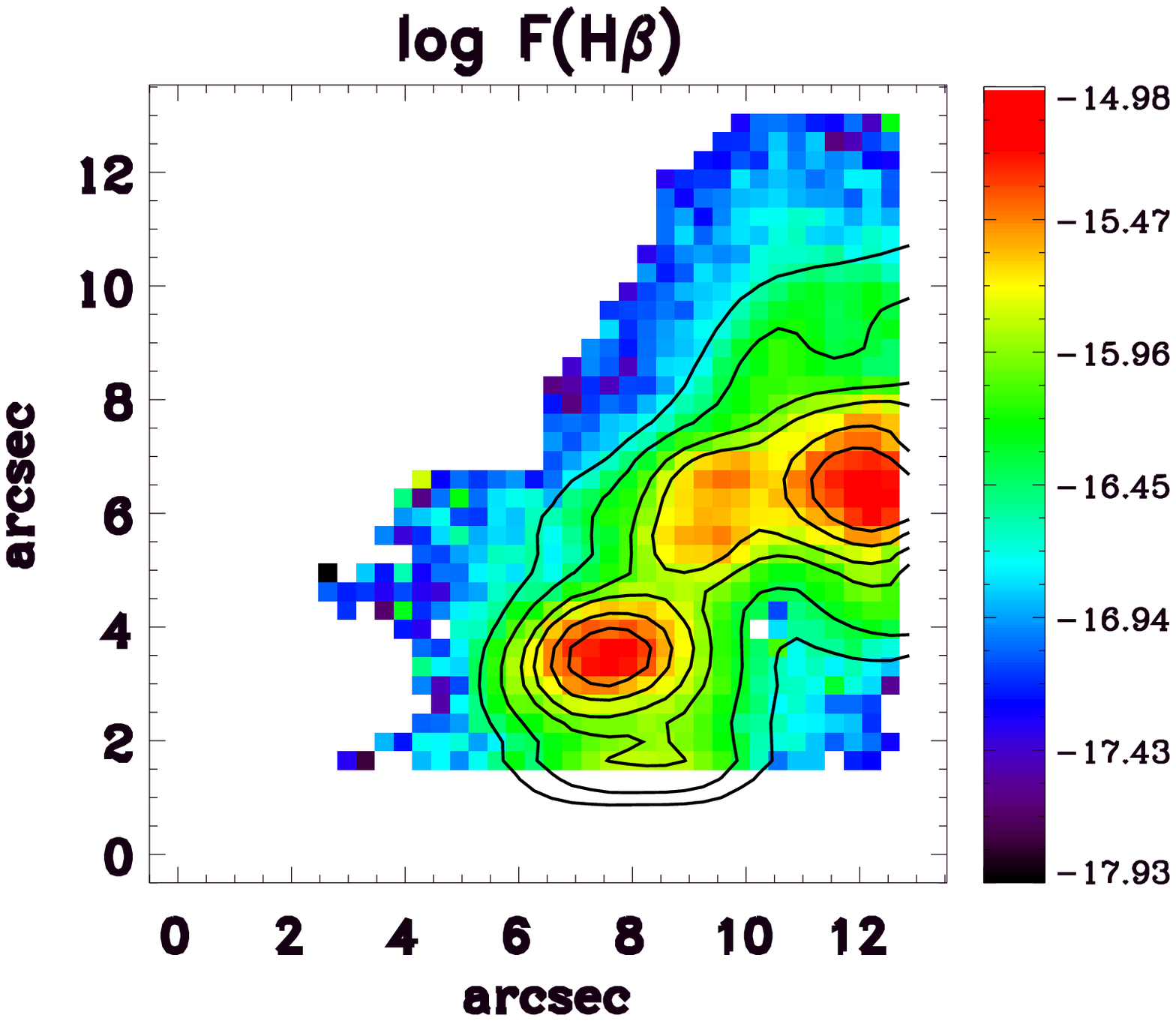}
\includegraphics[width=58mm]{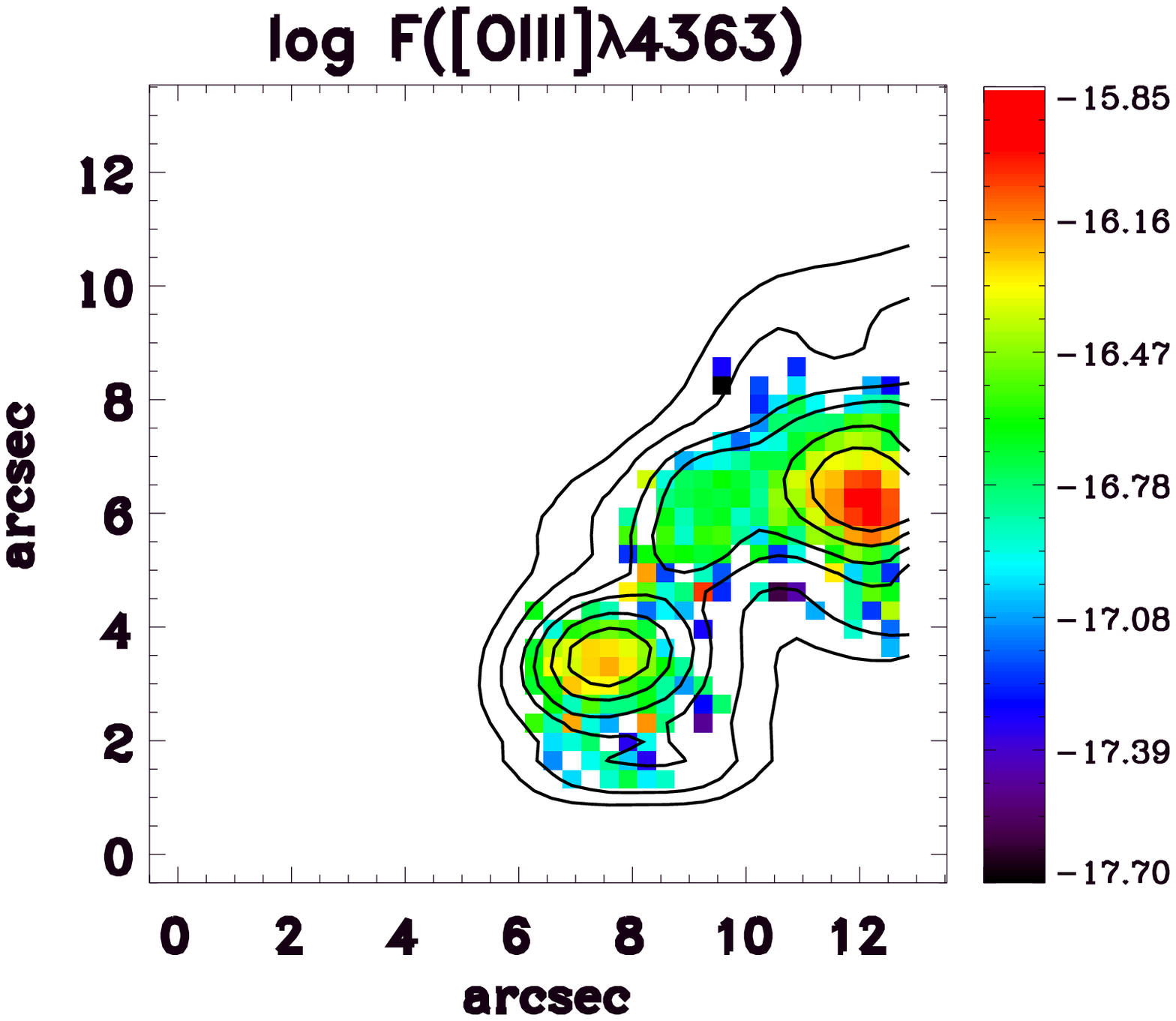}
\includegraphics[width=58mm]{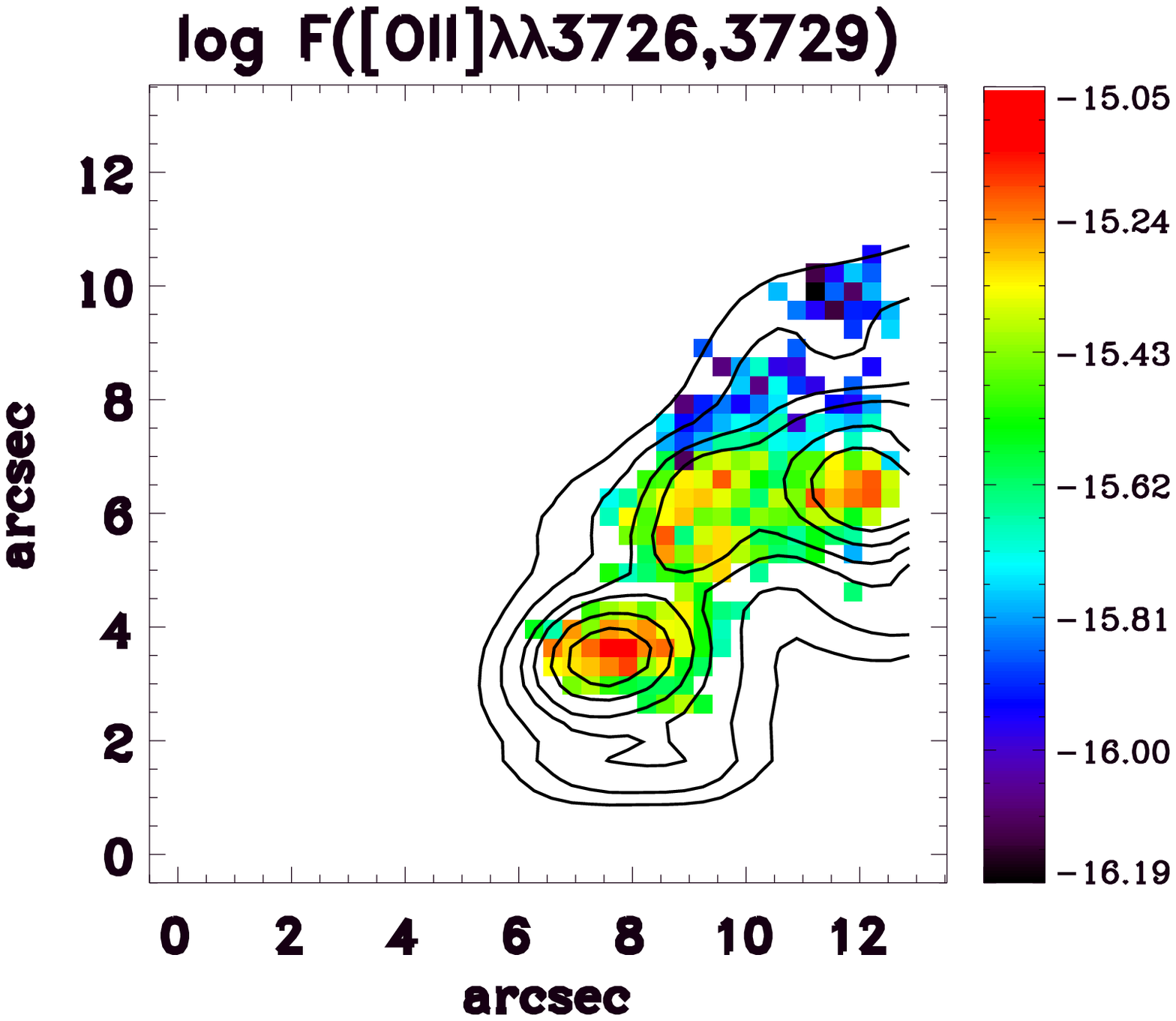}
  \caption{Mrk 600 - Emission line maps: [S II]$\lambda$6717, [N II]$\lambda$6584, [O III]$\lambda$5007, H$\beta$,
  [O III]$\lambda$4363 and [O II]$\lambda\lambda$3726,3729. 
  H$\alpha$ emission line contours are overlaid on each map. North is up and east is to the left.
}
\label{figure_emission_lines600}
\end{figure*}

We used the flux measurements described in Section \ref{sect_obser_reduc} to
produce the following emission line maps: [S\,{\sc ii}]$\lambda$6731, [S\,{\sc ii}]$\lambda$6717, [N\,{\sc ii}]$\lambda$6584, 
H$\alpha$, [O\,{\sc iii}]$\lambda$5007, [O\,{\sc iii}]$\lambda$4959, H$\beta$, [O\,{\sc iii}]$\lambda$4363 and 
[O\,{\sc ii}]$\lambda\lambda$3726,3729. In Figures \ref{figure_emission_lines461} (UM\,461) and \ref{figure_emission_lines600} (Mrk\,600) 
we  show a selection of those maps. We note that, when deriving the maps we only use spaxels with emission 
fluxes $>$ 3$\sigma$ in the background observations.
Our H$\alpha$ maps (Figure \ref{figure_image_campo_Halpha}) reveal that the nebular emission is concentrated  
in the main bodies of both UM\,461 and Mrk\,600 and the centres are coincident with the continuum emission maxima 
\citep[see][]{Lagos2007}.
However, extended diffuse ionized gas emission surrounding the GH\,{\sc ii}Rs is also observed within the VIMOS-IFU FoVs 
for both galaxies. In UM\,461, we resolve two regions or GH\,{\sc ii}Rs labelled as regions no. 1 and no. 2 
in Figure \ref{figure_image_campo_Halpha} (upper right panel) as well as several adjacent faint structures.
We compared the UM\,461 [O\,{\sc ii}], [O\,{\sc iii}], [S\,{\sc ii}], [N\,{\sc ii}] forbidden emission line morphologies to
that of the  H$\alpha$ emission. The morphologies for emission lines closely match each another, 
although due to our sensitivity limit the extent of the H$\alpha$ and
[O\,{\sc iii}]$\lambda$5007 emission is larger than those of [O\,{\sc ii}]$\lambda\lambda$3726,3729, [S\,{\sc ii}]$\lambda\lambda$6717,6731, 
[O\,{\sc iii}]$\lambda$4363 and [N\,{\sc ii}]$\lambda$6584 (see Figure \ref{figure_emission_lines461}).

Overall, the H$\alpha$ emission of Mrk\,600 displays an elongated morphology and four GH\,{\sc ii}Rs are labelled 
regions no. 1 to 4 on the \halpha\ map (Figure \ref{figure_image_campo_Halpha} lower right panel).  
As for UM\,461, the spatial distribution of recombination lines (H$\alpha$, H$\beta$, etc.) in Mrk 600 is very similar 
to the emission from forbidden lines (see Figure \ref{figure_emission_lines600}). 
Interestingly, we observe two extended structures or shells, adjacent to region no. 1 
(see Figure \ref{figure_image_campo_Halpha} lower right). 
H$\alpha$ narrow-band images presented by \cite{GildePaz2003} and \cite{JanowieckiSalzer2014} show that both 
structures are very well resolved. This confirms the presence of an extended shell, or bubble, 
that is $\sim$180 pc away from the H$\alpha$ peak of region no. 1.

\begin{figure*}
\includegraphics[width=58mm]{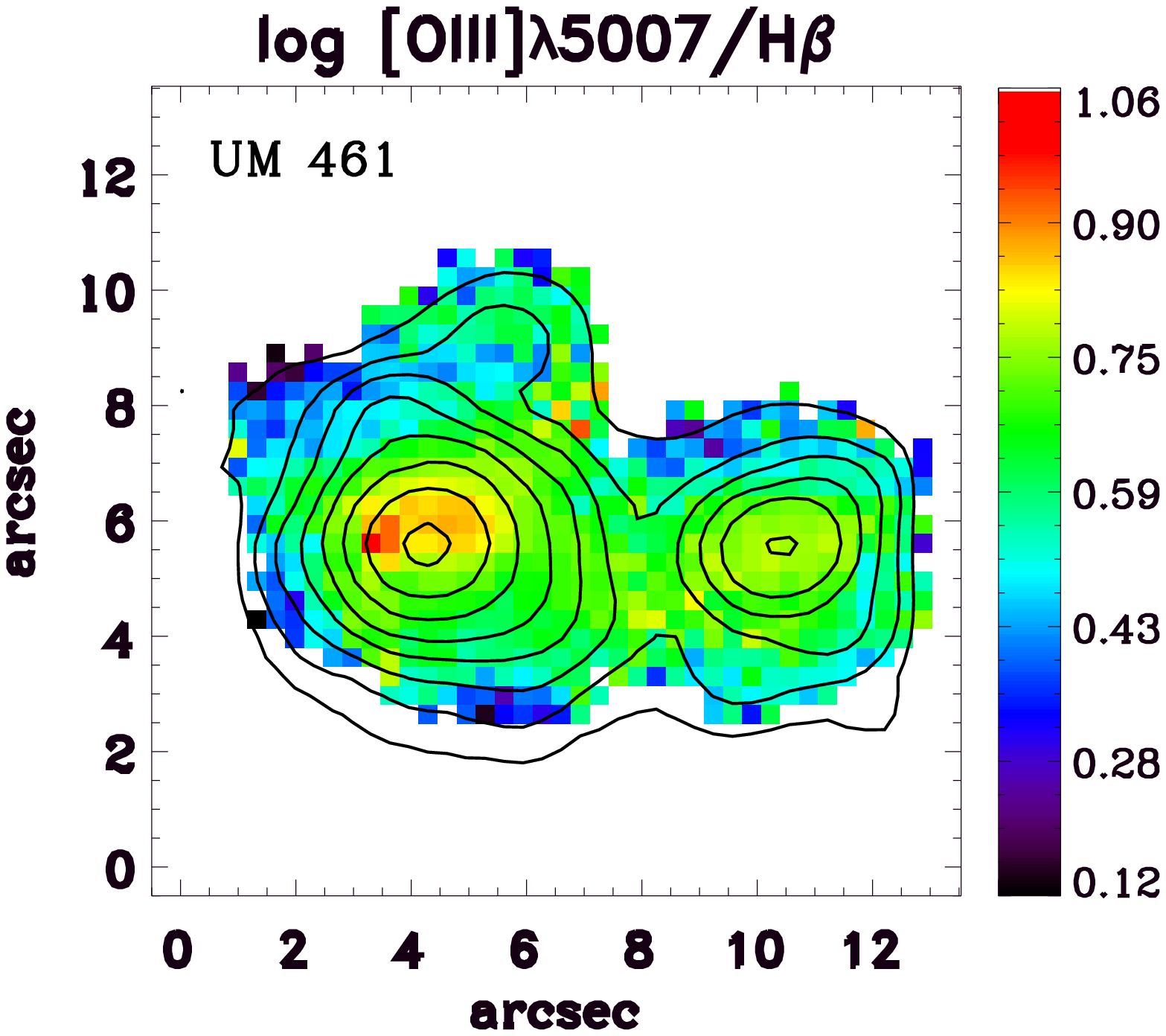}
\includegraphics[width=58mm]{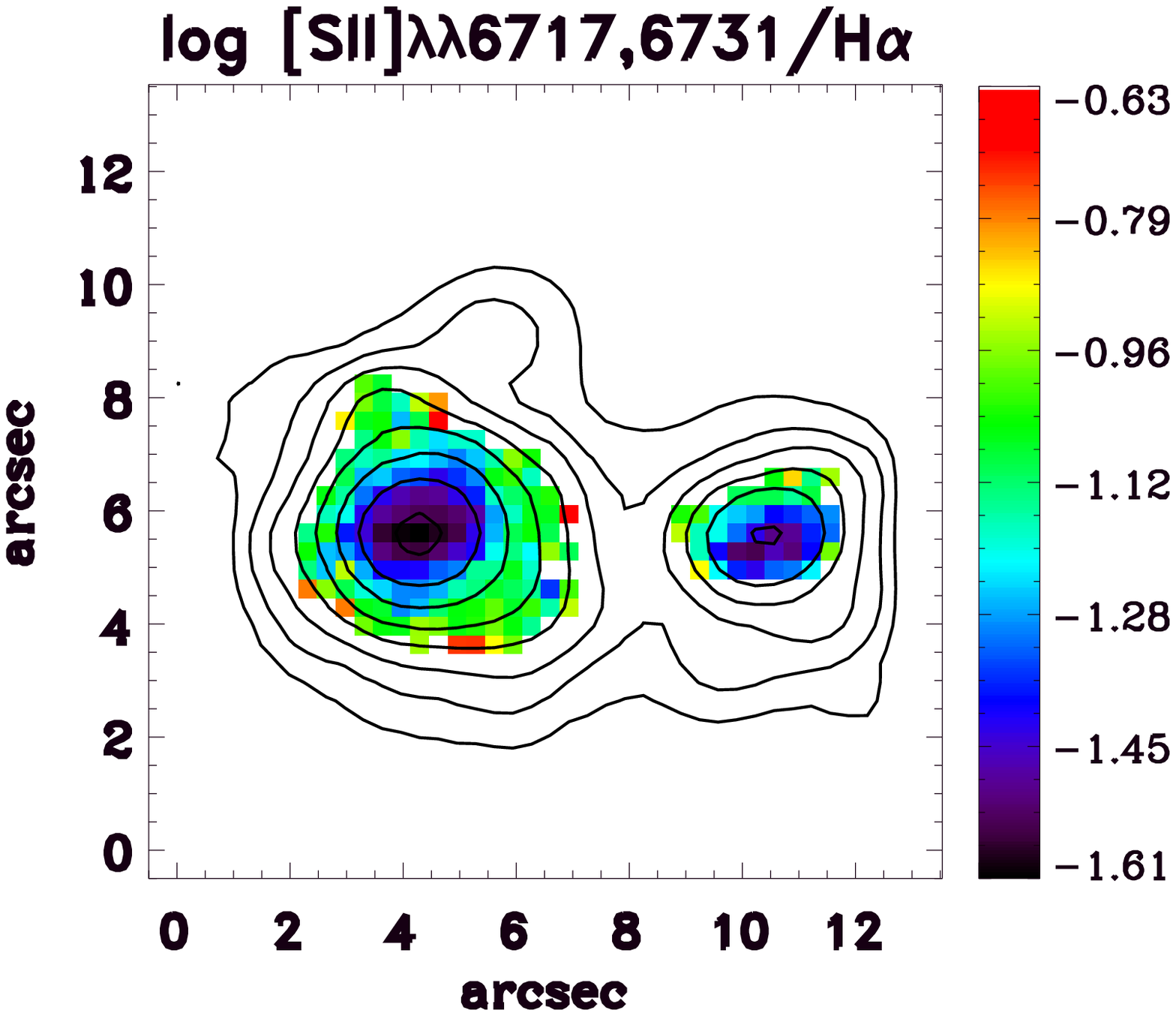}
\includegraphics[width=58mm]{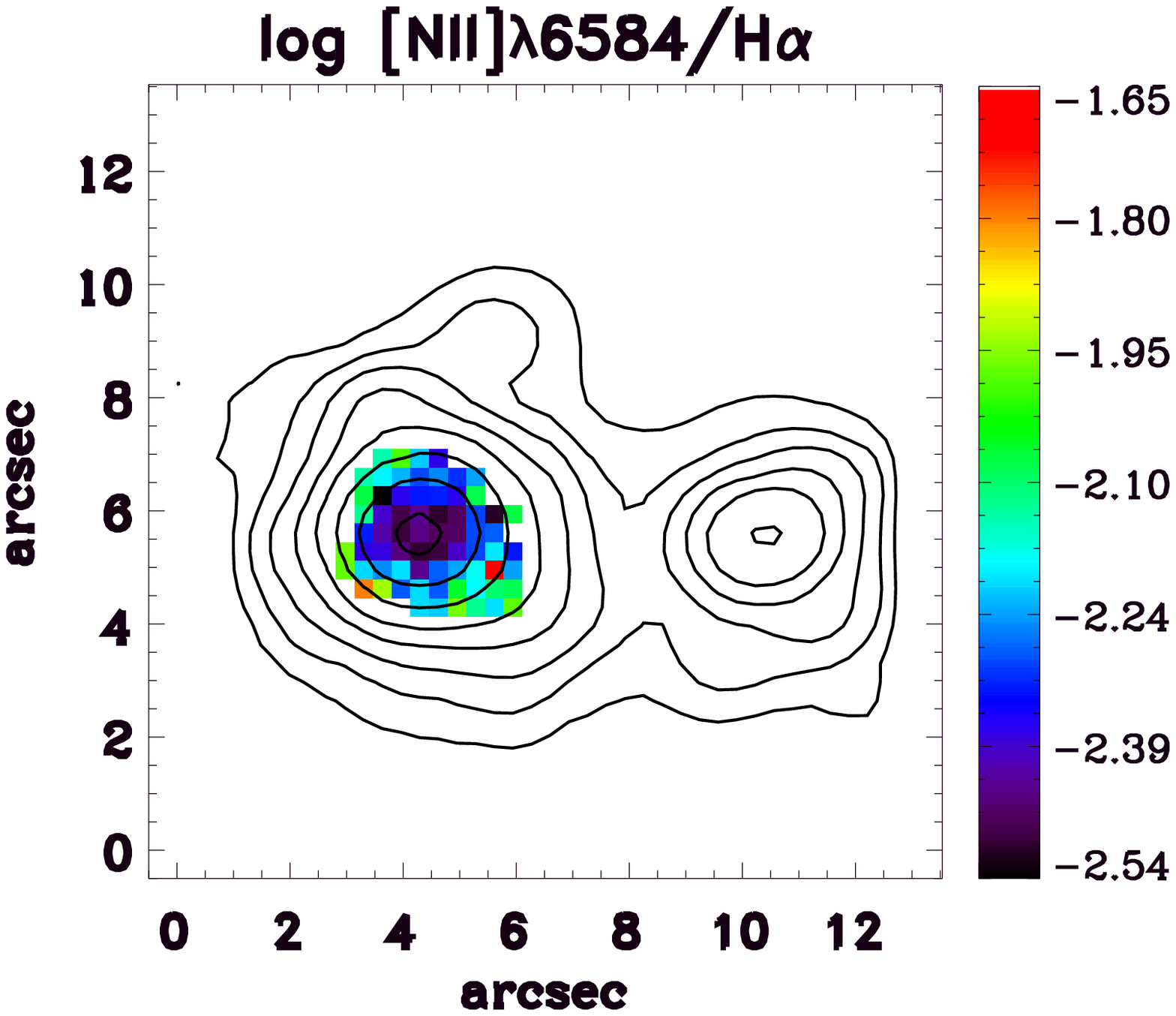}\\
\includegraphics[width=58mm]{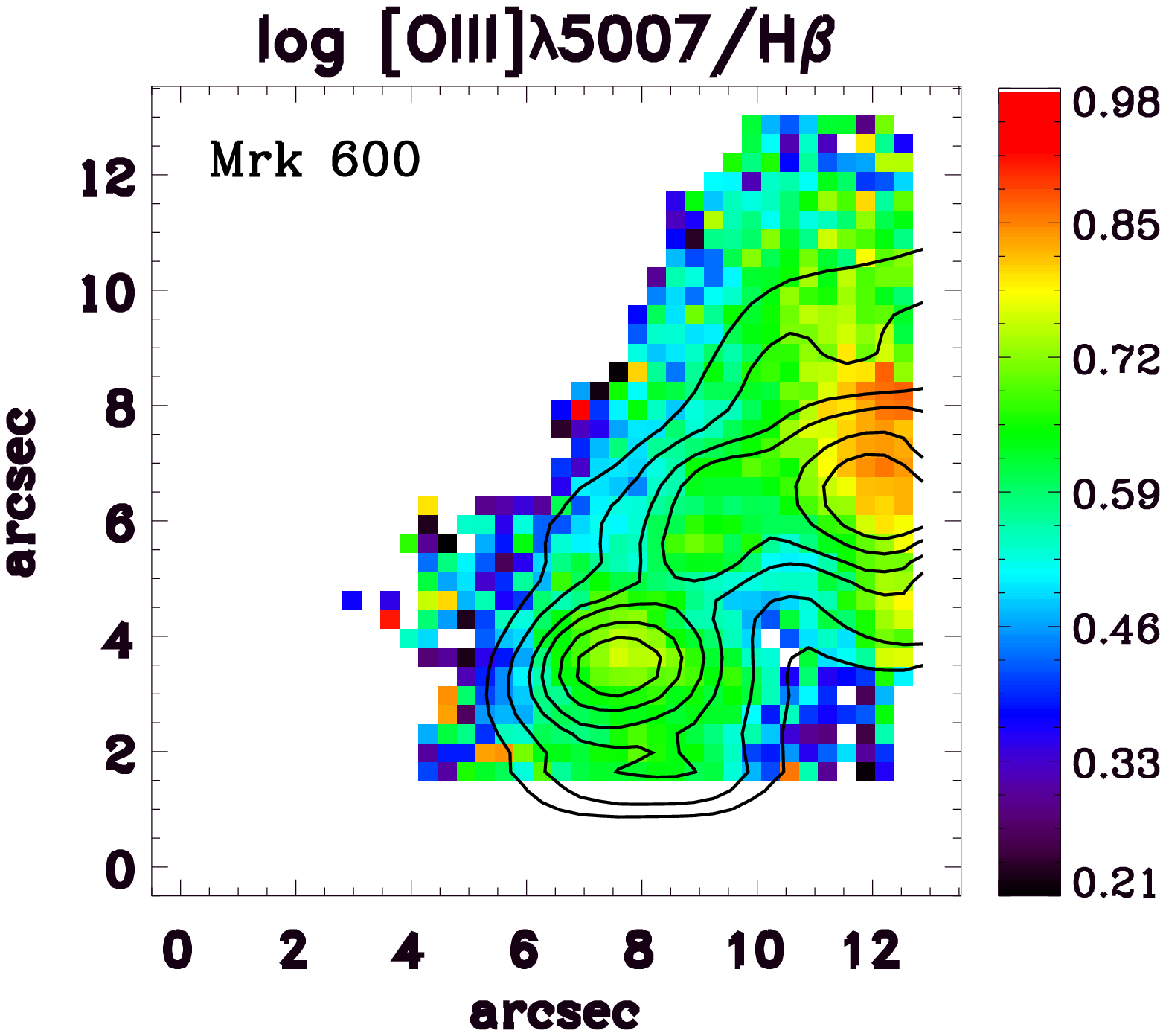}
\includegraphics[width=58mm]{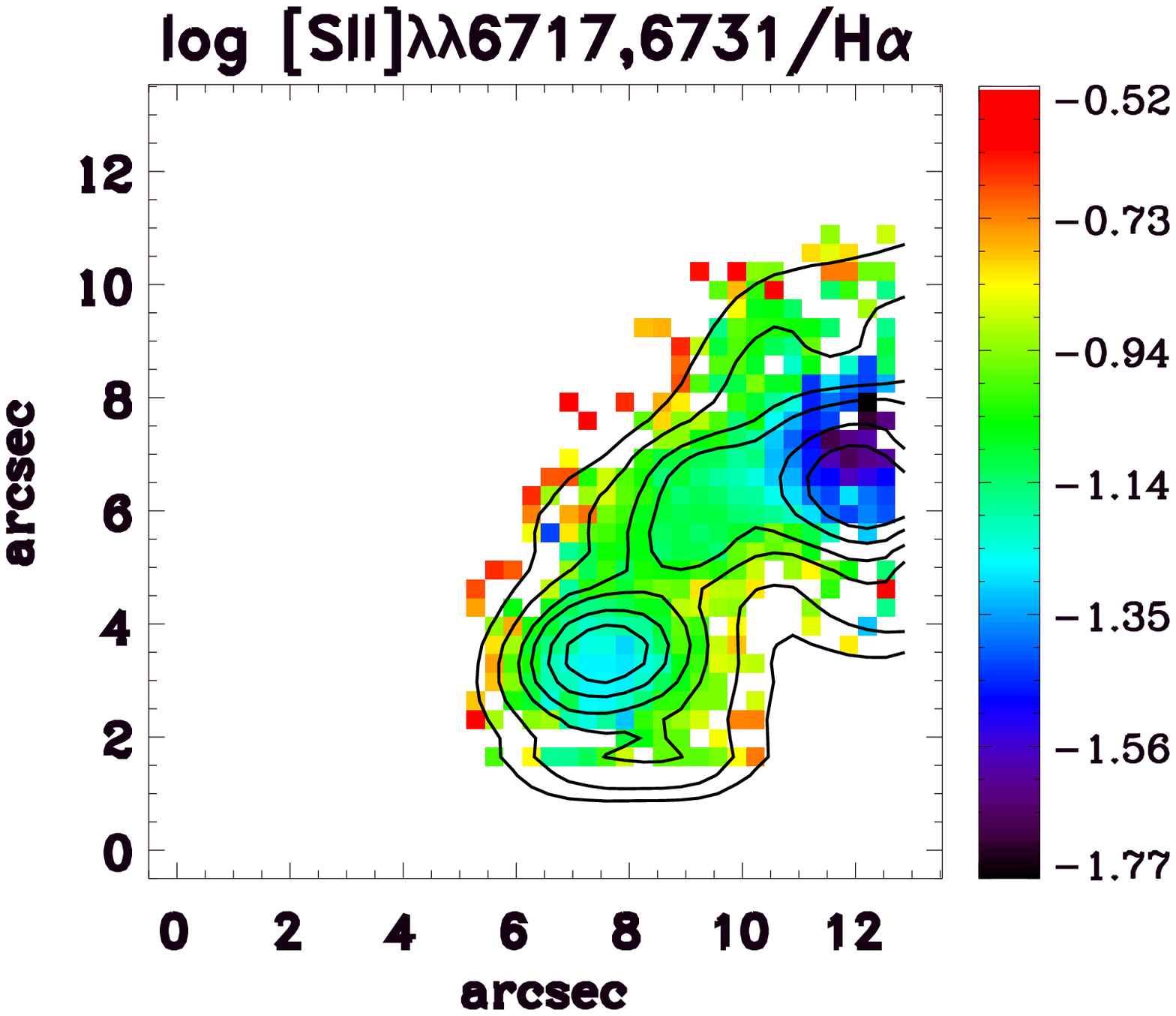}
\includegraphics[width=58mm]{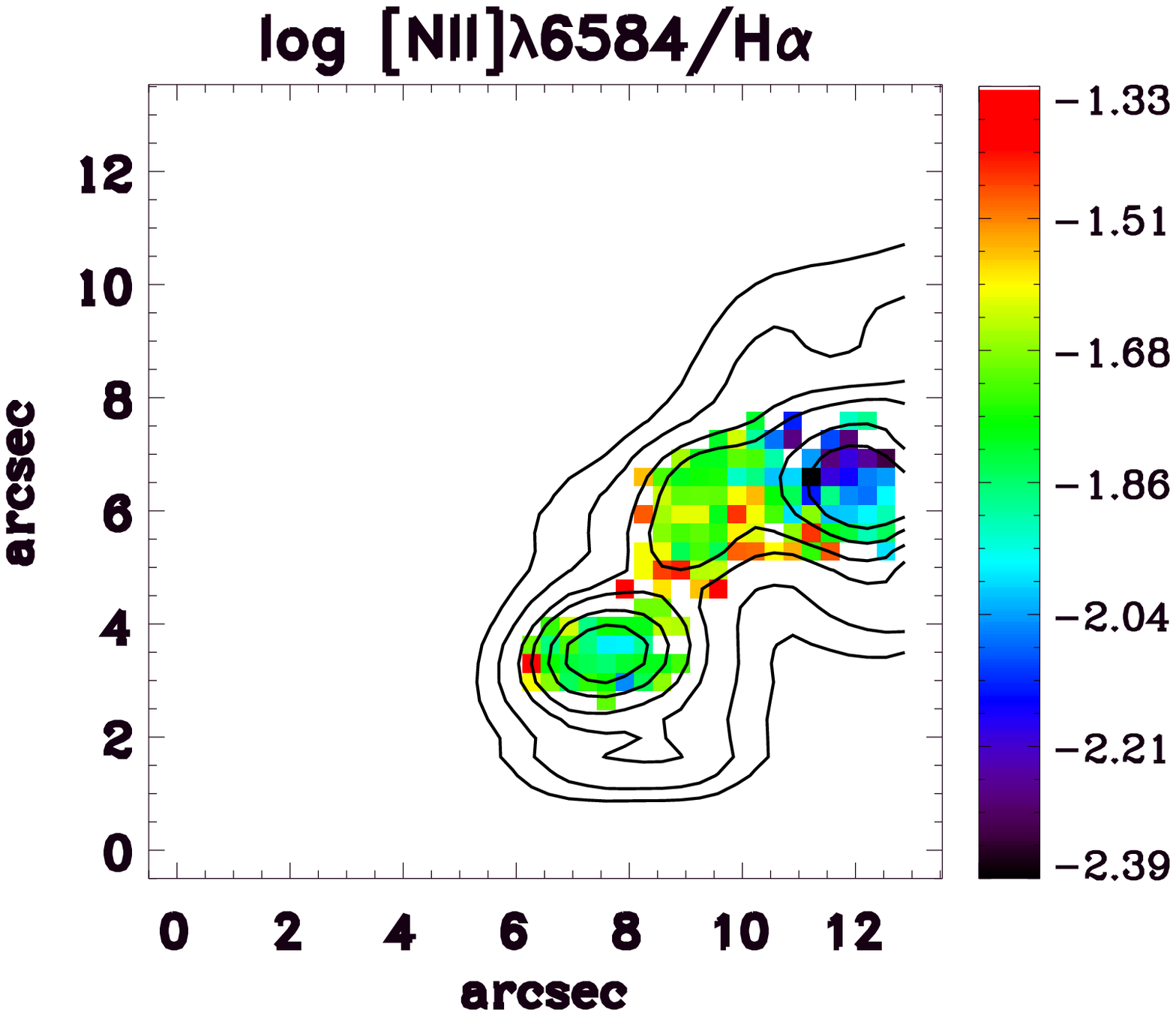}
  \caption{Emission line ratio maps: log [O III]$\lambda$5007/H$\beta$, 
log [S II]$\lambda\lambda$6717,6731/H$\alpha$ and log [N II]$\lambda$6584/H$\alpha$ 
for  UM 461 (upper panels) and Mrk 600 (lower panels).
H$\alpha$ emission line contours are overlaid on each map. North is up and east is to the left.
}
\label{figure_emission_line_ratios}
\end{figure*}

\subsubsection{Emission-line ratios}
For both UM 461 and Mrk 600, we employed the commonly used  BPT \citep{Baldwin1981} diagrams 
to infer the dominant ionization mechanism at spaxel scales using the following emission-line ratios: 
[O\,{\sc iii}]$\lambda$5007/H$\beta$, [S\,{\sc ii}]$\lambda\lambda$6717,6731/H$\alpha$ and 
[N\,{\sc ii}]$\lambda$6584/H$\alpha$ (see Figure \ref{figure_emission_line_ratios}). 
The spatial profiles of the emission line ratios differ significantly from one another, as shown
in Figure \ref{figure_emission_line_ratios}, between the peak of the H$\alpha$ emission and the edge of the VIMOS-IFU  FoV.
The ionization structure within the inner most part of the GH\,{\sc ii}Rs, for both galaxies, is rather constant 
as measured by [S\,{\sc ii}]$\lambda\lambda$6717,6731/H$\alpha$ and [N\,{\sc ii}]$\lambda$6584/H$\alpha$, 
but these ratios increase at greater distances from the GH\,{\sc ii}Rs. 
However, the [O\,{\sc iii}]$\lambda$5007/H$\beta$ ratios do not show a uniform distribution. 
In the case of UM 461 its values are highest in a curved structure, which surrounds the peak of H$\alpha$ emission.
Our [O\,{\sc iii}]$\lambda$5007/H$\beta$ ratio map, in this galaxy, is in excellent agreement with the map   
obtained by \cite{SampaioCarvalho2013} using Gemini Multi-Object Spectrograph (GMOS) IFU.
We do not show, in this paper, the BPT diagrams but all points fall in the locus predicted by models of photo-ionization 
by young stars in H\,{\sc ii} regions \citep{OsterbrockFerland2006},  
indicating that photoionization from stellar sources is the dominant excitation mechanism in UM 461 and  Mrk 600.
We compared the aforementioned integrated emission line ratios, for both galaxies, showed  
in Table \ref{table_integrated_fluxes_UM461_MRK600} with the values found in the literature. 
Our values  obtained in UM 461 are in agreement, within the uncertainties, 
with those reported by \cite{Perez-MonteroDiaz2003}, i.e., log([O\,{\sc iii}]$\lambda$5007/H$\beta$) = 0.78,  
log([S\,{\sc ii}]$\lambda\lambda$6717,6731/H$\alpha$) = -1.47 and log([N\,{\sc ii}]$\lambda$6584/H$\alpha$) = -2.12.
Finally, the emission line ratios in Mrk 600 (see Tables \ref{table_integrated_fluxes_UM461_MRK600} 
and \ref{table_regions_fluxes_MRK600}) are in agreement with those from \cite{Guseva2011}, i.e., 
log([O\,{\sc iii}]$\lambda$5007/H$\beta$) = 0.81, 
log([S\,{\sc ii}]$\lambda\lambda$6717,6731/H$\alpha$) = -1.46 and log([N\,{\sc ii}]$\lambda$6584/H$\alpha$) = -2.07.

\subsection[]{Abundance determinations}\label{sect_abundances}

For our abundance estimates, we first determined the electron temperature T$_e$ and electron 
density n$_e$, making use of the line ratios [O\,{\sc iii}]$\lambda$4959,5007/[O\,{\sc iii}]$\lambda$4363 and  
[S\,{\sc ii}]$\lambda$6716/[S\,{\sc ii}]$\lambda$6731, and the IRAF STS package \textit{nebular}. 

Oxygen and nitrogen ion abundances O$^{+}$, O$^{++}$ and N$^{+}$ were calculated using the five-level atomic model FIVEL 
implemented in the IRAF STS task \textit{abund}. The total oxygen abundance for each aperture is obtained assuming 
the contributions from O$^{+}$ and O$^{++}$, therefore we have:

\begin{equation}
\frac{O}{H}=\frac{O^{+}}{H^{+}} + \frac{O^{++}}{H^{+}},
\end{equation}
and
\begin{equation}
\frac{N}{H}=ICF(N) \frac{N^{+}}{H^{+}},
\end{equation}
with
\begin{equation}
ICF(N)= \frac{O^{+}+O^{2+}}{O^{+}}.
\end{equation}

\begin{figure*}
\includegraphics[width=85mm]{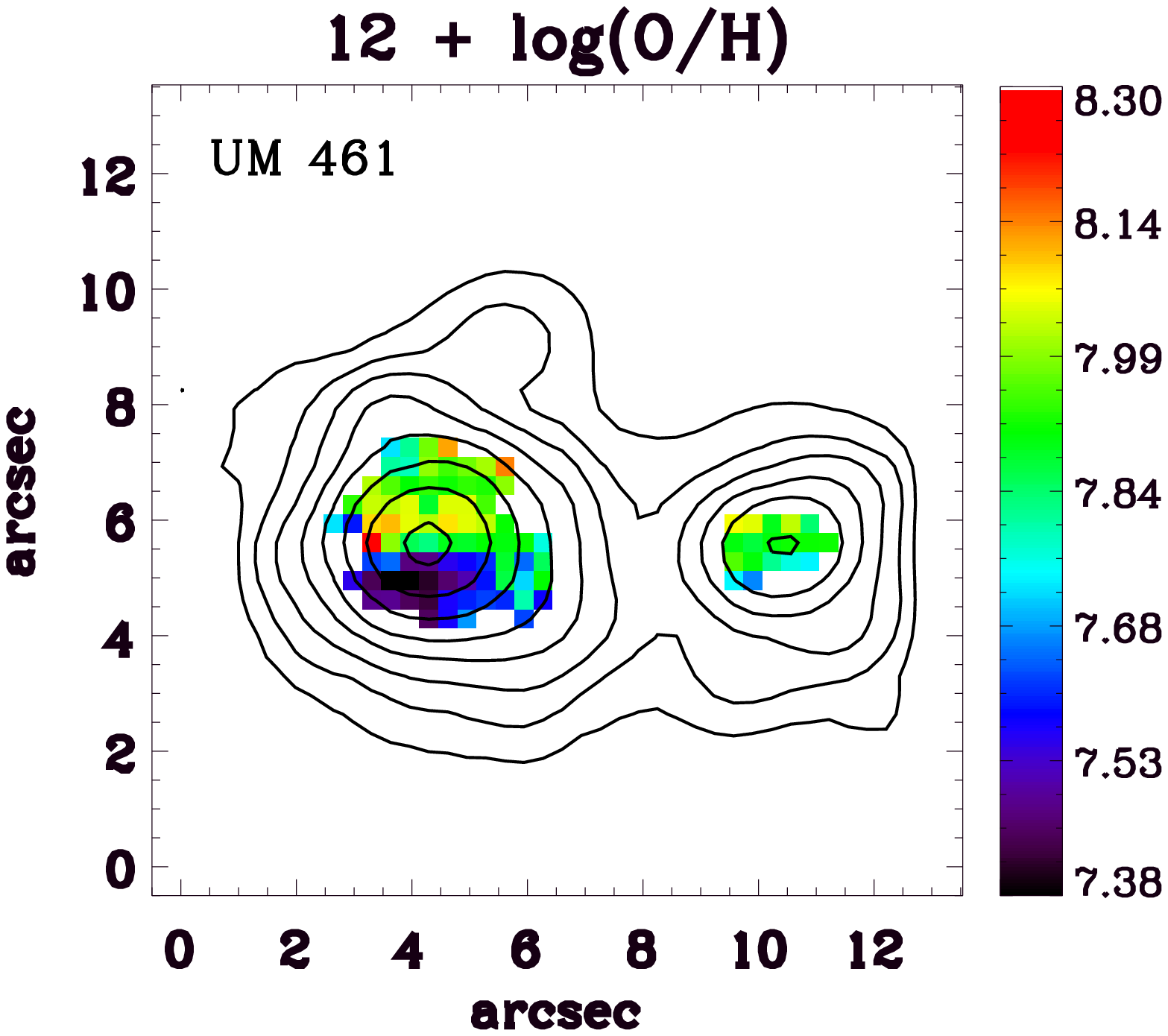}
\includegraphics[width=85mm]{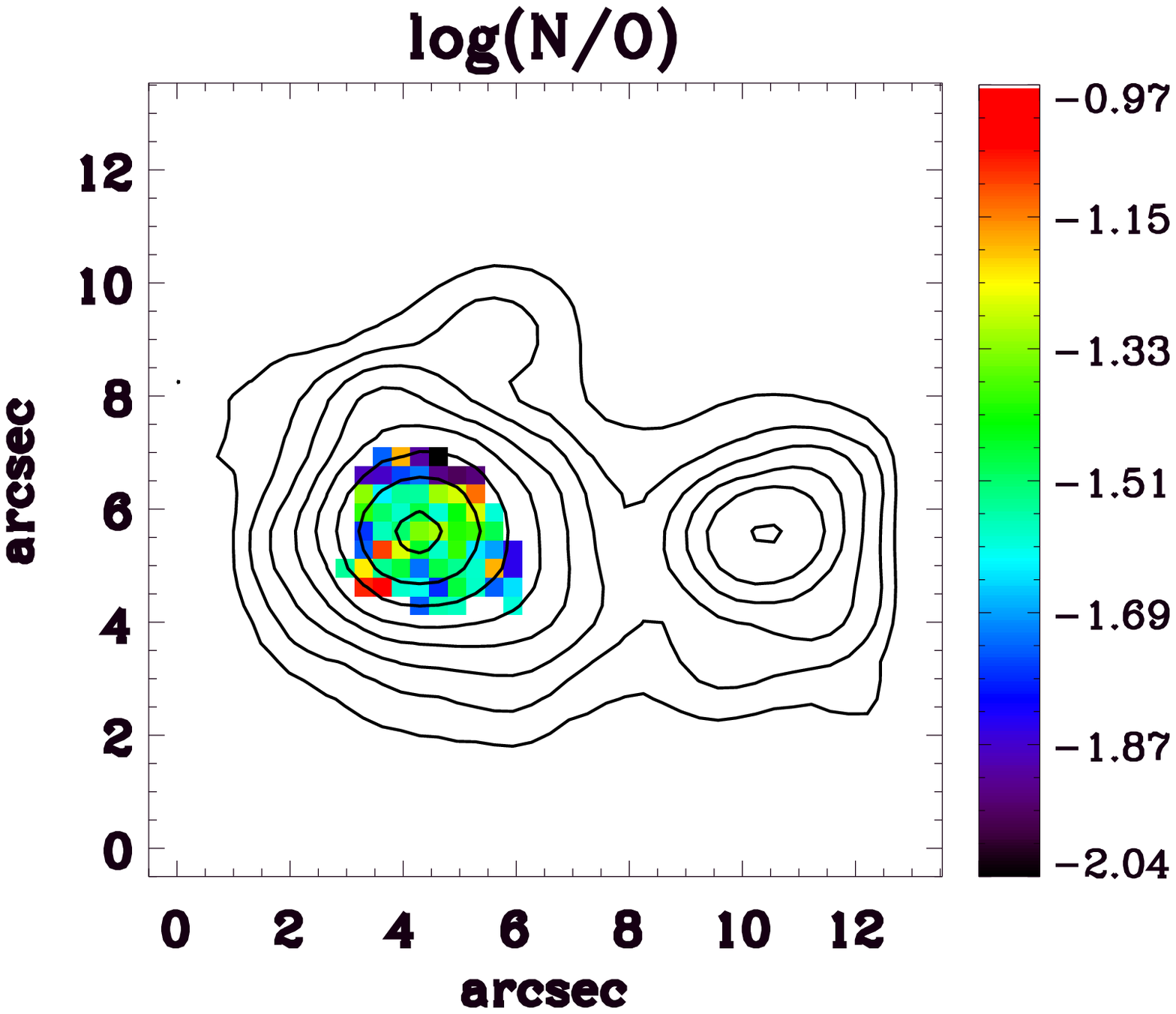}
\includegraphics[width=85mm]{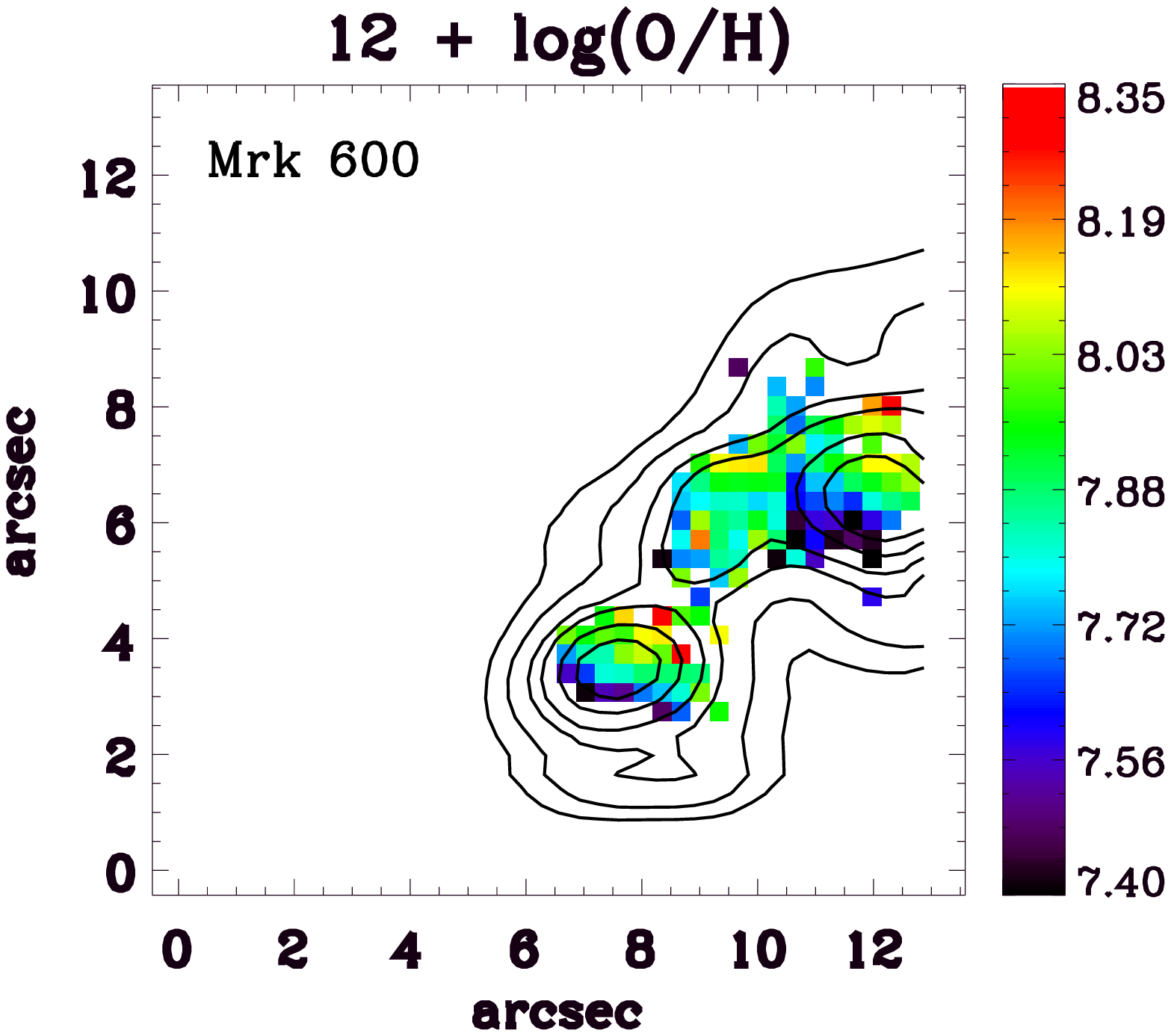}
\includegraphics[width=85mm]{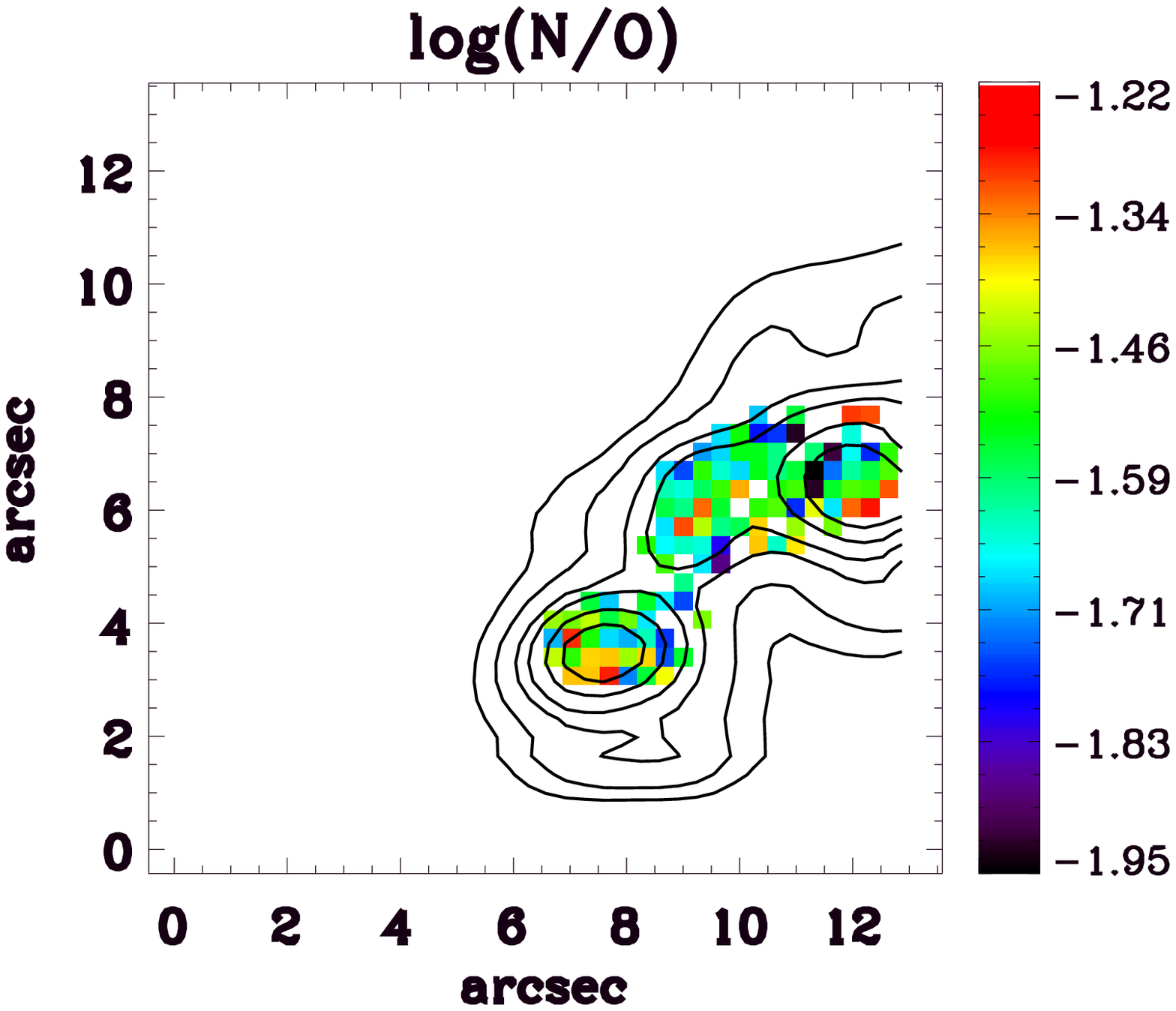}
  \caption{12 + log(O/H) and log(N/O) abundance maps for UM 461 (upper panels) and Mrk 600 (lower panels). 
  Oxygen abundances were determined using using the direct method.
  H$\alpha$ emission line contours are overlaid on each map. North is up and east is to the left.}
\label{figure_OH_NO_maps}
  \end{figure*}

In Figure \ref{figure_OH_NO_maps}, we show the oxygen abundance and the log(N/O) ratio maps in the left and right panels,
for both UM 461 and Mrk 600. 
Tables \ref{table_abundances_UM461} and \ref{table_abundances_Mrk600}, respectively, show the abundances measured for 
the individual apertures within the FoVs of UM\,461 and Mrk\,600 (see Figure \ref{figure_image_campo_Halpha}). 
Our integrated oxygen abundances of 12 + log(O/H) = 7.84$\pm$0.08 and 7.85$\pm$0.09 are in agreement  
with those  in \cite{IzotovThuan1998} of 7.78$\pm$0.03 and 7.83$\pm$0.01 for UM 461 and Mrk 600, 
respectively. For UM\,461 a larger discrepancy however occurs with the value of 7.32$\pm$0.15 from \cite{Sanchez2015}.
In the case of Mrk 600, the difference between the oxygen abundances of the GH\,{\sc ii}Rs 2, 3 and 4 and region no. 1 is 
$\Delta$(O/H) $\leq$ 0.04 dex, while in UM 461 we find a difference of $\Delta$(O/H) = 0.06 dex between the two main regions. 
Therefore, within the uncertainties we can consider that the oxygen abundances amongst and between  
the GH\,{\sc ii}Rs of each galaxy are similar. 
The 12 + log(O/H) values in Figure \ref{figure_OH_NO_maps} range from 7.38 to 8.30 for UM 461 and from 
7.40 to 8.35 in Mrk 600. In Mrk 600 the lowest values of 12 + log(O/H) ($<$ 7.6) are found surrounding 
the brightest regions (no. 1 and no. 4). For UM\,461 the lowest values of 12 + log(O/H) are in the southern part of region no. 1 
with an extent of $\sim$0.7 kpc oriented towards the faint SW stellar tail (see Figure \ref{figure_image_campo_Halpha} upper right). 
This region has a mean 12 + log(O/H) value of $\sim$7.52 and a difference of $\Delta$(O/H) = 0.46 dex between the lowest abundance
and the integrated value. 
Interestingly, this region is not coincident with the peak of H$\alpha$ emission 
or  other star clusters resolved by \cite{Lagos2011} within our region no. 1.

\begin{figure}
\includegraphics[width=80mm]{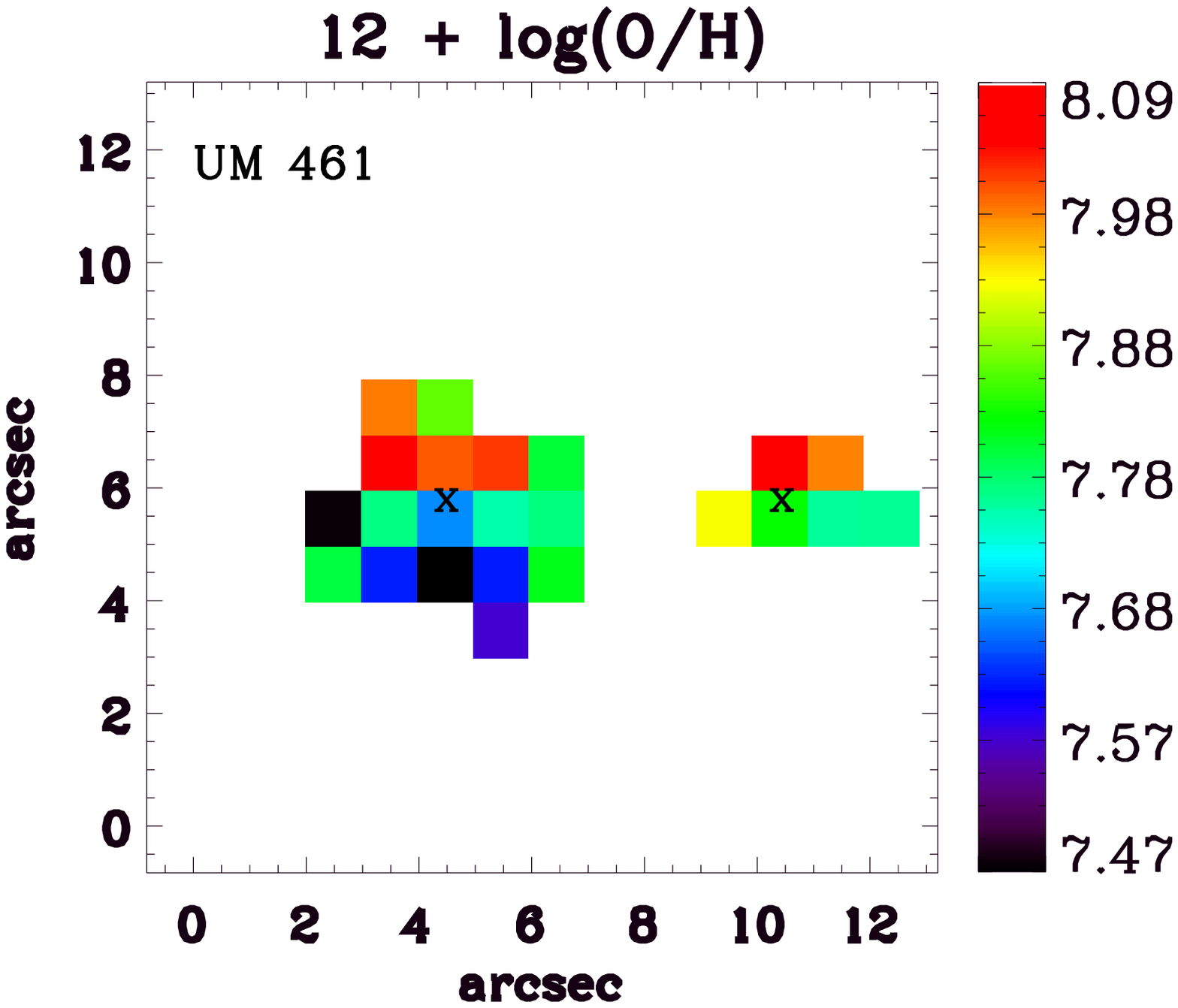}
\includegraphics[width=80mm]{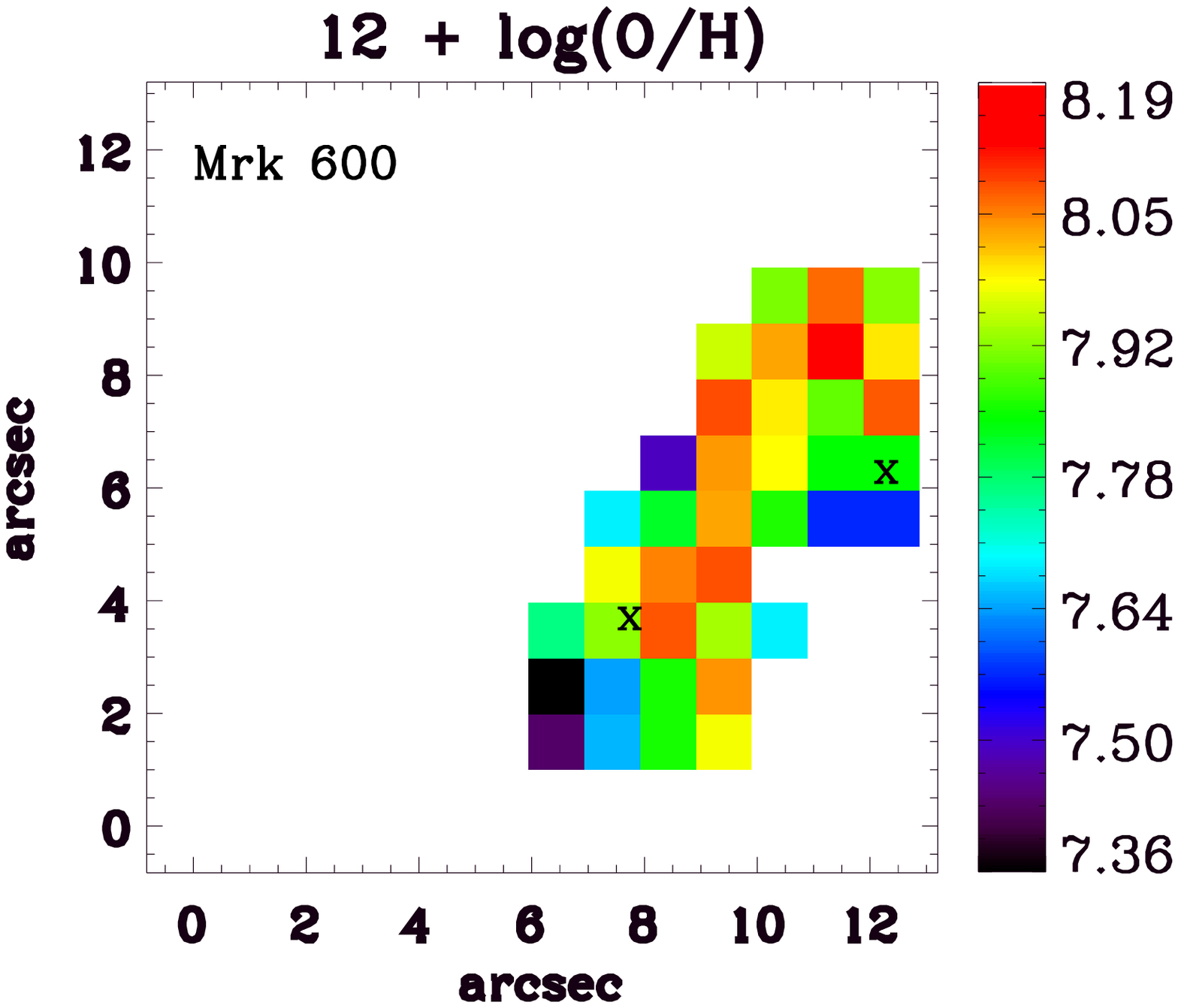}
  \caption{Spatial distribution of the binned ($\sim$1$\arcsec$ spaxels) 12 + log(O/H) abundances for both galaxies UM 461
  (upper panel) and Mrk 600 (lower panel). The maximum H$\alpha$ emission of the main regions are indicated in the maps by a X symbol.
  North is up and east is to the left.}
  \label{figure_OH_3x3}
  \end{figure}

In order to test the accuracy in the detection of variations of oxygen abundance, we introduced several offsets 
of 0.33$\arcsec$ (one spaxel) in the [O\,{\sc iii}]$\lambda$4363 maps.  
In most cases (75\%), we found that the spatial variations are preserved within 0.1 dex. 
Given that the seeing during our observations was $\sim$1.0$\arcsec$ (3 spaxels), 
the pixel-to-pixel variation in our maps are potentially due to  measurement uncertainties rather than real variations. 
This point is commonly ignored in most of the IFU studies.
To further test whether the variations were real or not, we binned the data cube from 0.33$\arcsec$ to $\sim$1.0$\arcsec$ (3 spaxels). 
In Figure \ref{figure_OH_3x3} we show the spatial distribution of the binned ($\sim$1$\arcsec$ spaxel)
12 + log(O/H) abundances. 
Again, in both cases the abundance patterns are preserved. Nevertheless, it is more practical 
for the analysis to use the 0.33$\arcsec$ pixel scale.

We find an integrated value of 12 + log(N/H) = 6.31$\pm$0.18 for UM 461 and 6.03$\pm$0.22
for Mrk 600. The nitrogen-to-oxygen ratio in these galaxies is log(N/O) = -1.54$\pm$0.27 and 
-1.83$\pm$0.30 for UM 461 and Mrk 600, respectively.
In the case of UM\,461, the integrated log(N/O) is consistent with those found in XMP galaxies 
\citep[log(N/O) $\sim$ -1.60; e.g.][]{EdmundsPagel1978,Alloin1979,IzotovThuan1999}.
Interestingly, the region no. 1 in UM 461 and the area of low metallicity, in the same region, show a similar log(N/O)
value $\sim$ -1.50.
Therefore, the log(N/O) ratio is uniform at large scales in the brightest region of this galaxy.
Finally, we found that the integrated log(N/O) values agree, within the uncertainties, with those 
obtained by \cite{IzotovThuan1998}, i.e., log(N/O) = -1.50 for UM 461 and of -1.67 for Mrk 600.

\begin{table*}
 \centering
 \begin{minipage}{85mm}
  \caption{UM 461: ionic abundances and integrated properties.}
 \begin{tabular}{@{}lccc@{}}
  \hline
                                &  Integrated   &   Region no.1 & Region no.2\\                              
 \hline
Te(OIII) K                      &15184$\pm$548  &15487$\pm$288  &13838$\pm$576\\
Ne(SII) cm$^{-3}$               & 431$\pm$223   & 208$\pm$172   & 662$\pm$286 \\
O$^{+}$/H$^{+} \times$10$^{5}$  & 0.50$\pm$0.03 & 0.34$\pm$0.01 & 0.87$\pm$0.06\\
O$^{++}$/H$^{+} \times$10$^{5}$ & 6.49$\pm$0.59 & 6.71$\pm$0.31 & 7.32$\pm$0.83\\
O/H $\times$10$^{5}$            & 6.99$\pm$0.62 & 7.04$\pm$0.32 & 8.19$\pm$0.88\\
12 + log(O/H)                   & 7.84$\pm$0.08 & 7.85$\pm$0.05 & 7.91$\pm$0.10\\
N$^{+}$/H$^{+} \times$10$^{6}$  & 0.14$\pm$0.01 & 0.11$\pm$0.01 & $\ldots$\\
ICF(N)                          &14.08$\pm$2.11 &21.74$\pm$1.52 & 9.46$\pm$1.65\\
N/H $\times$10$^{6}$            & 2.02$\pm$0.36 &2.24$\pm$0.20  & $\ldots$\\
12+log(N/H)                     & 6.31$\pm$0.18 & 6.35$\pm$0.09 & $\ldots$\\
log(N/O)                        &-1.54$\pm$0.27 &-1.50$\pm$0.14 & $\ldots$\\
\hline
\end{tabular}
\label{table_abundances_UM461}
\end{minipage}
\end{table*}

\begin{table*}
 \centering
 \begin{minipage}{130mm}
  \caption{Mrk 600: ionic abundances and integrated properties.}
 \begin{tabular}{@{}lccccc@{}}
  \hline
                                & Integrated            & Region no.1           & Region no.2          & Region no.3  & Region no.4\\
 \hline
Te(OIII) K                      &14825$\pm$670 &15491$\pm$462 &14075$\pm$580 &14200$\pm$558 &14156$\pm$438\\
Ne(SII) cm$^{-3}$               &502$\pm$134   &30$\pm$110    &$\sim$100     &170$\pm$110   &222$\pm$93     \\
O$^{+}$/H$^{+} \times$10$^{5}$  &1.78$\pm$0.12 &0.63$\pm$0.03 &1.63$\pm$0.11 &1.99$\pm$0.13 &1.51$\pm$0.07\\
O$^{++}$/H$^{+} \times$10$^{5}$ &5.37$\pm$0.62 &6.12$\pm$0.45 &5.33$\pm$0.59 &5.25$\pm$0.55 &5.97$\pm$0.49\\
O/H $\times$10$^{5}$            &7.16$\pm$0.75 &6.76$\pm$0.48 &6.96$\pm$0.70 &7.24$\pm$0.67 &7.48$\pm$0.56\\
12 + log(O/H)                   &7.85$\pm$0.09 &7.83$\pm$0.07 &7.84$\pm$0.09 &7.86$\pm$0.09 &7.87$\pm$0.09\\
N$^{+}$/H$^{+} \times$10$^{6}$  &0.26$\pm$0.01 &0.13$\pm$0.01 &0.28$\pm$0.01 &0.26$\pm$0.01 &0.23$\pm$0.01\\
ICF(N)                          &4.01$\pm$0.69 &10.66$\pm$1.24&4.27$\pm$0.71 &3.64$\pm$0.57 &4.95$\pm$0.62\\
N/H $\times$10$^{6}$            &1.06$\pm$0.24 &1.39$\pm$0.20 &1.21$\pm$0.26 &0.95$\pm$0.19 &1.15$\pm$0.18\\
12+log(N/H)                     &6.03$\pm$0.22 &6.14$\pm$0.15 &6.08$\pm$0.22 &5.98$\pm$0.20 &6.06$\pm$0.16\\
log(N/O)                        &-1.83$\pm$0.30&-1.69$\pm$0.22&-1.76$\pm$0.30&-1.88$\pm$0.29&-1.81$\pm$0.24\\
\hline
\end{tabular}
\label{table_abundances_Mrk600}
\end{minipage}
\end{table*}

\subsection[]{Velocity fields}\label{sect_velocity_field}

\begin{figure*}
\includegraphics[width=85mm]{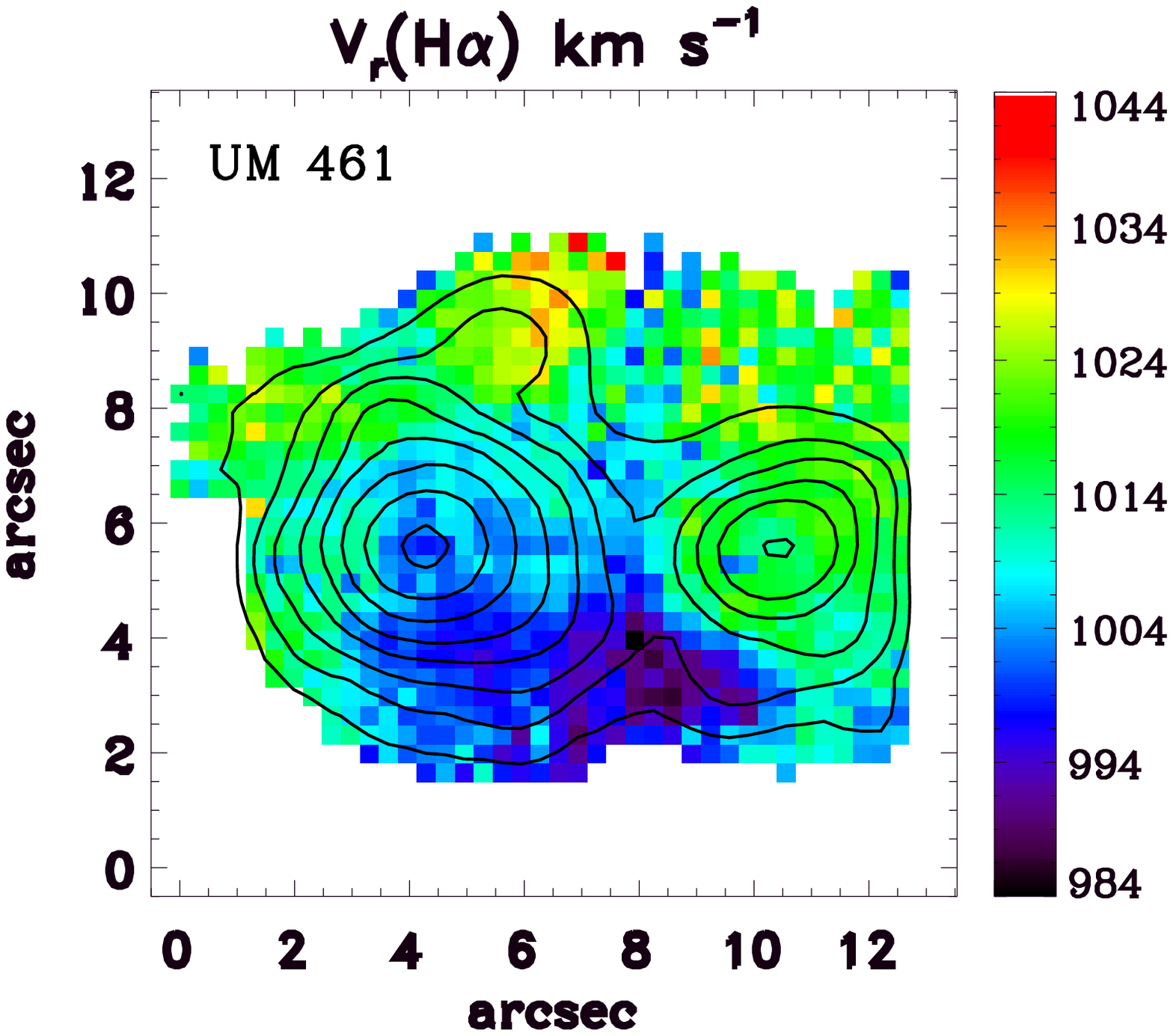}
\includegraphics[width=85mm]{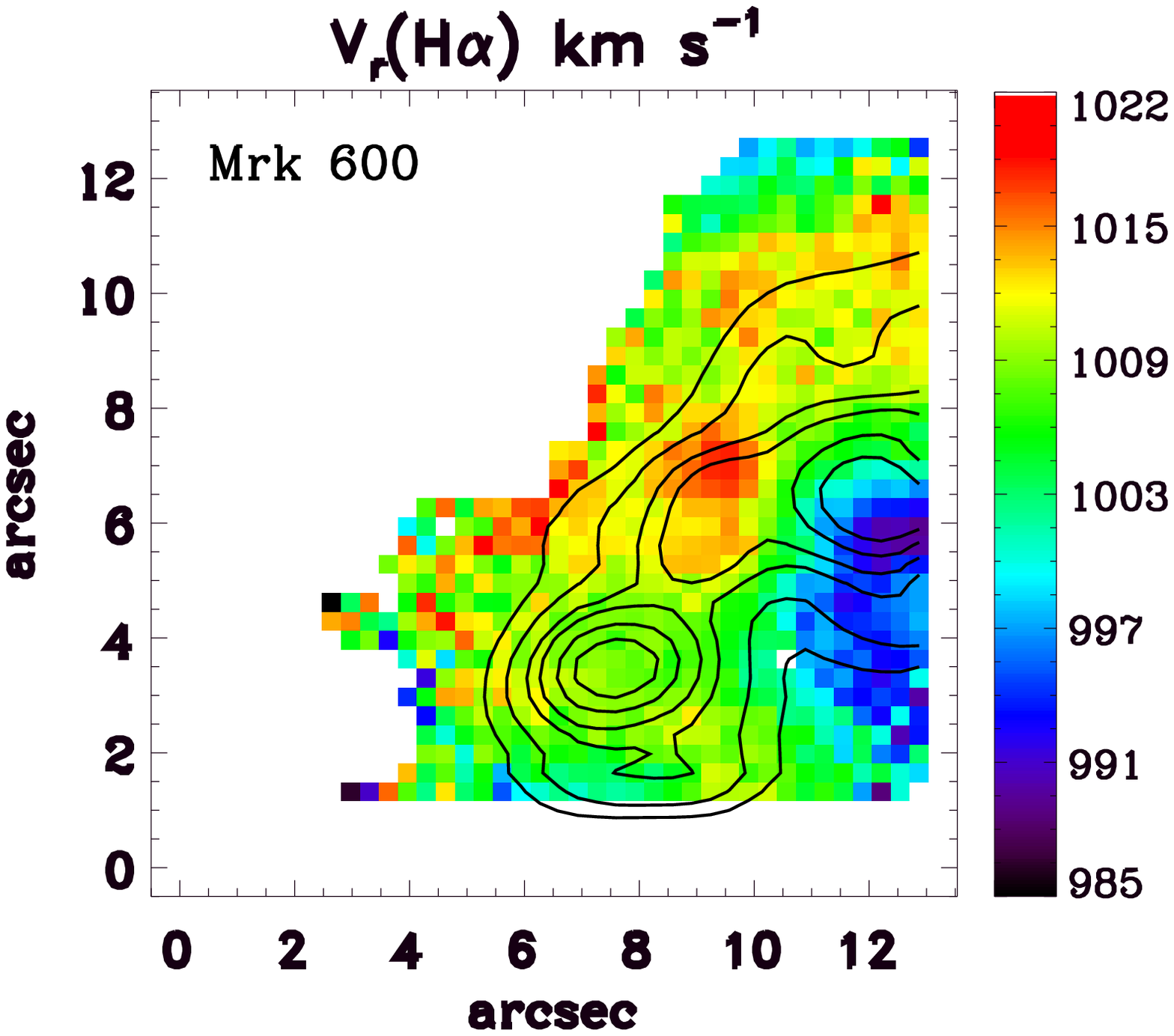}
  \caption{Radial velocity of the H$\alpha$ emission line in units of km s$^{-1}$ for UM 461 and Mrk 600.
  Contours display the H$\alpha$ morphology of the galaxies. North is up and east is to the left.}
\label{figure_velocity}
  \end{figure*}

We obtained the radial velocity v$_r$(H$\alpha$) by fitting a single Gaussian to the H$\alpha$
emission line profiles. The v$_r$(H$\alpha$) velocity fields showed in Figure \ref{figure_velocity}  
are rather complex. For UM\,461 the velocity field shows an apparently systemic 
trend with the northern part redshifted, while the southern part is blueshifted, 
with a systemic velocity of $\sim$1040 km s$^{-1}$. As mentioned in Section \ref{sect_intro}, 
this galaxy is part of a binary system with UM 462. 
Interestingly, the velocity distribution in UM 462 shows no spatial correlation with 
the H$\alpha$ emission \citep[see Figure 2 in][]{James2010}. A similar lack of correlation 
is observed in UM\,461. The  range of radial velocities displayed in the UM\,461 map 
is about 60 km s$^{-1}$, while the velocity difference between regions no. 1 and no. 2 
is $\sim$ 13 km s$^{-1}$. The UM\,461 V$_{r}$(H$\alpha$) velocity field shows 
the same overall pattern and detailed variations as the velocity field reported by 
\cite{SampaioCarvalho2013} from GMOS-IFU observations. Our IFU observations did not detect 
asymmetric line profiles or multiple  components in the base of H$\alpha$ profile 
as observed by \cite{OlmoGarcia2017} from their 1$\arcsec$-width long-slit observation. 

In the case of Mrk\,600, despite of the small VIMOS FoV, we observe that the southwestern 
part of the galaxy is slightly blueshifted with respect to the systemic velocity of 1016 km s$^{-1}$. 
The variation in velocity within the Mrk\,600 VIMOS-IFU FoV is $\sim$30 km s$^{-1}$. 
The v$_r$(H$\alpha$) maximum, in Mrk\,600, is located very closed to region no. 2, 
the position where an expanding shell has been reported, while the minimum is near to region no. 1.

\section{Discussion}\label{sect_discussion}

\subsection{Spatial variation of oxygen abundance}\label{subsect_distri_oxygen}
\begin{figure}
\includegraphics[width=75mm]{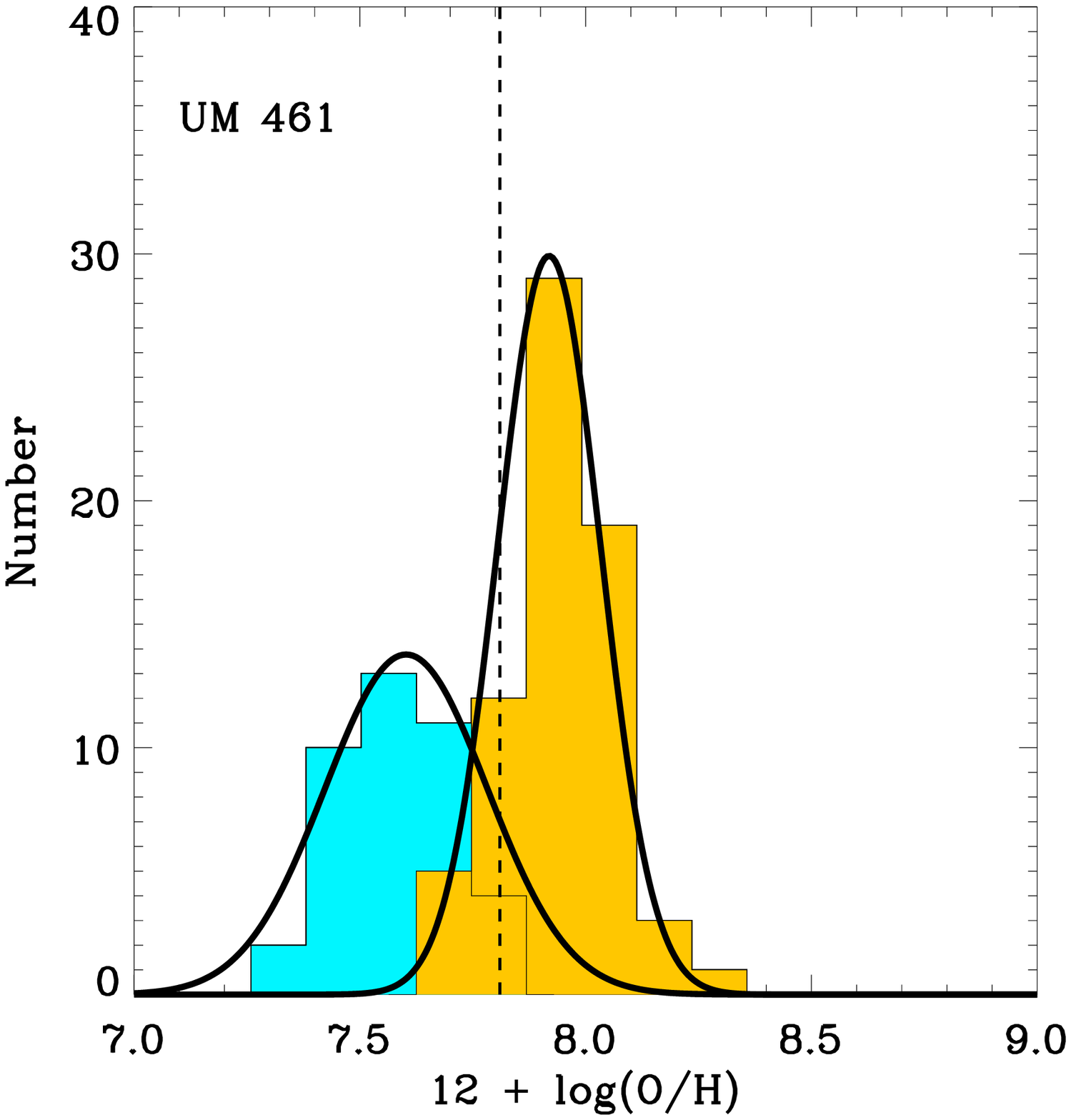}
\includegraphics[width=75mm]{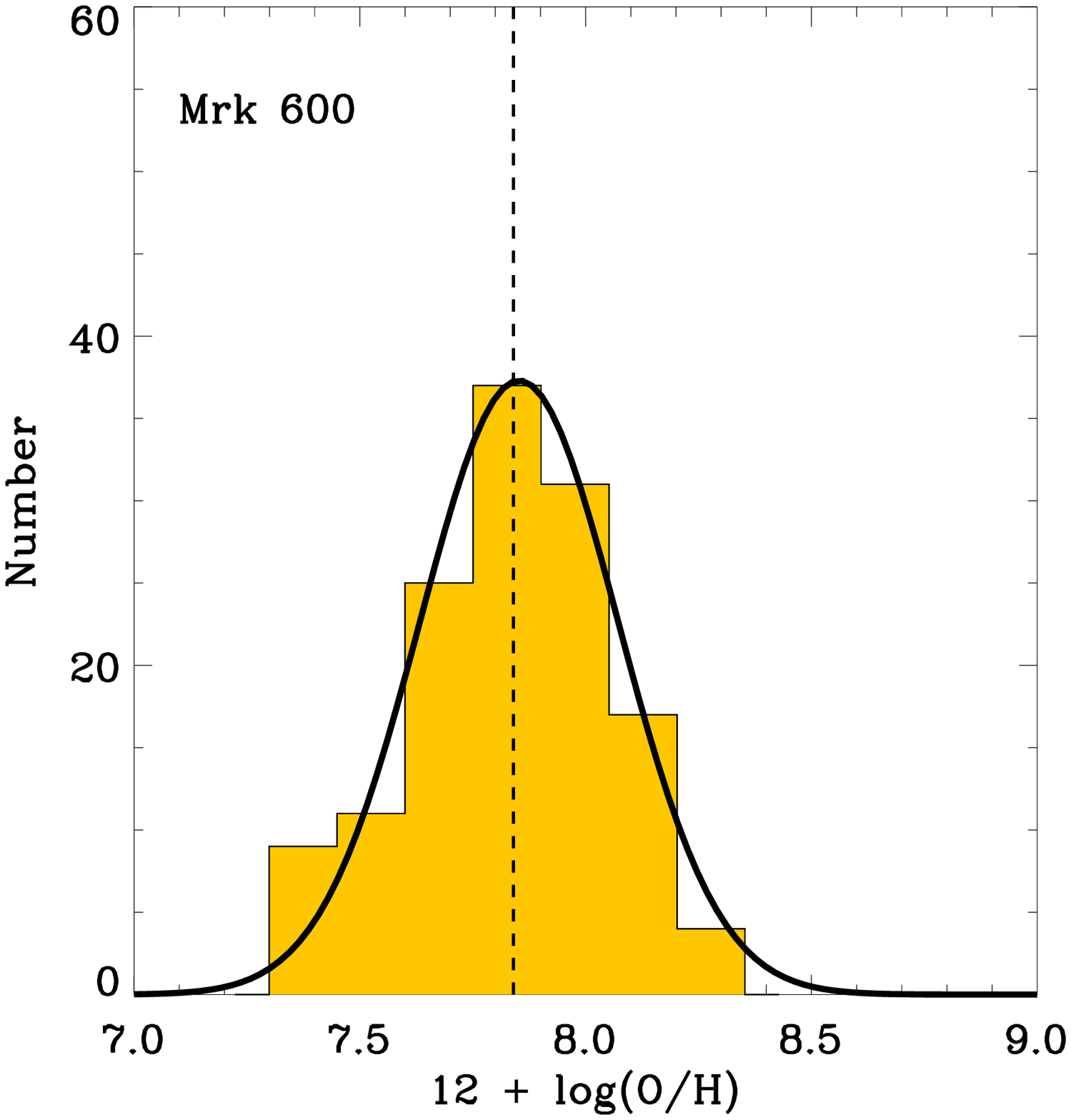}
  \caption{Histograms showing  the distribution of 12 + log(O/H) spaxel values for UM 461 (upper panel) 
  and Mrk 600 (lower panel). For each galaxy, the mean of the 12 + log(O/H) spaxel values 
  (7.81 and 7.84, respectively) is indicated with a dotted line. 
 The Figure also shows Gaussian fits to the distributions.}
\label{figure_OH_dist}
  \end{figure}

Here, we characterize the spatial distribution and variation of oxygen abundance
found in Section \ref{sect_abundances}.
In Figure \ref{figure_OH_dist} we show histograms of the distribution of the 12 + log(O/H) spaxel values for UM\,461 (upper panel) 
and Mrk\,600 (lower panel). The dotted lines in the Figure indicate the mean 12 + log(O/H) 
values of 7.81 and 7.84; with the same standard
deviations of $\sim$0.21 for UM 461 and Mrk 600, respectively. 
We note that the mean value of these distributions agree at the 1$\sigma$ level  
with the integrated 12 + log(O/H) values for the galaxies in Tables \ref{table_abundances_UM461} and \ref{table_abundances_Mrk600}. 
Interestingly, in the case of Mrk 600 the distribution can be fitted by a single Gaussian, while in the case of UM\,461 
the distribution is well fitted by two Gaussian components.

Following the statistical analysis in \cite{Perez-Montero2011} and \cite{Kehrig2016} 
we consider the two conditions for oxygen abundance to be considered homogeneous:
i) the derived values of 12 + log(O/H) should be fitted by a normal
distribution according to the Lilliefors test and ii) the observed variations 
of the data distribution around the mean values
$\sigma_{Gaussian}$ should be lower or of the order of the typical uncertainty
of the property considered.
The dispersion of the normal distribution $\sigma_{Gaussian}$ in Mrk 600 is of the order of the uncertainty
of the oxygen abundance, estimated as the square root of the weighted variance of the data points $\sigma_{weighted} \sim0.21$ .
While, in UM 461 $\sigma_{Gaussian}$ (=0.20) $> \sigma_{weighted}$ (=0.17). 
Individual statistical analysis of the Gaussian components in UM 461 shows that $\sigma_{Gaussian} \lesssim \sigma_{weighted}$.
Those results indicate that at large scales the ISM is chemically homogeneous.
In addition, we checked the null hypothesis that the data come from a normally distributed population by applying
the Lilliefors test. From this,  we do not have enough evidence to conclude that the data in Mrk 600 
were not drawn from a normal distribution (p-value $\sim$0.5). However, we find that the p-value of the Lilliefors test for UM 461 is
$\sim$0.0001, then it is significantly non-normal.
Finally, given that for each galaxy the mean of the 12 + log(O/H) spaxel value distribution agrees with the integrated value, the spatial 
variations observed in UM 461 cannot be understood as only statistical fluctuations \citep{Perez-Montero2011}.

We conclude that the spatial variation and extended low metallicity region in UM 461 
appear to be real, within our uncertainties, and it 
could indicate the recent infall of non-pristine metal-poor gas into  the galaxy. 
Alternatively, it could be produced by the outflow of a large amount of enriched gas, consequently diminishing 
the metal content in this region. Below in this Section we discuss those scenarios in the context of the main properties of the ISM 
and the triggering of star-formation.

\subsection{Oxygen abundance derivation using different diagnostics}\label{metal_gradient}

In this Section we compute  oxygen abundances using several different diagnostics and calibrations. 
First, the 12 + log(O/H) abundance was derived by applying the relation between the line ratio of 
[N\,{\sc ii}]$\lambda$6584/H$\alpha$ with the oxygen abundance \cite[N2;][]{Denicolo2002}, i.e., 
12 + log(O/H) = 9.12 + 0.73 $\times$ N2, with N2 = log ([N\,{\sc ii}] $\lambda$6584/H$\alpha$).
One of the most common methods used for estimating the oxygen abundance of metal-rich galaxies 
(12 + log(O/H) $\gtrsim$ 8.4) and also metal-poor galaxies (12 + log(O/H) $\lesssim$ 8.4) utilizes the 
R23 = (([O\,{\sc ii}]$\lambda$3727 + [O\,{\sc iii}]$\lambda\lambda$4959,5007)/H$\beta$) parameter, which is the 
ratio of the flux in the strong optical oxygen lines relative to  H$\beta$. 
Applying this method the oxygen abundance, in our case, is given by the metal-poor branch 
12 + log(O/H) = 7.056 + 0.767 x + 0.602 x$^2$ - y(0.29 + 0.332 x - 0.331 x$^2$),
where x = log(R23) and y = log(O32) = log([O\,{\sc iii}]$\lambda\lambda$4959,5007)/[O\,{\sc ii}]$\lambda\lambda$3726,3729) 
\citep[R23;][]{Kobulnicky1999}. 
Another widely used indicator of oxygen abundance is given by 12 + log(O/H) = 8.73 - 0.32 $\times$ O3N2, where 
O3N2 = log([O\,{\sc iii}] $\lambda$5007/H$\beta$ $\times$ H$\alpha$/[N\,{\sc ii}] $\lambda$6584) \citep[O3N2;][]{PettiniPagel2004}.
Additionally, we use a new calibrator which has a weak dependence on the ionization parameter given by 
12 + log(O/H) = 8.77 + Y, where Y = log([N\,{\sc ii}] $\lambda$6584/[S\,{\sc ii}]$\lambda\lambda$6717,6731) 
+ 0.264 $\times$ N2 \citep[D2016;][]{Dopita2016}.
In order to cross check our results with previous determinations 
in the literature \citep[e.g.][]{Sanchez2015} we use the code HII--CHI--mistry version 2.1 \citep[][]{Perez-Montero2014}
which is a python program that calculates the 12 + log(O/H) abundance for gaseous nebulae ionized by massive stars 
using a set of emission-line optical intensities, i.e., [O\,{\sc ii}]$\lambda$3727/H$\beta$, [O\,{\sc iii}]$\lambda$4363/H$\beta$, 
[O\,{\sc iii}]$\lambda$5007/H$\beta$, [N\,{\sc ii}]$\lambda$6584/H$\beta$ and [S\,{\sc ii}]$\lambda\lambda$6717,6731/H$\beta$.

\begin{figure*}
\includegraphics[width=200mm]{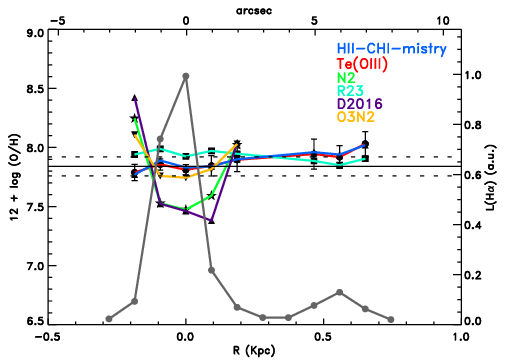}\\
  \caption{UM\,461: radial distribution of oxygen abundance for simulated long slit observations across 
           the main body of the galaxy. Six  determinations of oxygen abundance are indicated in this Figure: 
           i) the direct method (circles, red line), ii) N2 (stars, green line), iii) R23 (squares, cyan line), 
           iv) O3N2 (upside down triangle, orange line), v) D2016 (triangles, purple line) calibrators and vi) 
           HII--CHI--mistry (triangles, blue line).
           The integrated 12 + log(O/H) = 7.84$\pm$0.08 obtained for UM 461 is represented 
           by the horizontal black line, with its 1$\sigma$ error indicated with dotted lines. We also include 
           the normalized H$\alpha$ luminosity (L) profile (gray line and dots), with the maxima at regions 
           no. 1 and no. 2.}
  \label{figure_O-R_UM461}
\end{figure*}

For UM\,461 we simulated long-slit observations along the main body of the galaxy, assuming 
a slit width of $\sim$1$\arcsec$. Figure \ref{figure_O-R_UM461} shows the radial 
profile of the oxygen abundance with respect to the UM\,461 peak of H$\alpha$ emission using the direct method 
(determination of oxygen abundance using the T$_e$) compared to oxygen abundances determined using 
the other methods described above. 

The integrated oxygen abundance in region no. 2 of UM\,461, based on T$_e$(O\,{\sc iii}), was determined to be 0.06 dex
higher than in region no. 1. This difference is reflected in the slight gradient observed in Figure \ref{figure_O-R_UM461}.
In this Figure we see that the R23 method provides similar values for 12 + log(O/H) as the direct 
method within the uncertainties.
On the other hand, the values based on the empirical N2 calibration in region no. 1 are $\sim$0.4 dex lower than 
those obtained from the direct method. Using D2016 gives the same relative values compared to N2.   
The oxygen abundances obtained using the O3N2 calibrator provide a similar lower but less extreme profile compared to those  
obtained using the N2 and D2016 calibrations. However, most of its values agree within the uncertainties
with the ones found by the direct method. We find a good agreement between abundances computed using 
HII--CHI--mistry and by the direct method.
It is clear that using the N2 and D2016 methods alone to study the spatial variation of oxygen abundance underestimates
the abundance profile in region no. 1. 

\begin{figure*}
\centering
\includegraphics[width=4.3cm]{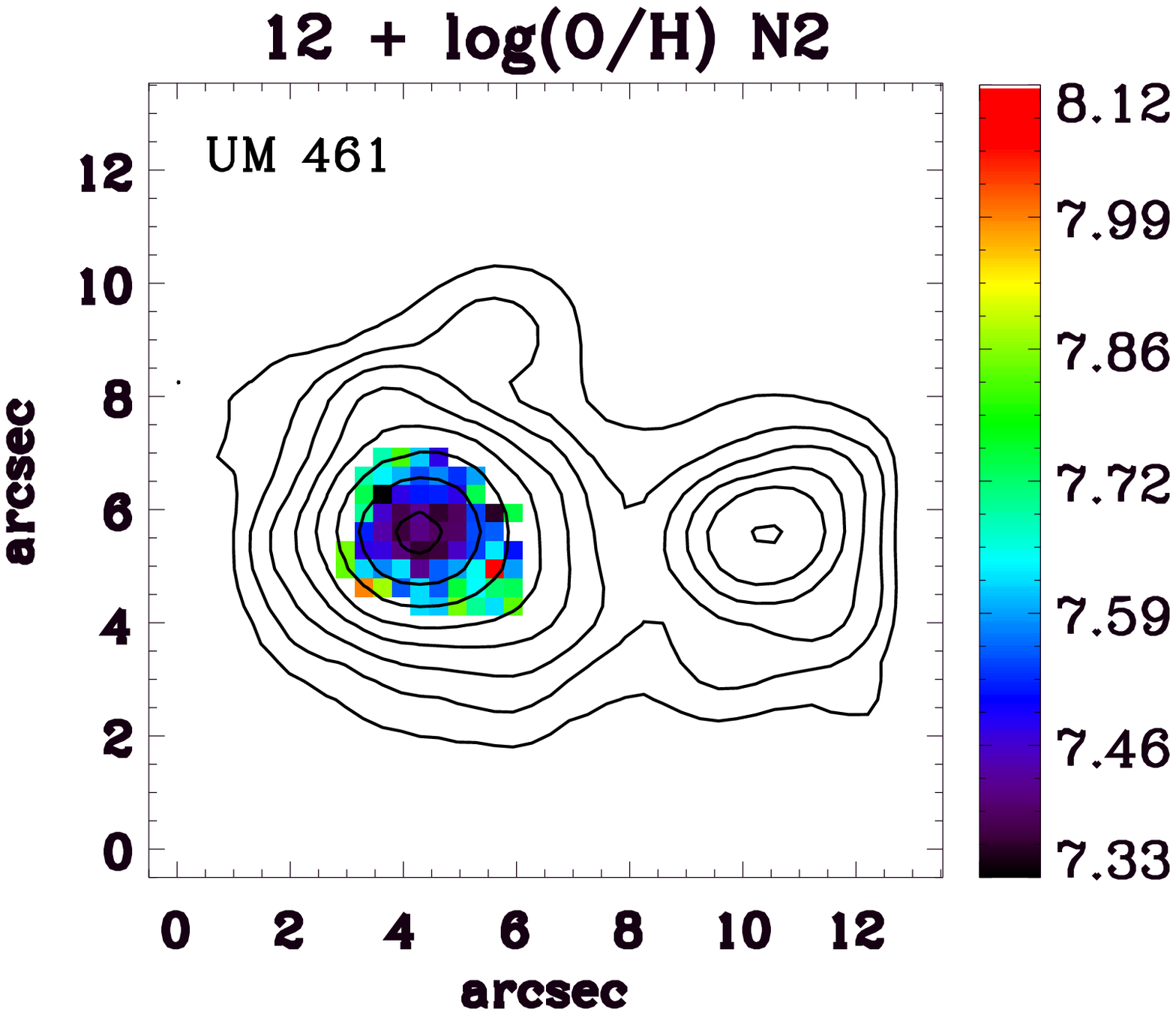}
\includegraphics[width=4.3cm]{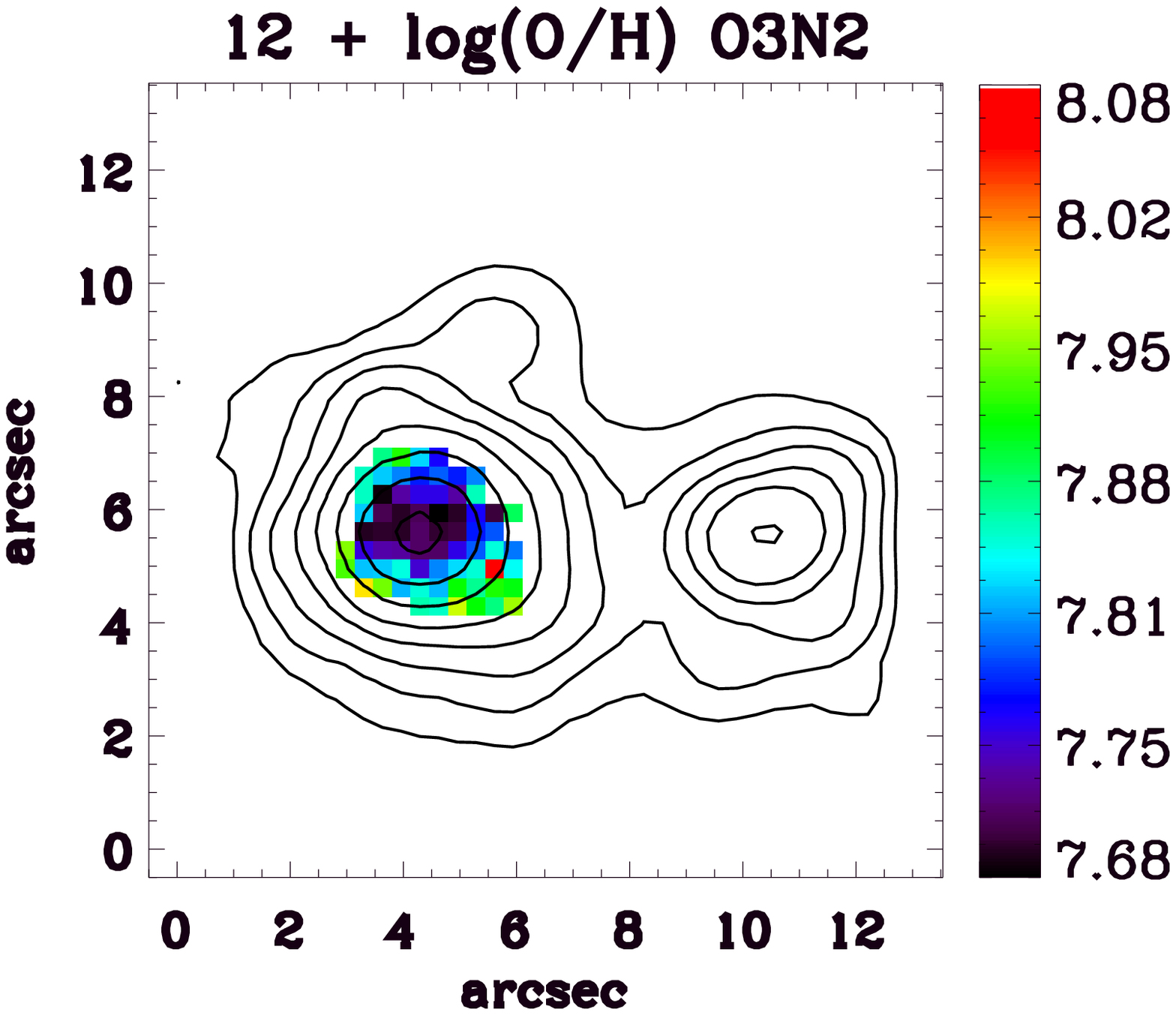}
\includegraphics[width=4.3cm]{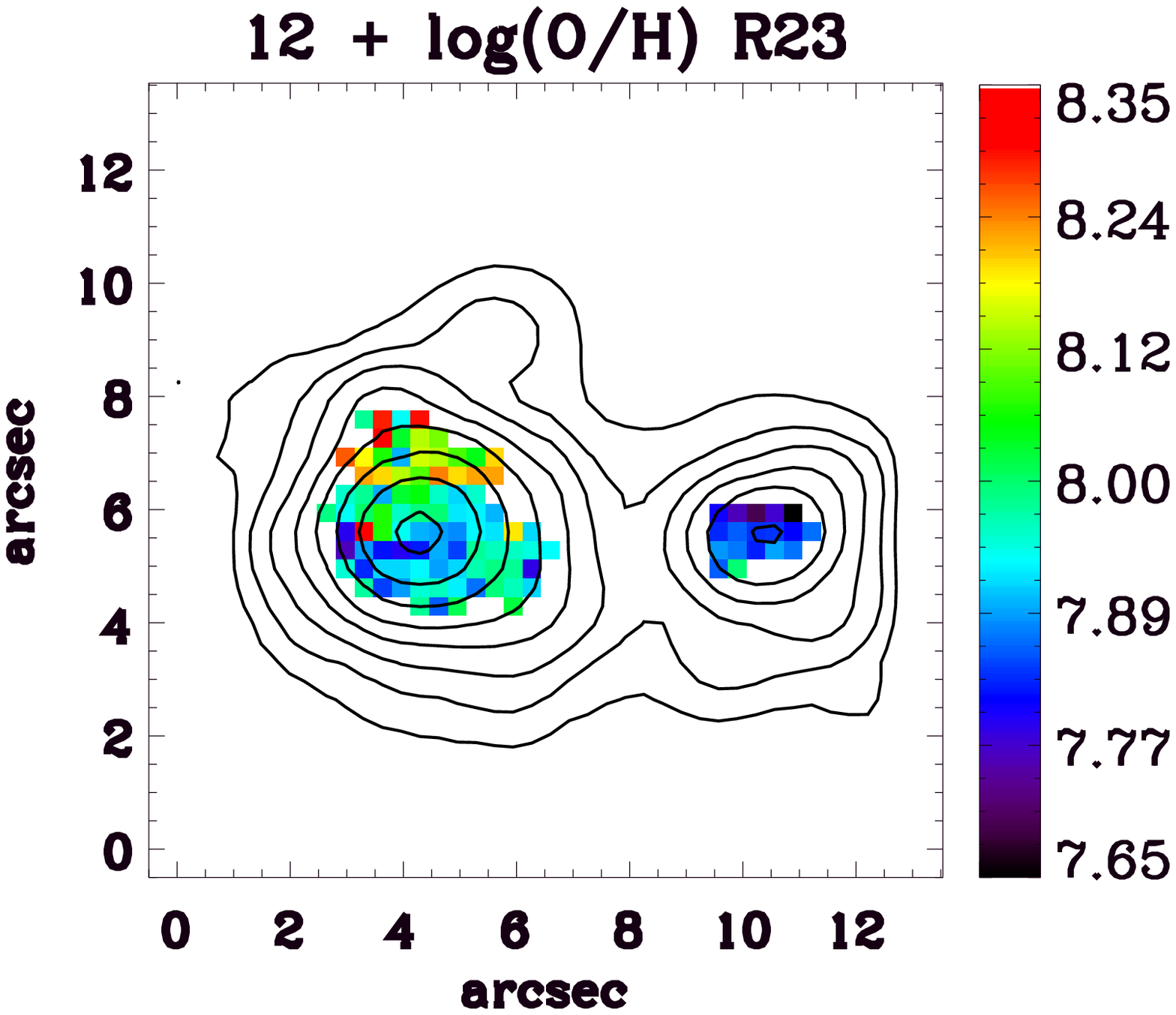}
\includegraphics[width=4.3cm]{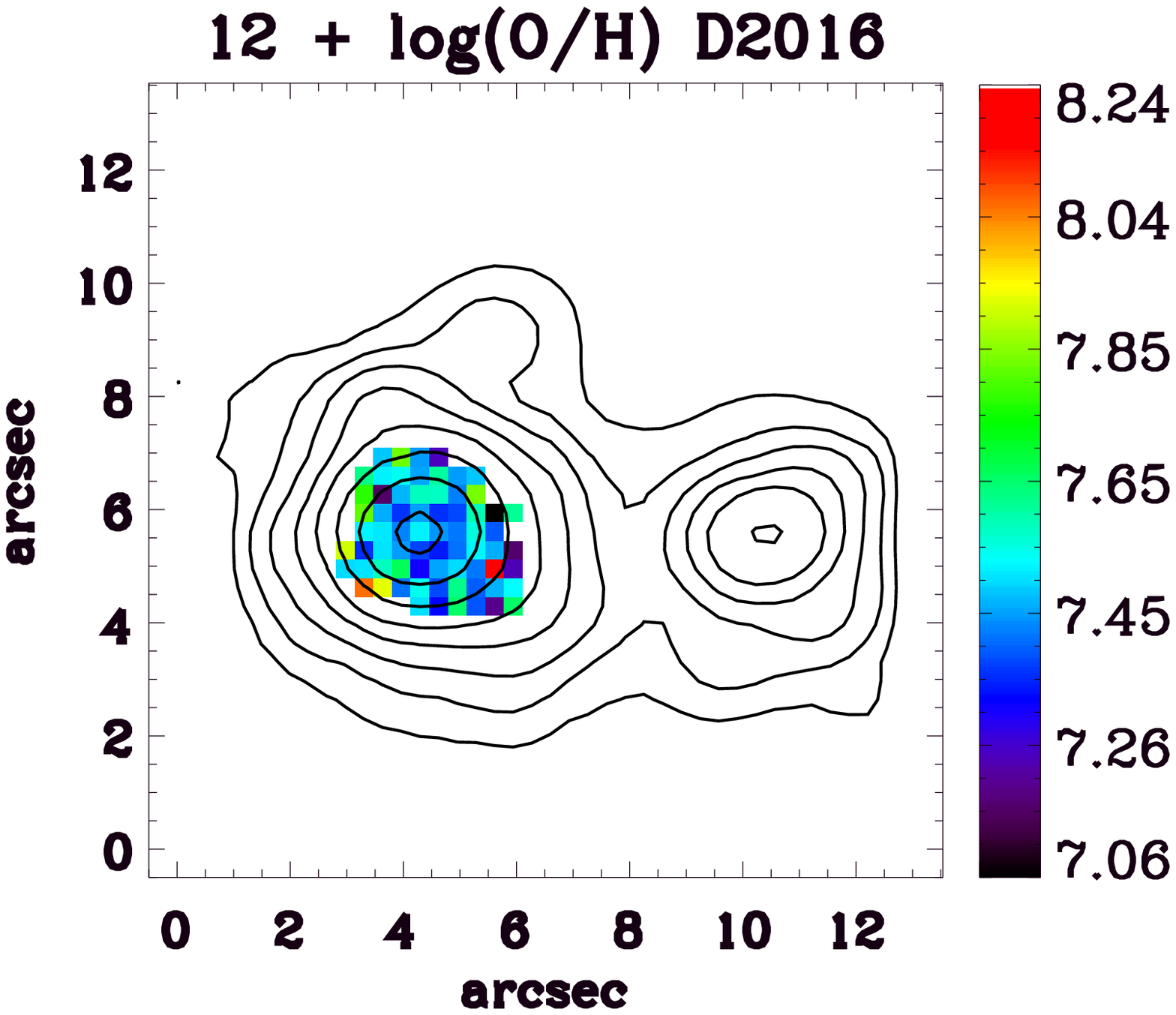}\\
\includegraphics[width=4.3cm]{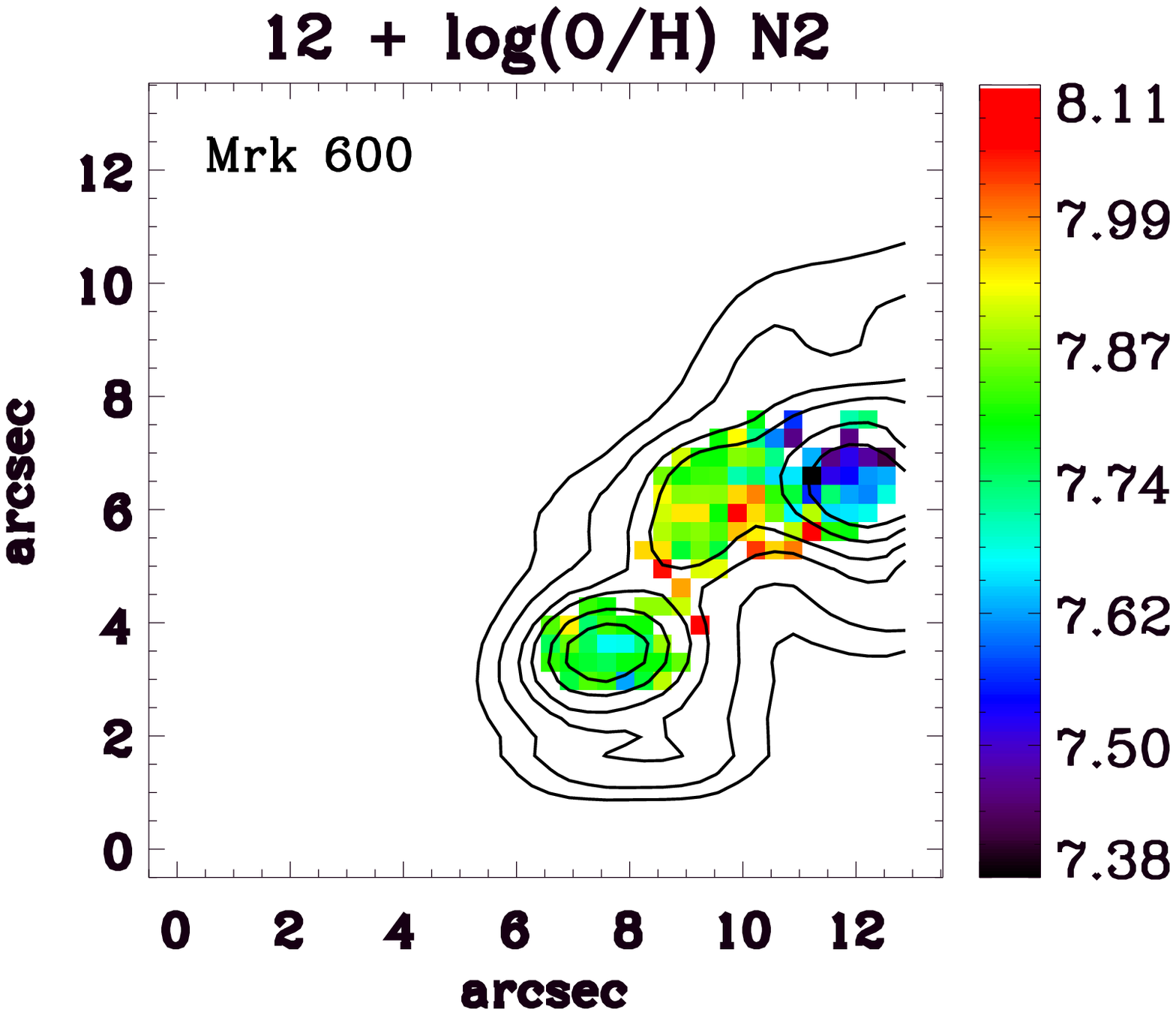}
\includegraphics[width=4.3cm]{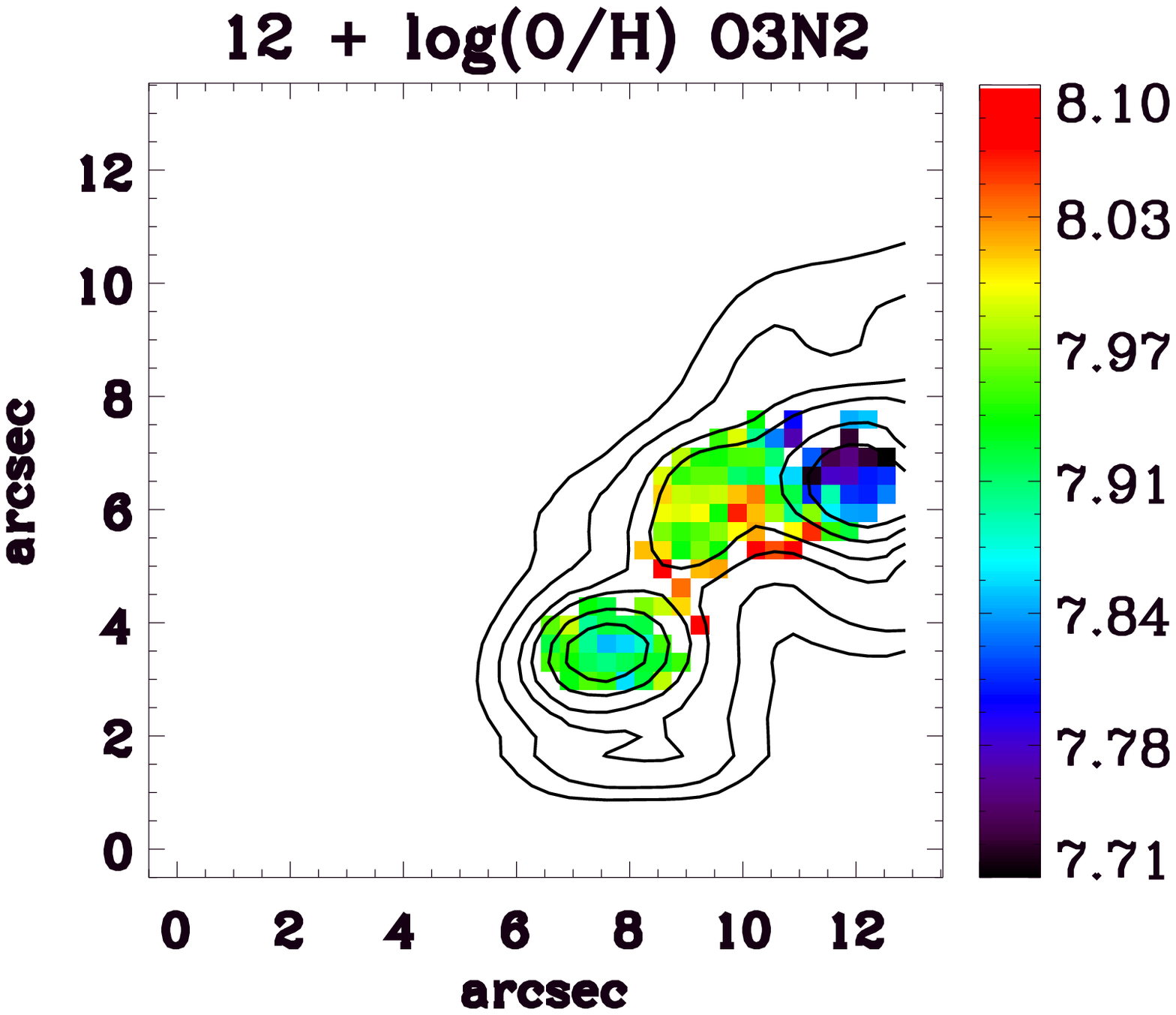}
\includegraphics[width=4.3cm]{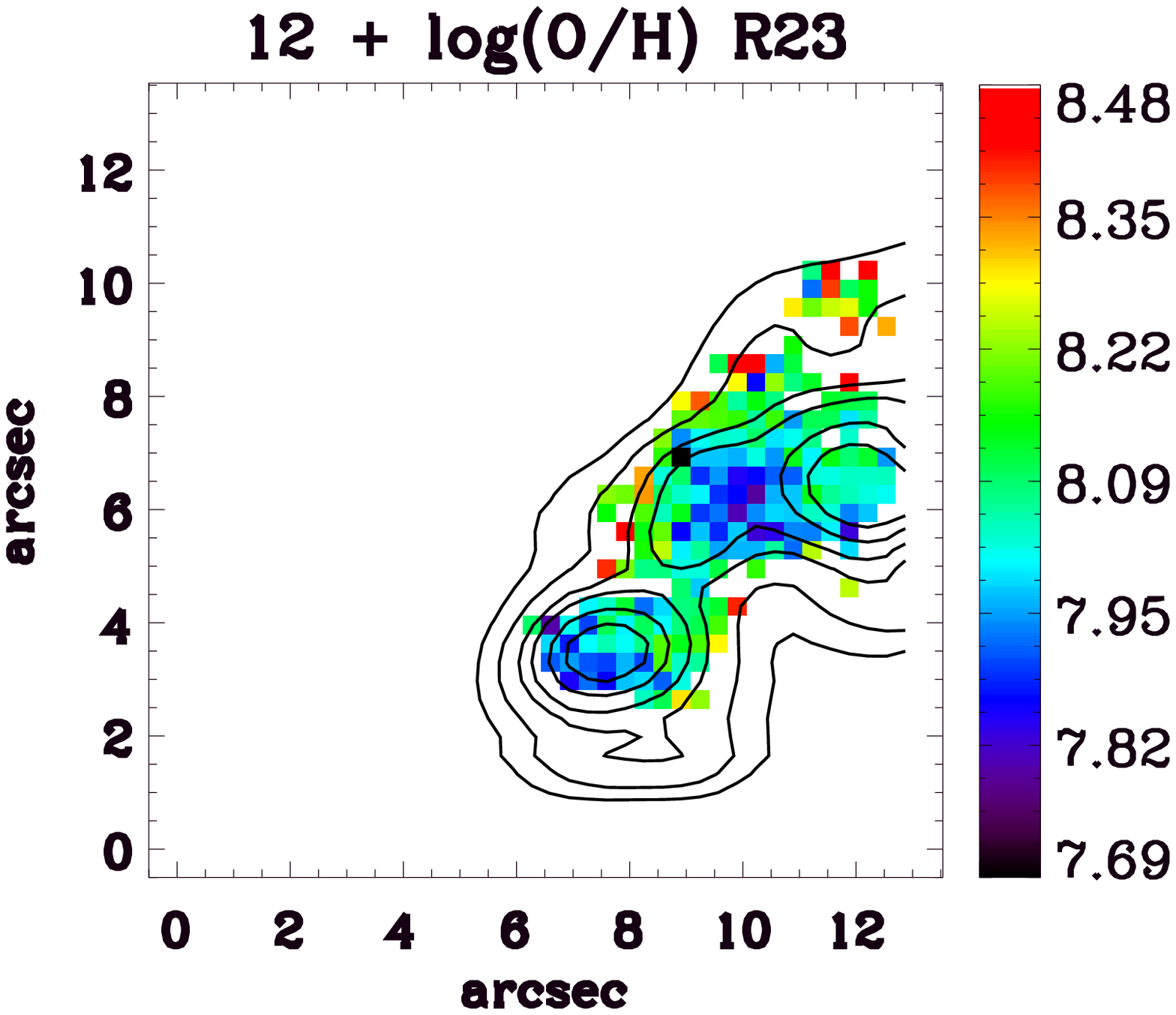}
\includegraphics[width=4.3cm]{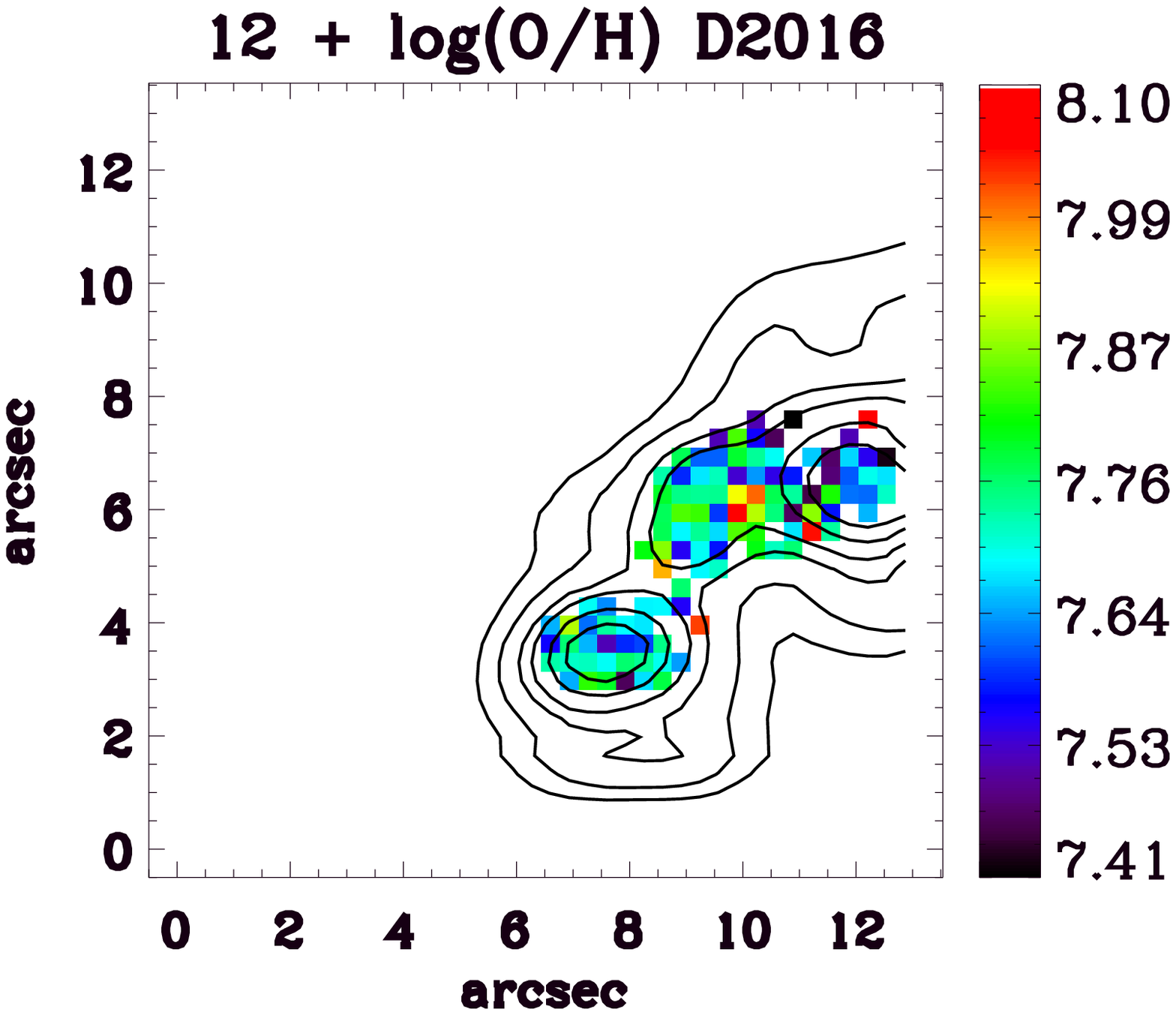}
\caption{12 + log(O/H) maps for UM 461 (Top row) and Mrk 600 (Bottom row) obtained using 
         the N2, O3N2, R23 and D2016 calibrators.
         Contours display the H$\alpha$ morphology of the galaxies. North is up and east is to the left.}
\label{figure_2D_calibrators}%
\end{figure*}

\begin{figure}
\centering
\includegraphics[width=4.1cm]{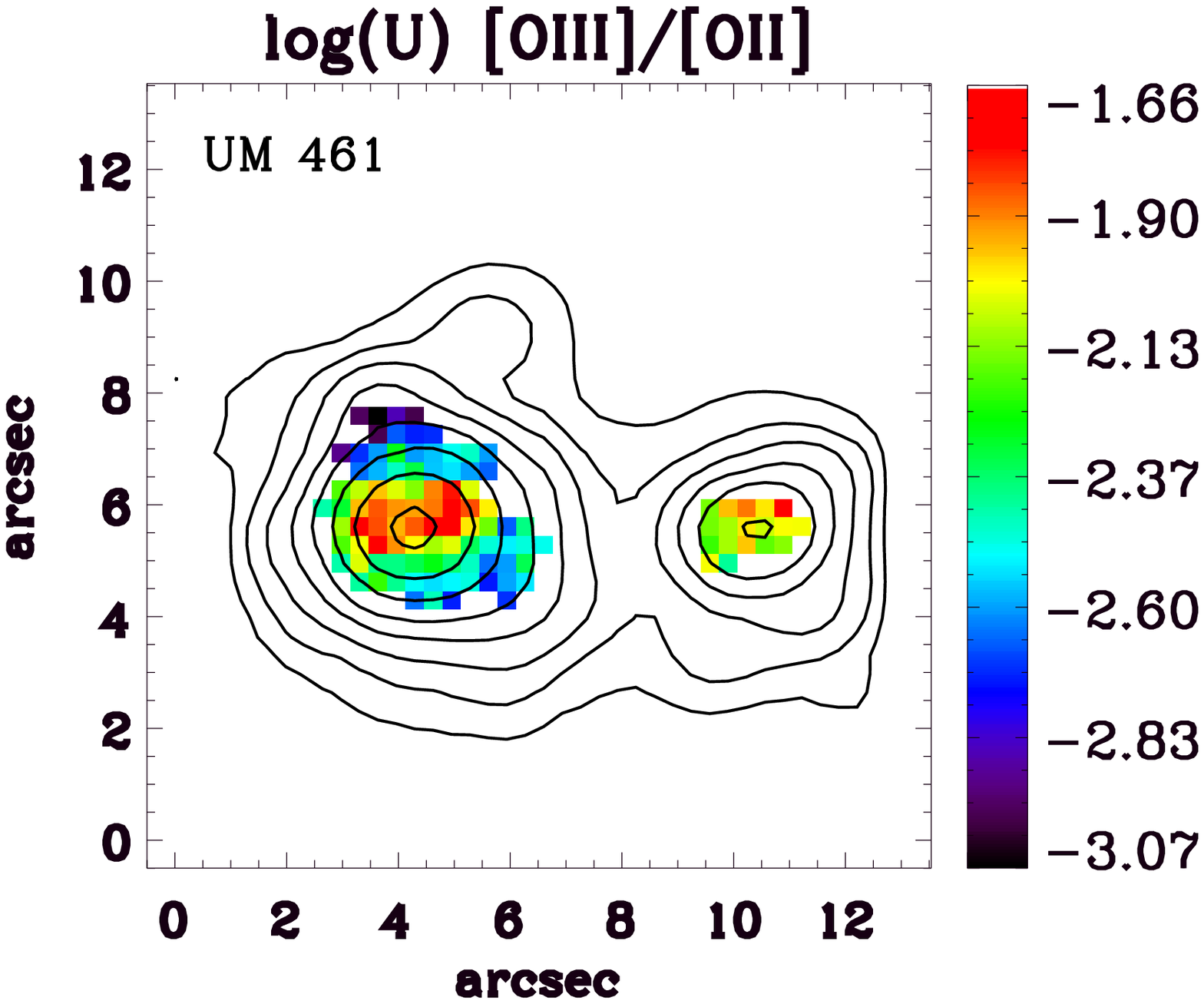}
\includegraphics[width=4.1cm]{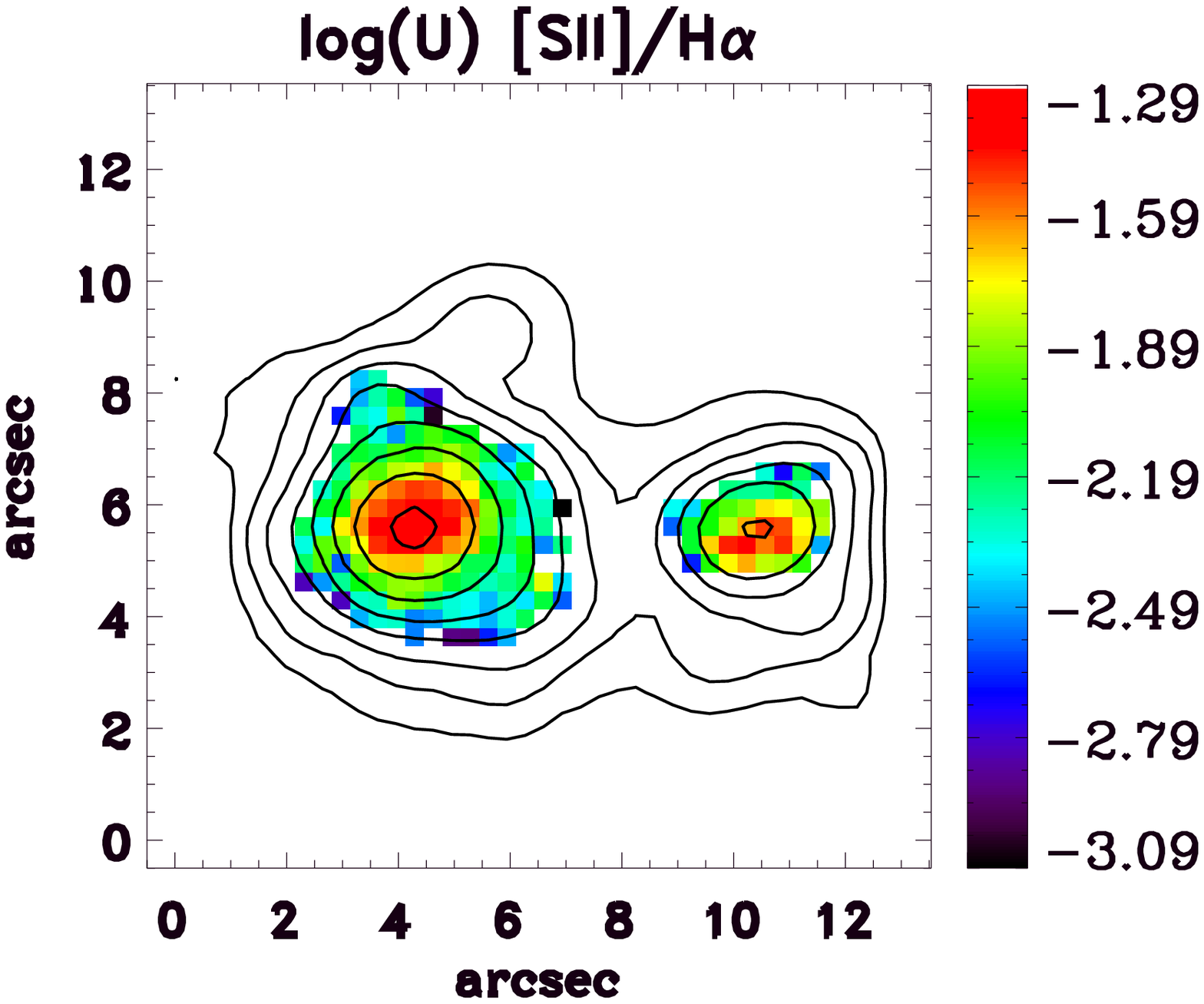}\\
\includegraphics[width=4.1cm]{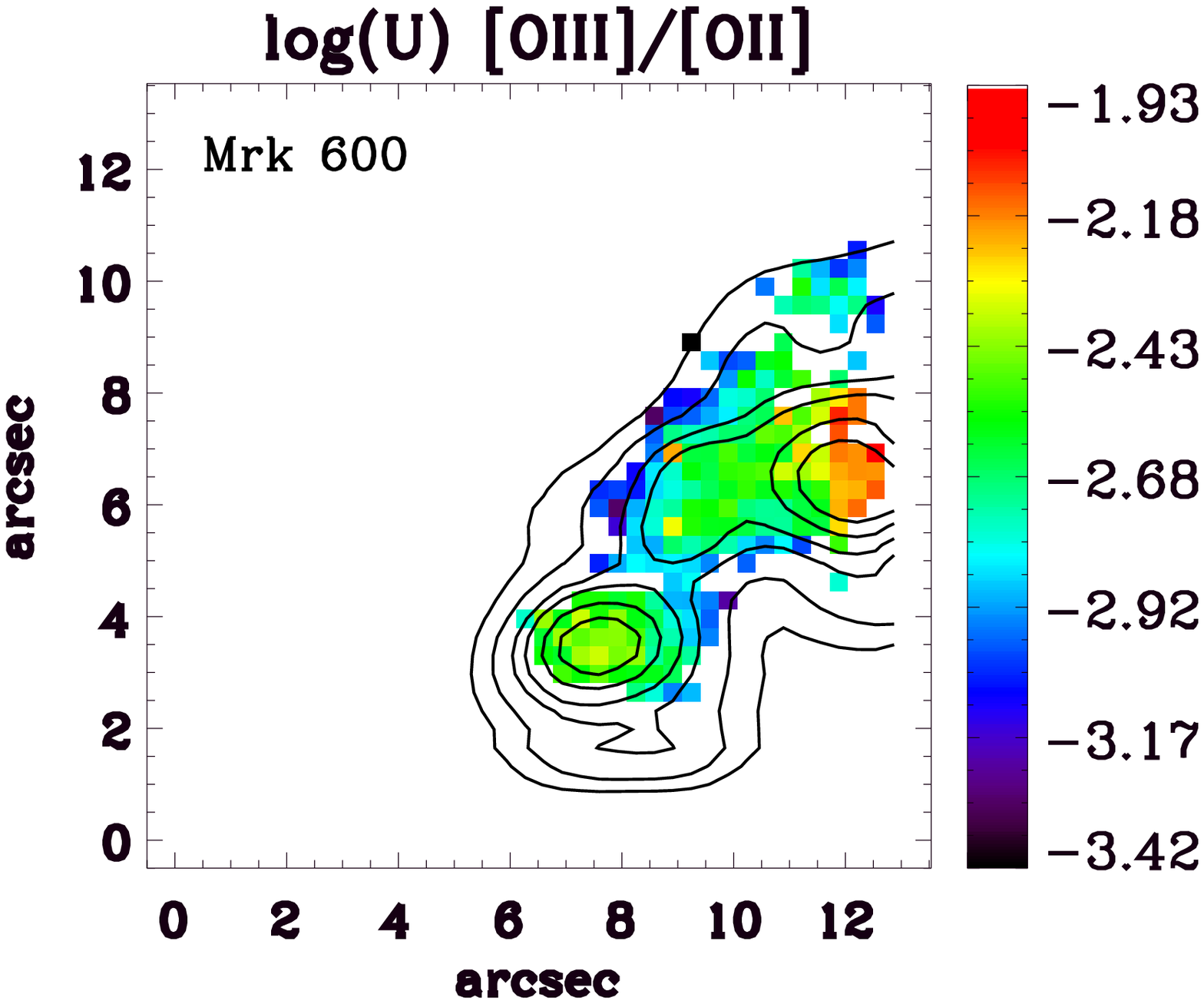}
\includegraphics[width=4.1cm]{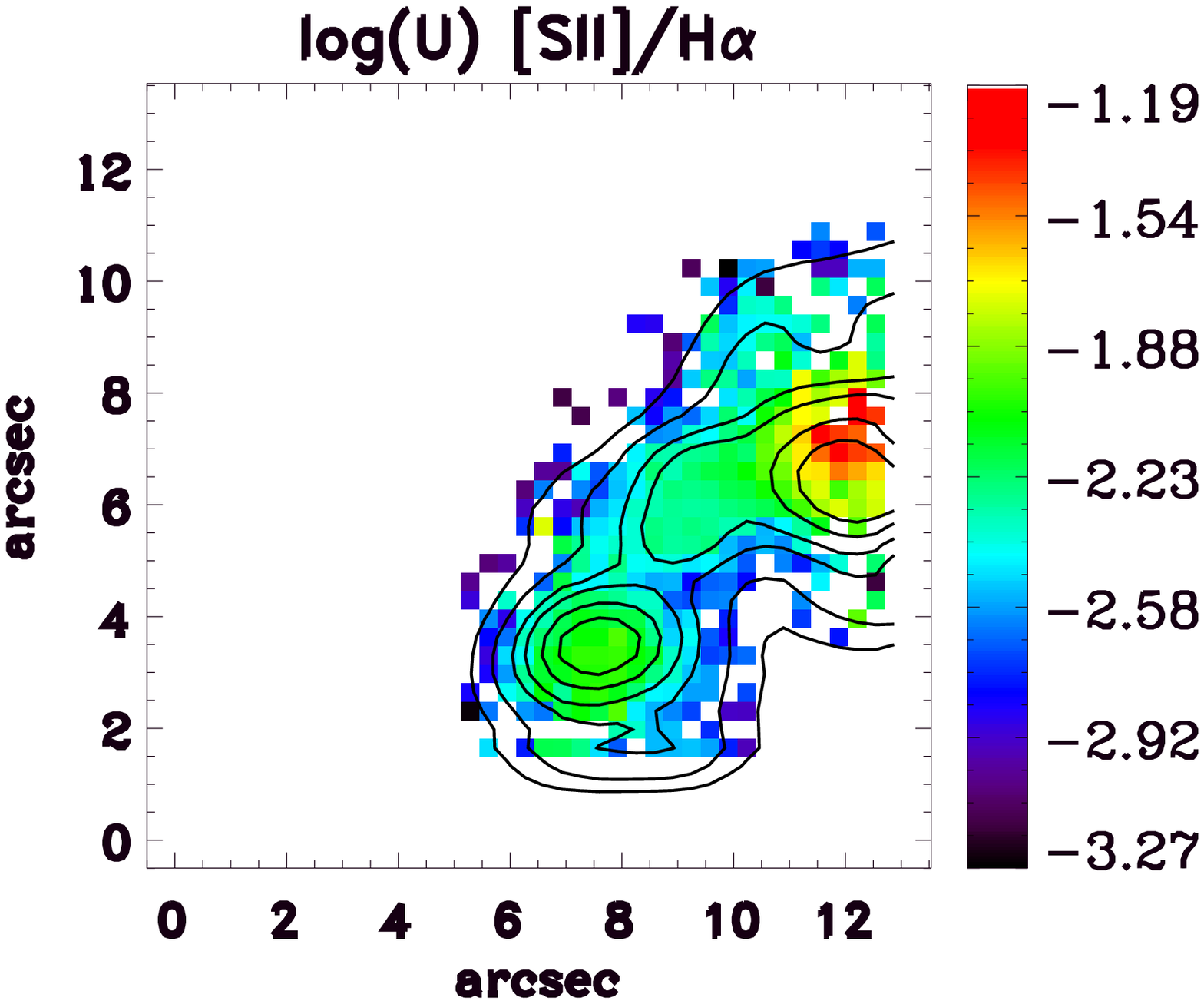}
\caption{Ionisation parameter $U$ as mapped from D\'iaz et al. (2000) ([O\,{\sc iii}]/[O\,{\sc ii}]) 
         and Dors et al. (2011) ([S\,{\sc ii}]/H$\alpha$), for UM 461 (upper panels) and Mrk 600 (lower panels).
         Contours display the H$\alpha$ morphology of the galaxies.}
\label{figure_U_dist}
\end{figure}

In Figure \ref{figure_2D_calibrators} we show the 12 + log(O/H) maps, for both galaxies, 
obtained using the N2, O3N2, R23 and D2016 oxygen abundance calibrators.
From this Figure we  observe significant spatial differences between 
the absolute values in the direct method maps compared to maps derived using some of the calibrators.
The N2 and O3N2 abundance maps  show spatial trends that are opposite to those shown 
by the other methods. 
Interestingly, both N2 and O3N2 show the lowest values in the regions of higher star-formation.
Despite the fact that a detailed analysis is beyond the scope of the present paper,
these results suggest a dependence on the ionization parameter $U$\footnote{The ratio of the ionizing photon
density  to  the  particle  density.} \citep[e.g.][]{KewleyDopita2002}. 
This value can be measured from the ratio of high ionisation to low ionisation species 
[O\,{\sc iii}]$\lambda$5007/[O\,{\sc ii}]$\lambda\lambda$3726,3729
using the parametrisation presented by \cite{Diaz2000}, i.e., log($U$) = -0.8$\times$log([OII]/[OIII]) - 3.02 and also 
by the emission line ratio [S\,{\sc ii}]$\lambda\lambda$6717,6731/H$\alpha$, 
log($U$) = -1.66$\times$log([SII]/H$\alpha$) - 4.13 \citep{Dors2011}. 
In Figure \ref{figure_U_dist} we show the log($U$) maps for both galaxies.
Clearly, the log($U$) is highest at the position of the GH\,{\sc ii}Rs and decreases radially outwards.
The spatially resolved shape of abundances based on N2 and O3N2 correlates with the ionisation parameter.  
The shape and relative values of R23 maps agrees reasonably well with the direct method indicating only 
a weak dependence on $U$ in our sample of H\,{\sc ii}/BCDs \citep[e.g.][and references therein]{Kehrig2016}. 
While, the D2016 maps does not closely correlate with the direct method and the log($U$) maps.
The latter assumes the ISM conditions in high-z galaxies, 
which differ from those found in local galaxies. In Table \ref{table_mean_OH} we show the mean values of 12 + log (O/H)
obtained from the different methods.
We find a difference between the mean value of R23 and the direct method of 0.16 dex and 0.23 dex
for UM 461 and Mrk 600, respectively. While, the N2 and D2016 methods underestimate the mean oxygen abundance in UM 461.
In summary, we conclude that most of the above methods provide a reasonable estimation of the integrated oxygen abundance,
but some are less suited for a detailed study of spatial variations within the ISM of our BCDs.
The observed variation of line ratios could be due to variations of $U$ instead of real metallicity variations.
Below, we consider HII--CHI--mistry as a reliable tracer of the spatially resolved oxygen abundance when compared to the direct method.

\begin{table}
 \centering
 \begin{minipage}{80mm}
  \caption{Statistical properties of 12 + log (O/H) using different methods.}
 \begin{tabular}{@{}ccccc@{}}
  \hline
                                &  \multicolumn{2}{c}{UM 461}   &  \multicolumn{2}{c}{Mrk 600}\\   
 \hline
                                & Mean & STD\footnote{Standard deviation} & Mean & STD\\
 \hline
T$_e$      & 7.81 & 0.21 & 7.84 & 0.21\\
N2         & 7.59 & 0.16 & 7.85 & 0.17\\
O3N2       & 7.81 & 0.10 & 7.96 & 0.10\\
R23        & 7.97 & 0.14 & 8.07 & 0.14\\
D2016      & 7.51 & 0.20 & 7.75 & 0.16\\
HCM\footnote{HII--CHI--mistry} with [O\,{\sc iii}]$\lambda$4363    & 7.83 & 0.21 & 7.86 & 0.21\\
HCM without [O\,{\sc iii}]$\lambda$4363 & 7.94 & 0.15 & 8.02 & 0.13\\

\hline
\end{tabular}
\label{table_mean_OH}
\end{minipage}
\end{table}

\begin{figure*}
\centering
\includegraphics[width=15cm]{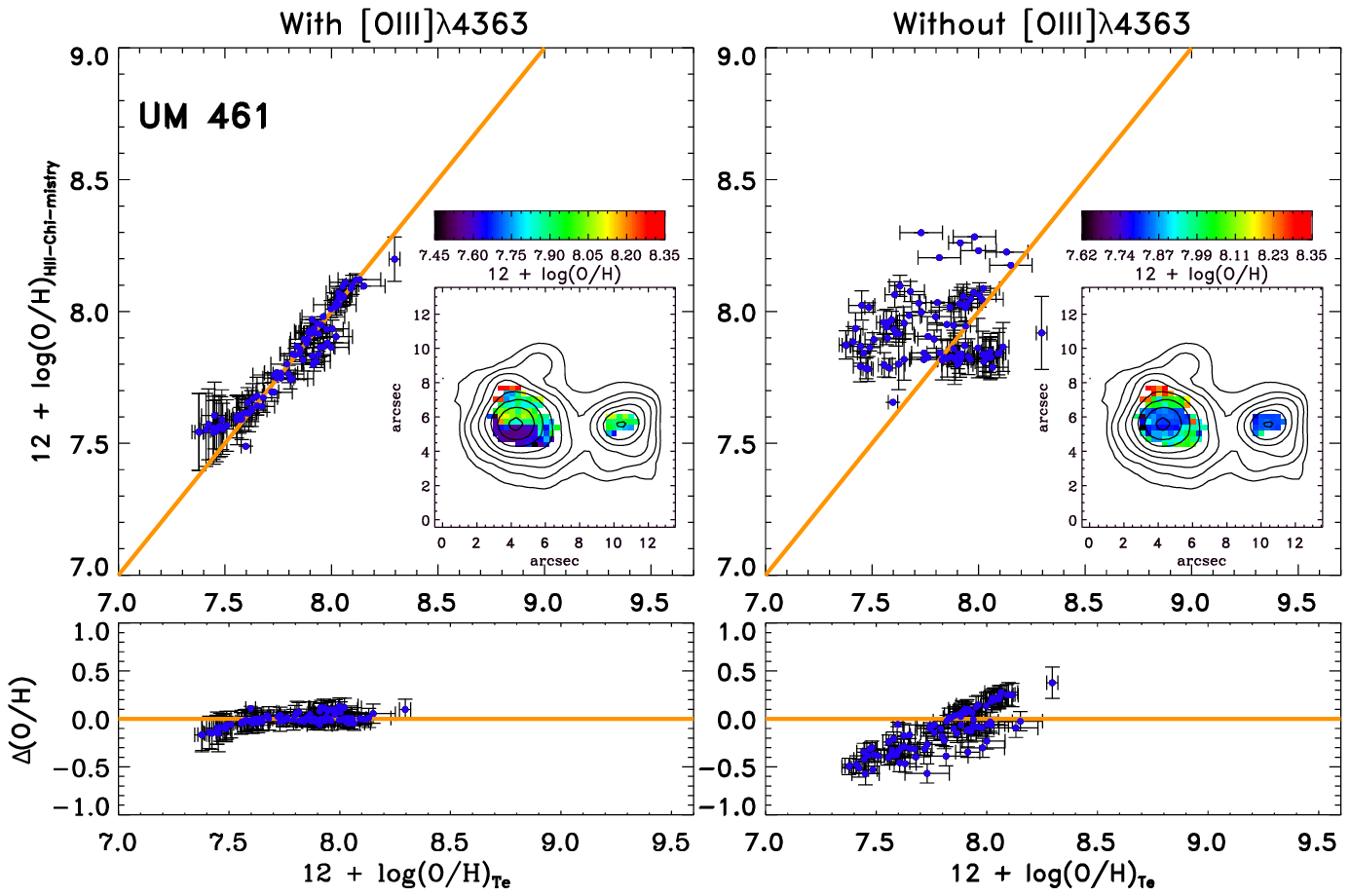}
\includegraphics[width=15cm]{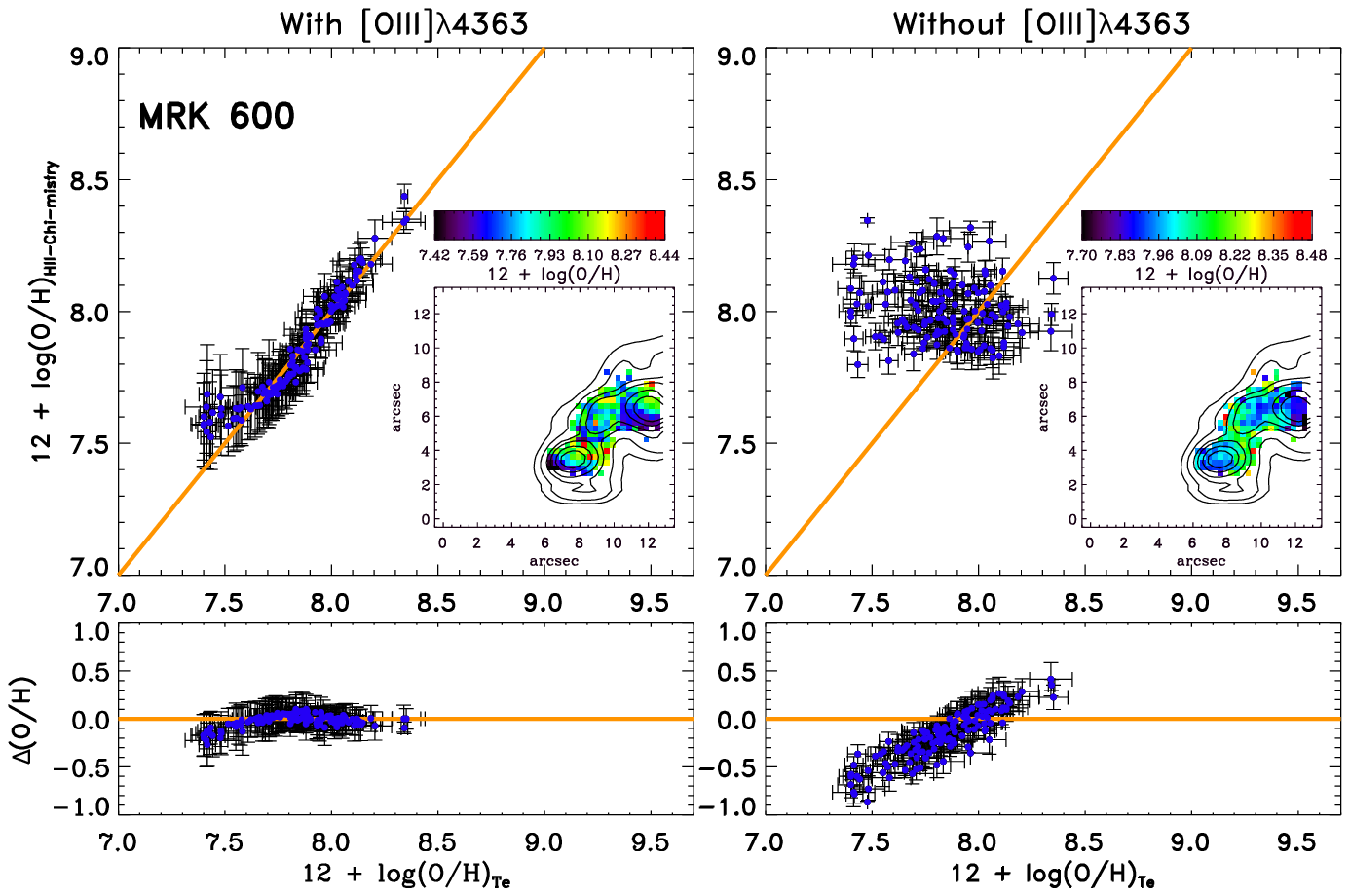}
\caption{Spaxel by spaxel comparison between 12 + log(O/H) derived using HII-CHI-mistry \citep{Perez-Montero2014} and the T$_{e}$ 
         method for the galaxies UM 461 (upper panels) and Mrk 600 (lower panels). 
         $\Delta$(O/H) is defined as log(O/H)$_{T_e}$ - log(O/H)$_{HII-CHI-mistry}$. 
         Contours display the H$\alpha$ morphology of the galaxies.}
\label{figure_comp_oxygen}%
\end{figure*}

We create the oxygen abundance maps (see Figure \ref{figure_comp_oxygen}, inset panels) of the galaxies 
using the HII--CHI--mistry code as indicated above.
It is important to note that the results obtained by using HII--CHI--mistry 
provide abundances that are consistent with the direct method,
only when the [O\,{\sc iii}]$\lambda$4363/H$\beta$ intensity is included.
In order to better illustrate this, in Figure \ref{figure_comp_oxygen} we show 
the spaxel by spaxel comparison between 12 + log(O/H) derived using the direct method and 
HII--CHI--mistry \citep{Perez-Montero2014} with (left panels) and without (right panels) [O\,{\sc iii}]$\lambda$4363 emission
for UM 461 (upper panels) and Mrk 600 (lower panels), respectively.
From this, we find a mean difference between the direct method and HII-CHI-mistry of $\mid\Delta$(O/H)$\mid =$0.02 dex
for both galaxies (see Table \ref{table_mean_OH}) when we include the [O\,{\sc iii}]$\lambda$4363 intensity.
While the mean difference without [O\,{\sc iii}]$\lambda$4363 is 0.13 dex and 0.18 dex for UM 461 and Mrk 600, respectively.
These checks show consistency with the values obtained in Section \ref{sect_abundances}
and also in Figure \ref{figure_O-R_UM461}, when  [O\,{\sc iii}]$\lambda$4363 emission is included as an input for the code. 
We note that when we use [O\,{\sc iii}]$\lambda$4363/H$\beta$ the code overestimate the abundances \citep{Sanchez2016} 
as compared to the direct method at 12 + log (O/H) $\lesssim$ 7.6. These differences are small compared with their uncertainties.
However, the 12 + log(O/H) abundance maps created with and without [O\,{\sc iii}]$\lambda$4363 emission do not show 
any spatially consistency each other.

In the case of UM 461 our results (see Figure \ref{figure_O-R_UM461}) differ from those in \cite{OlmoGarcia2017}, whose values appear 
consistent with using HII--CHI--mistry, excluding the [O\,{\sc iii}]$\lambda$4363 intensity. 
We therefore conclude that using HII--CHI--mistry recovers the oxygen abundance values, within the errors, and spatial variation 
of abundances obtained  with the use of the direct method, only when the [O\,{\sc iii}]$\lambda$4363/H$\beta$ intensity is included. 
We emphasise that our results clearly show that metallicity can appear to drop in regions 
of high star-formation activity (see Figure \ref{figure_comp_oxygen}) if  [O\,{\sc iii}]$\lambda$4363 emission 
is not considered, which can lead to a misinterpretation of the real variation of oxygen abundances across the objects. 
Therefore, special attention must be paid to which emission lines are used when the HII--CHI--mistry method 
is applied to the study of spatial variation of abundances.

\subsection[]{Spatially resolved star-formation, starburst properties and chemical abundances}\label{sect_star-formation}

\begin{figure*}
\includegraphics[width=85mm]{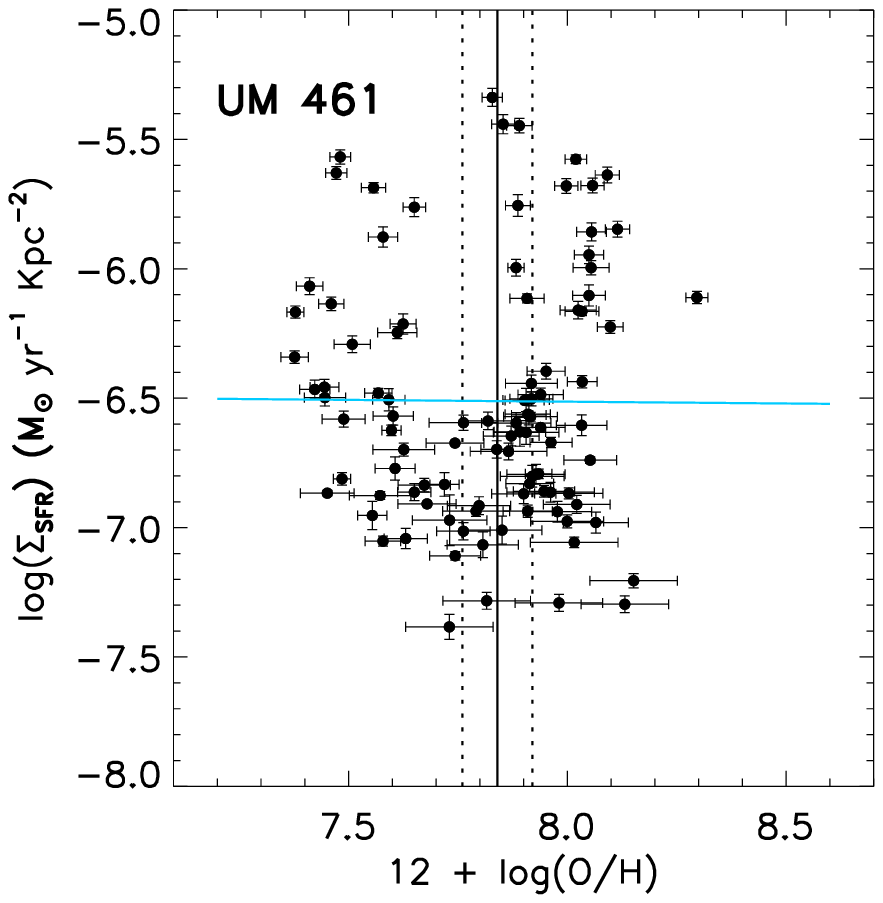}
\includegraphics[width=85mm]{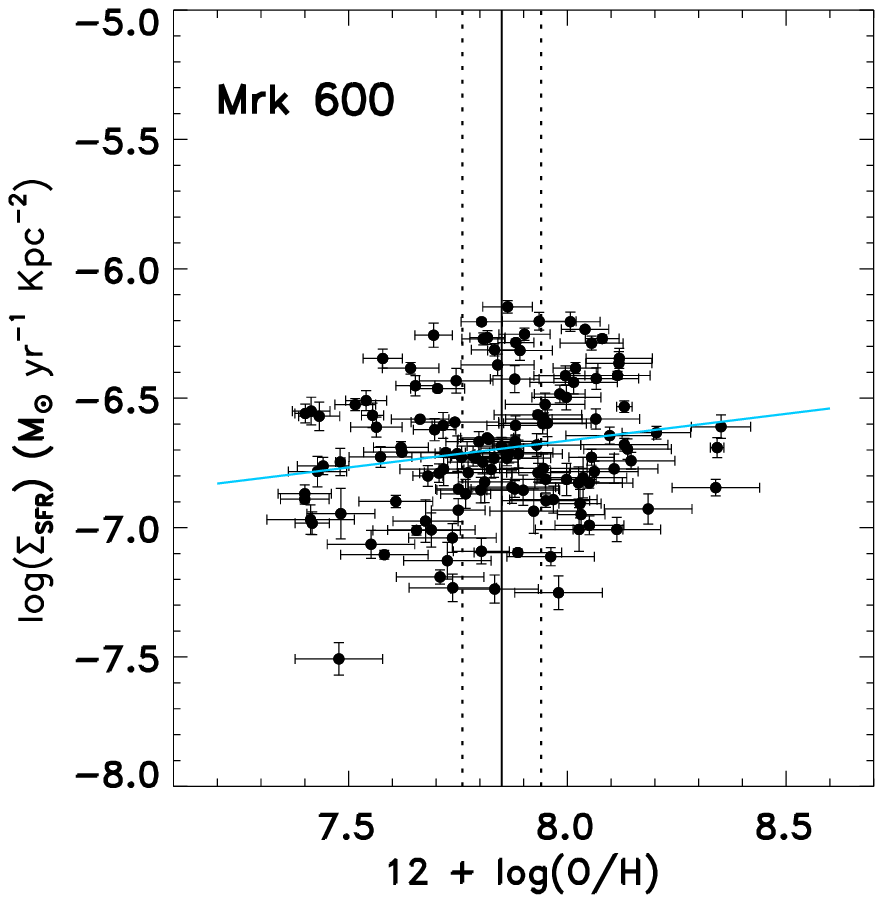}
  \caption{Relation between the star-formation rate Log($\Sigma_{SFR}$) and oxygen abundance 12 + log(O/H) 
  at spaxel scales. The cyan line represents the linear fit to this relation. 
  The integrated 12 + log(O/H) = 7.84$\pm$0.08 and 7.85$\pm$0.09 obtained for UM 461 and Mrk 600 are represented 
  by the vertical solid lines, while the errors at the 1$\sigma$ level are shown as dotted lines.}
\label{figure_SFR_OH}
  \end{figure*}

In this Section, we discuss whether or not the observed star-formation traced by \halpha\ correlates 
with the estimated oxygen abundance at spaxel scales.  
The current star-formation rate (SFR) was inferred from the extinction-corrected H$\alpha$ emission 
and the \cite{Kennicutt1998} formula, after correction for a Kroupa initial mass function \citep{Calzetti2007}. 
Accordingly, we found a SFR = 0.077 M$_{\odot}$ yr$^{-1}$ and $>$0.017 M$_{\odot}$ yr$^{-1}$ for UM 461 and Mrk600, respectively.
In the case of Mrk 600, an important fraction of the ISM is outsize the FoV (see Figure 1), then 
we found a lower limit for the SFR in this object.
Note that, following common practice, SFRs are estimated from the H$\alpha$ luminosity and assuming solar metallicity. 
We caution that this standard conversion relies on two certainly overly simplistic 
assumptions commonly made, namely that a) star-forming activity is occurring continuously at a constant SFR 
for at least 100 Myr and b) Lyman photon escape is negligible. 
Using the 12 + log(O/H) from Section \ref{sect_abundances}, in Figure \ref{figure_SFR_OH} we show 
the spatially resolved relation between the Log($\Sigma_{SFR}$) and the oxygen abundance 
in UM\,461 and Mrk\,600. 

\textit{UM\,461}. No correlation is found between the SFR and 12 + log(O/H) at spaxel scales in this galaxy. 
However, the spatial distribution of its oxygen abundance shows an extended area 
with a low value (12 + log(O/H) $<$ 7.6) in the southern part of region no. 1, as noted in Section \ref{sect_abundances}. 
The metallicity in this region appears to decrease with increasing distance from cluster no. 2 and it increases 
when approaching cluster no. 3 (Figure \ref{figure_image_campo_Halpha} top right). 
Therefore, the lowest abundance of this off-centre region does not correlate 
with any of the aforementioned star clusters. The same behaviour is observed in the cometary galaxy Tol 65 \citep{Lagos2016}. 

If we assume that UM 461 has recently experienced a single intense 
starburst, or series of starbursts, then the relatively oxygen deficient region could be the result of intensive starburst,
then ejecting part of the pre-enriched gas \citep[][and references therein]{Veilleux2005}.
Since different elements are produced on different time scales\footnote{Oxygen is predominately synthesised in high-mass stars 
($>$8M$_{\odot}$) and subsequently released to the ISM by stellar winds and supernovae explosion. 
While nitrogen is produced by low and intermediate mass stars.}, it is expected that such a sequence of bursts would 
decrease the N/O ratio when massive stars die. 
However, the N/O is quite homogeneous indicating that the outflowing material is uniform and well mixed.
In that case the N/O ratio is unchanging \citep[][]{vanZeeHaynes2006} within the uncertainties.
This interpretation is consistent with the N/O ratio map in Figure \ref{figure_OH_NO_maps}.
The age of the aforementioned clusters are $\sim$1 Myr and $\sim$4 Myr for clusters no. 2 and 3 respectively \citep{Lagos2011}.  
If we assume that the velocity, $v_{exp} = D / t_{exp}$, of the expanding material is constant, we obtain 
an outflow velocity of $\sim$340 km s$^{-1}$ after 1 Myr of expansion. 
This assumes that the radius D of the low metallicity region is an approximation of
the distance from the star cluster no. 2 to the shock front. This scenario is plausible 
and indicates that supernovae (SNe) and stellar outflows, in UM 461, are capable of depleting the surrounding gas 
during the current starburst. In this circumstance metals ejected out of the ISM by supernovae-driven outflows are not 
completely lost \citep{SilichTenorioTagle2001} into the intergalactic medium. 
The presence of broad components in the line profiles of the strongest emission lines would provide evidence of such fast motions 
\citep[e.g.][]{BordaloTelles2011}, but these profiles are not detected in UM\,461.
 
On the other hand interpreting the UM\,461 metal-poor region as the consequence of a recent infall of metal-poor gas 
\citep[e.g.][]{KoppenHensler2005} implies the scatter in Figure \ref{figure_SFR_OH} (left panel) arises because the metal-poor gas 
is not fully dispersed and mixed into the ISM. In this view, the star-formation activity in UM\,461 started recently, in agreement with
the findings by \cite{Lagos2011}. Infalling gas from the outskirts of the galaxy could have triggered this star-formation 
activity \citep[e.g.][]{EktaChengalur2010} as well as diluting the oxygen abundance. 
The latter effect is the most likely to diminish the oxygen abundance, keeping the N/O ratio constant,
since the current star cluster formation efficiency in UM\,461 is very low \citep{Lagos2011}.  

\textit{Mrk 600}. In Figure \ref{figure_SFR_OH} (right panel) we observe a marginal gradient 
of increasing 12 + log(O/H) abundance and SFR, indicating that parcels of gas with higher metallicity 
are the locus of stronger star-forming activity than those found in low metallicity environments. 
However, the Pearson's correlation coefficient between these two quantities is $\sim$0.2. 
The critical value for the two tailed non-directional test (0.02 significance) 
exceeds the Pearson's correlation coefficient, supporting the hypothesis that the variables are not linearly correlated.
We discard infalling metal-poor gas as the trigger for the ongoing starbursts in Mrk 600, since
the ISM is chemically homogeneous as seen in Section \ref{subsect_distri_oxygen}.
Interestingly, the 12 + log(O/H) map based on R23 shows higher values in the area in between
the resolved shells in region no. 1 (see Figure \ref{figure_2D_calibrators}). 
Consequently, there may have been an outflow of oxygen-enriched
gas due to SNe. However, we cannot draw firm conclusions on this because no direct estimations of abundances 
were obtained.

In summary, the dispersion in the Log($\Sigma_{SFR}$) versus 12 + log(O/H) relation, at spaxel scales, 
for both galaxies is relatively high reflecting their star-formation histories. 
Given that dwarfs galaxies have shallow potential wells, both the ejection of metal rich material and 
accretion/interactions have a huge impact on their evolution.
However, the current burst in UM\,461 is unlikely to diminish its metal content, given that star-formation
is inefficient at driving outflows. Therefore, an additional mechanism, such as cold accretion \citep{Keres2005} 
of metal-poor gas, must be at work in order to explain its observed properties and morphology.

\subsection[]{Relationship between neutral and ionised gas}\label{sect_evolution_UM461/62}

\hi\ is highly sensitive to interactions even with minor satellite galaxies \citep{Martinez2009,Scott2014} while 
\hi\ kinematic and morphological perturbations from major interactions 
can remain detectable for between 0.4 and 0.7 Gyr following a tidal interaction \citep[e.g.][]{Holwerda2011}. 
UM\,461 \citep[M$_{HI}$ = 1.71 $\times$ 10$^8$ \msolar;][]{vanZee1998} 
has a near neighbour, UM\,462, which is projected $\sim$17 arcmin  (62 kpc) to the SE,  with a $\Delta$ V$_{optical}$ 
of only 18 \km. A $\sim$ 6 arcmin (22 kpc) \hi\ tail seen extending SE of UM\,461 toward UM\,462 in a VLA\footnote{Very Large Array} 
D--array \hi\ map was originally interpreted as evidence of a recent tidal interaction between the pair 
\citep[][their figure 8a]{Taylor1995}. However, subsequent higher resolution VLA B and C--array \hi\ mapping of UM\,461 
by \cite{vanZee1998} failed to detect this tail, with those authors arguing the earlier apparent \hi\  
tail was probably an artefact produced by solar interference. 
Further evidence against a recent interaction between the pair comes from the regular \hi\  morphology and velocity field 
for UM\,462 \citep[][their figure 10]{vanZee1998}. Additionally, \cite{James2010} found no evidence for significant nitrogen 
or oxygen variations across UM\,462 at the 0.2 dex level. This result shows that from a chemical point of view, if there 
has been a recent interaction,  it is not currently producing significant metallicity deviations or gradients in UM\,462 at large scales.

While the \cite{vanZee1998} \hi\ velocity field for UM\,461 (their figure 9d) shows an overall rotation pattern 
with a NW--SE rotation axis, it also reveals a strong asymmetric warp  at velocities below 1040 \km\  
projected  S and SW of G\hii R no. 1. This highly warped region is referred to hereafter as the ``disturbed \hi\ region".  
The faint broad optical SW tail seen in Figure \ref{figure_image_campo_Halpha} is projected at the western end 
of the disturbed \hi\ region. At the eastern end of the disturbed \hi\ region, directly S of the G\hii R no. 1 
is the location of the anomalously metal-poor gas clump  (Figs. \ref{figure_OH_NO_maps}). 
The v$_r$(H$\alpha$) minimum of $\sim$ 995 \km\ (Figure  \ref{figure_velocity}) is offset slightly further to the SW, 
but still within the disturbed \hi\ region. 
The highest resolution \citep{vanZee1998} \hi\ map ($\sim$ 5 arcsec resolution) also shows the SW side of the \hi\ disk 
is asymmetrically extended into the disturbed \hi\ region. 
The combination of the disturbed \hi\ region's properties
and the VIMOS-IFU data are consistent with the recent infall from the SW of a low mass metal-poor dwarf or \hi\ cloud 
into the region now exhibiting the lowest metallicity, and localised perturbed neutral and ionized gas kinematics. 
We may be observing the impact of an event similar to that in CIG\,85, where it is proposed that a small dwarf 
is in the process of being subsumed into a larger galaxy \citep{Segupta2012}. 
We note that the faint broad optical tail in CIG\,85 is attributed to the interaction (possibly multiple times) 
with a  minor satellite. DDO 68 is another low metallicity dwarf galaxy with evidence of the recent accretion 
of a smaller satellite galaxy \citep{Annibali2016,Sacchi2016}.

In the case of Mrk\,600, we did not find evidence of optical companions using NED. 
However, \cite{Noeske2005} suggest that the distribution of the star-forming knots in this galaxy may be 
the result of an ongoing or recent interaction. Noeske et al. argue that the U--B colours of -0.64, -0.72 and -0.86 
of our regions no. 1 and no.2 and their region c, respectively, suggest  propagating star-formation activity,
while the colours of the underlying stellar component are indicative of a population of several Gyr old. 
However, they failed to detect an optical or near-IR counterpart to  \hi\ companion detected $\sim$ 1.25 arcmin SW 
of Mrk\,600  by \cite{Taylor1993} in their VLA D--array \hi\ map. The companion's reported M(H\,{\sc i}) 
was 2.2 $\times$ 10$^7$ M$_{\odot}$, i.e. $\sim$ 10\% of the  Mrk\,600 M(H\,{\sc i}), and it has a maximum column density of 
$\sim$1.5 $\times$ 10$^{20}$ atoms cm$^{-2}$ \citep{Taylor1993}. 
Nevertheless, the higher resolution VLA C--array \hi\ map \citep{Taylor1994} does not show a separate structure 
at the position of the preciously reported \hi\ companion. The velocity field from VLA C--array observations for Mrk\,600 
revealed an  overall, although rather irregular, \hi\ rotation patten. 
If future \hi\ observations confirm the \hi\ companion this will indicate Mrk \,600 is at an earlier stage of accreting  
a \hi\ cloud with a significant mass.

\subsection[]{Gas metallicity}\label{sect_model}
Minor mergers or interactions could potentially  provide a supply of infalling gas  and the energy 
transfer to drive the internal motions of the parent galaxy. In this Section we will analyse this scenario 
in the context of the chemical evolution of UM\,461 and Mrk\,600. 
In a closed box model, the gas metal mass fraction Z \citep{Schmidt1963,SearleSargent1972} is determined entirely 
by the yield ($y$) and gas fraction $f_{gas}$ = M$_{gas}$/(M$_{gas}$+M$_{stars}$) as

\begin{equation}
Z_{gas} = y \times ln(f_{gas}^{-1}).
\end{equation}

\begin{figure}
\includegraphics[width=80mm]{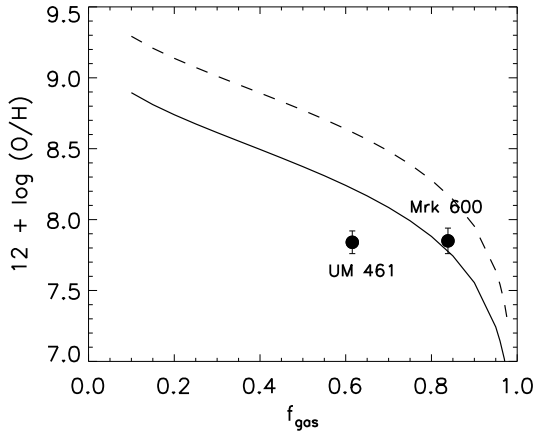}
\caption{Relation between 12 + log(O/H) as function of gas fraction $f_{gas}$ assuming a closed-box model assuming
a true yield $y_{O}$ = 0.004 \citep[solid line;][]{Dalcanton2007} and $y_{O}$ = 0.01 \citep[dashed line;][]{Tremonti2004},
typical oxygen yields for star-forming galaxies. Data points correspond to our measured values for UM 461 and Mrk 600.
}
\label{figure_OH_fgas}
\end{figure}

If we express Z$_{gas}$ in terms of the oxygen abundance we obtain:

\begin{equation}
12 + log(O/H) = 12 + log (y_{O}/11.728) + log (ln(f_{gas}^{-1})),
\end{equation}

where $y_O$ is the oxygen yield by mass and 11.728 is the factor to convert abundance by mass to abundance 
by number \citep{Lee2006}. Here, we consider M$_{gas}$ = 1.24$\times$M$_{HI}$ and the true yield 
log($y_{O}$) = -2.4 \citep{Dalcanton2007}. Therefore, we find that the measured oxygen abundance in UM\,461 is 
$\Delta$(O/H) $\sim$ 0.38 dex lower than the expected value assuming a closed box model, while 
in Mrk 600 the oxygen abundance is $\Delta$(O/H) $\sim$ 0.07 dex higher.
In terms of the effective yield $y_{eff}$ = Z$_{gas}$/ln($f_{gas}^{-1}$) (Z$_{gas}$ = 12$\times$O/H), we found that 
log($y_{eff}$) = -2.77 and -2.32 for UM 461 and Mrk 600, respectively.
In Figure \ref{figure_OH_fgas} we show 12 + log(O/H) as a function of the gas fraction assuming a closed box model. 
The data points, in the same Figure, correspond to the measured values of UM 461 and Mrk 600.  
Note that, in the case of Mrk\,600, if we assume a true yield  $y_{O}$=0.01 \citep[]{Tremonti2004}, the data point 
is well explained, within the uncertainties, by the closed box model. 
However, the effective yield of UM\,461 remains  lower than the closed box model when we use either true yield prescription.

It is assumed that deviations from the closed box model indicate the presence of an outflow and/or inflow.
In principle, we cannot rule out any of those mechanisms to decrease the effective yield found 
in UM 461 and the slightly higher yield in Mrk 600. However, and according to our previous analysis, 
Mrk 600 is currently less affected by outflows or inflows, which makes it well explained by a closed box model. Mrk 600  
also presents a flat metallicity gradient within the uncertainties.
In the case of UM 461, the low metallicity region, with 12 + log(O/H) $\lesssim$ 7.6, could be the result of inflow of metal-poor gas. 
According to \cite{Thuan2016}, the deviations in the effective yields can be understood as a gas outflow,
in which a high fraction of metal-enriched gas is lost, and/or inflow of 
metal-poor gas in objects where $y_{eff} \lesssim y_{true}$ and a relatively metal-free H\,{\sc i} envelope for objects with 
$y_{eff} \gtrsim y_{true}$.
Therefore, the deviation from the closed-box model in UM 461 can be explained as the result 
of the competing effects of starburst driven-outflows \citep[e.g.][]{Tremonti2004} and the inflow of metal-poor gas. 
However, the latter is the most likely factor to explain the low effective yields observed in UM\,461, because
starburst driven-outflows are unlikely to be effective in removing large amounts of gas from the disk in low-mass galaxies 
\citep[e.g.][]{Dalcanton2007}, as discussed in Section \ref{sect_star-formation}. 
In this scenario, during the infall the oxygen abundance is reduced 
due to dilution of the pre-existing gas, without affecting the log(N/O) ratio, followed by the evolution of the system towards 
the closed-box relation \citep{KoppenHensler2005}.

The idea of an infall of pristine gas is unlikely to explain the effective yield of UM\,461 because 
this infalling gas would increase its value significantly \citep{Thuan2016}. 
Alternatively, based on its high H\,{\sc i} mass, the infall of metal-poor clouds \citep[$\sim$10$^7$M$_{\odot}$;][]{Verbeke2014} 
towards the centre could produce the observed low metallicity region in UM 461. 
H\,{\sc i} clouds have previously been found in the surroundings of some BCD galaxies \citep[e.g.][]{Thuan2004,Lelli2014}.
It seems that the H\,{\sc i} companion to Mrk 600, if it really exists, has not yet been accreted into the galactic disk 
and we speculate that it would available to fuel a future starburst episodes and  produce temporary metallicity 
and ionized gas kinematic inhomogeneities in the galaxy's disk.

If UM\,461 had been tidally disrupted due to an interaction(s) with UM\,462, this may have promoted an efficient 
flattening of the metallicity gradient and dilution by low-metallicity gas infalling into the galaxy centre. 
Interestingly, the difference in oxygen abundance between UM 461 (7.84 dex; this work) and UM 462 \citep[8.03 dex;][]{James2010} 
is 0.19 dex. This argues in favour of a coeval evolution of this pair. In fact, the current star-formation episodes 
in both galaxies are very young, not older than a few Myr with most of their underlying stellar populations formed
$\sim$1 Gyr ago \citep{Lagos2011}. See \cite{Lagos2011} and \cite{Vanzi2003}
for a detailed study of the star cluster population in UM 461 and UM 462, respectively.
However, the formation process of cometary galaxies near and far is unclear and different mechanisms may be at work 
during the evolution of those systems, i.e., propagating star-formation 
in local XMP BCDs \citep{Papaderos2008}, infall of metal-poor gas \cite[e.g.][]{EktaChengalur2010,Verbeke2014} and 
interactions \citep[e.g.][]{Noeske2001,Pustilnik2001}. If the accretion/inflow of gas is the main mechanism 
to trigger star-formation in cometary-like galaxies, this implies that both galaxies, in this study, are 
at different evolutionary stages.

By using IFU spectroscopy we have been able to investigate the chemical homogeneity in star-forming dwarf galaxies 
\citep[e.g.][]{LagosPapaderos2013,Lagos2014,Lagos2016}, obtaining precise abundance determinations 
in a sample of objects with clear detections of [O\,{\sc iii}]$\lambda$4363 line emission. 
The detection of chemical inhomogeneities in XMP BCDs using the direct method, likely produced 
by the infall of metal-poor gas-clouds onto the ISM disc, is a key component for the study of the chemical evolution of those systems.
Motivated by these results, we will explore with future IFU and H\,{\sc i} observations the relations 
among various spatially resolved quantities (e.g. star-formation, kinematics, abundances, etc).
This should give us insight into whether the infall of metal-poor gas-clouds is responsible for the detected 
low metallicity regions in some of those systems.

\section{Summary and Conclusions}\label{conclusions}

In this paper, we have analysed the ISM of the H\,{\sc ii}/BCD galaxies UM 461 and Mrk 600 using VIMOS-IFU spectroscopy. 
The following  points summarise the main results in this work:

\begin{itemize}
 
 \item We obtained integrated oxygen abundances, using the direct method, 12 + log(O/H) = 7.84 and 7.85 
       for UM\,461 and Mrk\,600, respectively. 
       We found a marginal difference between those integrated abundances and the ones found in the GH\,{\sc ii}Rs for both galaxies.
       Therefore, within the uncertainties we can consider that the oxygen abundance is fairly well mixed at large scales.  
       In Figure \ref{figure_OH_NO_maps} (left pannels) we showed the spaxel by spaxel
       12 + log(O/H) maps of the galaxies using the direct method.
       We note that in the case of Mrk 600 the distribution 
       of oxygen abundances from the spaxels can be fitted by a single Gaussian. While for UM\,461 
       the distribution is well fitted by two Gaussians (see Figure \ref{figure_OH_dist}). 
       The mean values of both distributions agree with the integrated ones indicating that, at large scales,
       the ISM is chemically homogeneous.
       However, we found evidences of an off-centre low metallicity region, located in the southern part of region no. 1
       in UM 461\,. This area has an extension of $\sim$0.7 kpc and a mean value of 12 + log(O/H) $\sim$ 7.52. 
       Whereas, Mrk 600, like other previously studied star-forming dwarf galaxies, is chemically homogeneous 
       \citep[see][]{LagosPapaderos2013}.

\item  We use BPT diagnostic diagrams to study the excitation conditions in both galaxies. We found
       that all points fall in the locus predicted by models of photo-ionization by young stars in H\,{\sc ii} regions  
       indicating that photoionization from stellar sources is the dominant excitation mechanism in UM 461 and Mrk 600.     
           
 \item We checked the spatial variation of 12 + log(O/H) abundances in both galaxies 
       using several calibrators (N2, O3N2, R23 and D2016), including the widely used  HII--CHI--mistry code. 
       The oxygen abundance maps N2 and O3N2 show spatial trends that are opposite to those shown by the direct method and 
       R23. The shape of R23 agrees with the direct method, indicating only a small dependence on $U$ as compared with N2 and O3N2,
       where this dependence could be significant as proved in the literature \citep[e.g.][]{KewleyDopita2002}. 
       The D2016 maps do not correlate with any of the previous determinations.
       Applying the  HII-CHI-mistry code \citep{Perez-Montero2014} to our spatially resolved data provides
       values for oxygen abundance consistent with the direct method only when the [O\,{\sc iii}]$\lambda$4363 emission line 
       is included as an input. Therefore, special attention must be paid when this method is used in studying 
       the real spatial variation of abundances across galaxies.

\item  In both objects we find that the \halpha\  velocity field (v$_r$) shows systemic motions. 
       But in the case of UM 461 the combination of a region of disturbed \hi\  \citep{vanZee1998} and our 
       VIMOS-IFU velocity fields at the same location are consistent with 
       the recent infall from the SW of a low mass metal-poor dwarf or \hi\ cloud 
       into the region now exhibiting the lowest metallicity and localised perturbed neutral and ionized gas kinematics. 
       
\item  The dispersion in the Log($\Sigma_{SFR}$) versus 12 + log(O/H) relations in our galaxies is quite high and
       no correlation is found at spaxel scales. 
       The effective yield for UM\,461 is lower than predicted by the closed box model. 
       This deviation from the model indicates the presence of inflows of non-pristine metal-poor gas which could explain 
       the region of anomalously low metallicity in this galaxy. 
       Whereas, the effective yield in Mrk\,600 is well explained, within the uncertainties, by the closed-box model. 
       
\end{itemize}

In summary, the spatially resolved properties of the galaxies are consistent with these systems being at different stages 
of accreting low metallicity \hi\ gas-clouds to their stellar disks. 
Therefore, the detection of a low-metallicity region with 12 + log(O/H) $<$ 7.6 in the brightest 
H\,{\sc ii} region of UM\,461 indicate that the low-metallicity cloud has been ``recently'' accreted.
Moreover, if the \hi\ companion to Mrk\,600 really exists, it has not yet been accreted into the galactic disk.

\section*{Acknowledgments}

Based on observations made with ESO Telescopes at the La Silla Paranal Observatory under programme 090.B-0242.
We thank the reviewer for his/her careful reading of the manuscript and helpful comments.
P.L. would like thanks to Enrique P\'erez Montero for his very useful comments.
P.L. acknowledges support by the Funda\c{c}\~{a}o para a Ci\^{e}ncia e a Tecnologia (FCT)
(Portugal) through the grant SFRH/BPD/72308/2010. 
T.S. acknowledges support for this project from FCT grant No. SFRH/BPD/103385/2014.
A.H. acknowledges FCT grant SFRH/BPD/107919/2015.
R.D. gratefully acknowledges the support provided by the BASAL Center for Astrophysics and Associated Technologies (CATA).
P.P. was supported by FCT through Investigador FCT contract IF/01220/2013/CP1191/CT0002.
This work was supported by FCT through national funds and by FEDER through COMPETE by the grants UID/FIS/04434/2013 
\& POCI-01-0145-FEDER-007672 and PTDC/FIS-AST/3214/2012 \& FCOMP-01-0124-FEDER-029170. 
We acknowledge support by European Community Programme ([FP7/2007-2013]) under grant agreement No. PIRSES-GA-2013-612701 (SELGIFS).
This research has made use of the NASA/IPAC Extragalactic Database (NED) which is operated 
by the Jet Propulsion Laboratory, California Institute of Technology, under contract with 
the National Aeronautics and Space Administration.


\end{document}